\documentclass[aps,reprint,twocolumn,showpacs,preprintnumbers,amsmath,amssymb,nofootinbib,superscriptaddress,showkeys]{revtex4-1}

\usepackage{bm}
\usepackage{epsfig}
\usepackage{slashed}
\usepackage{mathtools}

\usepackage{color}

\usepackage[utf8]{inputenc}

\begin{document}

\title{Heavy Baryon-Antibaryon Molecules in Effective Field Theory}
 \author{Jun-Xu Lu}

 \affiliation{School of Physics and Nuclear Energy Engineering, \\
   International Research Center for Nuclei and Particles in the Cosmos and \\
   Beijing Key Laboratory of Advanced Nuclear Materials and Physics, \\
   Beihang University, Beijing 100191, China
 }
 \affiliation{Institut de Physique Nucl\'eaire, CNRS-IN2P3, Univ. Paris-Sud, Universit\'e Paris-Saclay, F-91406 Orsay Cedex, France}
 
 \author{Li-Sheng Geng}\email{lisheng.geng@buaa.edu.cn}
 
 \author{Manuel Pavon Valderrama}\email{mpavon@buaa.edu.cn}
 \affiliation{School of Physics and Nuclear Energy Engineering, \\
International Research Center for Nuclei and Particles in the Cosmos and \\
Beijing Key Laboratory of Advanced Nuclear Materials and Physics, \\
Beihang University, Beijing 100191, China
}

\date{\today}

\begin{abstract}
\rule{0ex}{3ex}
We discuss the effective field theory description of bound states
composed of a heavy baryon and antibaryon.
This framework is a variation of the ones already developed
for heavy meson-antimeson states to describe the $X(3872)$
or the $Z_c$ and $Z_b$ resonances.
We consider the case of heavy baryons for which the light quark pair is
in S-wave and we explore how heavy quark spin symmetry constrains
the heavy baryon-antibaryon potential.
The one pion exchange potential mediates the low energy dynamics
of this system.
We determine the relative importance of pion exchanges, in particular
the tensor force.
We find that in general pion exchanges are probably non-perturbative
for the $\Sigma_Q \bar{\Sigma}_Q$, $\Sigma_Q^* \bar{\Sigma}_Q$
and $\Sigma_Q^* \bar{\Sigma}_Q^*$ systems, while for 
the $\Xi_Q' \bar{\Xi}_Q'$, $\Xi_Q^* \bar{\Xi}_Q'$ and
$\Xi_Q^* \bar{\Xi}_Q^*$ cases they are perturbative.
If we assume that the contact-range couplings of the effective field theory
are saturated by the exchange of vector mesons, we can estimate
for which quantum numbers it is more probable to find
a heavy baryonium state.
The most probable candidates to form bound states are
the isoscalar $\Lambda_Q \bar{\Lambda}_Q$, $\Sigma_Q \bar{\Sigma}_Q$,
$\Sigma_Q^* \bar{\Sigma}_Q$ and $\Sigma_Q^* \bar{\Sigma}_Q^*$ and 
the isovector $\Lambda_Q \bar{\Sigma}_Q$ and $\Lambda_Q \bar{\Sigma}_Q^*$
systems, both in the hidden-charm and hidden-bottom sectors.
Their doubly-charmed and -bottom counterparts ($\Lambda_Q {\Lambda}_Q$,
$\Lambda_Q {\Sigma}_Q^{(*)}$, $\Sigma_Q^{(*)} {\Sigma}_Q^{(*)}$)
are also good candidates for binding.
\end{abstract}

\pacs{03.65.Ge,13.75.Lb,14.40.Lb,14.40.Nd,14.40Pq,14.40Rt}

\maketitle

\section{Introduction}

Heavy hadron molecules -- bound states composed of heavy hadrons --
are a type of exotic hadron.
The theoretical basis for their existence is robust:
in analogy with the nuclear forces that bind the nucleons,
heavy hadrons can exchange light mesons,
generating exchange forces that might be strong enough to bind them~\cite{Voloshin:1976ap,Tornqvist:1991ks,Manohar:1992nd,Ericson:1993wy,Tornqvist:1993ng}.
The discovery of the $X(3872)$ more than a decade ago~\cite{Choi:2003ue}
probably provides the most paradigmatic candidate for a molecular state.
The $X(3872)$ turns out not to be alone: a series of similarly puzzling
hidden charm (hidden bottom) states that do not fit in the charmonium
(bottomonium) spectrum have been found in different experiments
since then.
They are usually referred to as XYZ states and a few are particularly
good candidates for molecular states.
In the hidden charm sector we have
the $Z_c(3900)$, $Z_c(4020)$~\cite{Ablikim:2013mio,Liu:2013dau}
which are suspected to be $D\bar{D}^*$, $D^*\bar{D}^*$
molecules~\cite{Wang:2013cya,Guo:2013sya},
and the $P_c(4380)^{+}$ and $P_c(4450)^{+}$
pentaquark states~\cite{Aaij:2015tga},
which might contain $\bar{D} \Sigma_c^*$, $\bar{D}^* \Sigma_c$,
$\bar{D}^* \Sigma_c^*$ and even $\bar{D} \Lambda_{c}(2590)$
molecular components~\cite{Chen:2015loa,Chen:2015moa,Roca:2015dva,He:2015cea,Xiao:2015fia,Burns:2015dwa,Geng:2017hxc}.
In the hidden bottom sector we have the $Z_b(10610)$ and
$Z_b(10650)$~\cite{Belle:2011aa,Adachi:2012im},
which might be $B\bar{B}^*$, $B^*\bar{B}^*$
molecules~\cite{Voloshin:2011qa,Cleven:2011gp}.
If we consider the open charm sector, 
the $D_{s0}(2317)$ and $D_{s1}(2460)$ mesons~\cite{Aubert:2003fg,Besson:2003cp}
were discovered before the $X(3872)$ and have been theorized to have a large
$D K$/$D^* K$ molecular component~\cite{Guo:2006fu,Guo:2006rp,Guo:2011dd,Altenbuchinger:2013vwa}.

We expect molecular states to be relatively narrow for states happening
above the open charm threshold.
For the moment the mass of the experimentally discovered states have
reached the heavy meson-meson and heavy meson-baryon threshold
($3.7$ and $4.1\,{\rm GeV}$/$4.3\,{\rm GeV}$ for $D\bar{D}$ and
$\Lambda_c\bar{D}$/$\Sigma_c\bar{D}$ respectively),
but barely the heavy baryon-baryon threshold
($4.5$, $4.7$, $4.9\,{\rm GeV}$ for $\Lambda_c \bar{\Lambda}_c$,
$\Lambda_c \bar{\Sigma}_c$ and $\Sigma_c \bar{\Sigma}_c$).
A narrow resonance near the heavy baryon-baryon threshold
would be an excellent candidate for a heavy baryon-antibaryon bound state.
Though these states have not been found yet, it is fairly straightforward
to extend the available descriptions of heavy meson-antimeson molecules
to them and explore the relevant dynamics behind these states.
In a few instances it might be possible to predict the location of heavy
baryonium states, with the $\Lambda_{c}(2590) \bar{\Sigma}_c$ systems being
an illustrative example~\cite{Geng:2017jzr}.

Heavy hadron-antihadron molecules are among the most interesting theoretical
objects of hadronic physics.
Owing to their heavy-light quark content, they are simultaneously subjected
to isospin, SU(3)-flavour, chiral and heavy quark symmetry,
a high degree of symmetry that can translate
into a fairly regular spectrum~\cite{AlFiky:2005jd,Voloshin:2011qa,Mehen:2011yh,Valderrama:2012jv,Nieves:2012tt,HidalgoDuque:2012pq,Guo:2013sya,Guo:2013xga}.
This spectrum will not be fully realized in nature: unless these states
are shallow they will be a mixture of molecule, charmonium and other
exotic components.
Yet these potential regularities in the molecular spectrum can be successfully
exploited to uncover the nature of a few of the XYZ states.
The most clear example probably is the $Z_c$'s and $Z_b$'s resonances,
which seem to be related by different realizations of
heavy quark symmetry~\cite{Guo:2013sya}.

Heavy hadron molecules possess another interesting quality:
they show a separation of scales.
On the one hand we have the size of the hadrons, which is of the order of
$0.5\,{\rm fm}$, while on the other we have the size of
the bound state, which should be bigger than
the individual hadrons within it.
As a consequence heavy hadron molecules are amenable to
an effective field theory (EFT) treatment,
where all quantities can be expressed as an expansion
of a light over a heavy energy scale.
EFT descriptions of heavy hadron molecules have been exploited successfully
in the past specifically to systems composed of
heavy mesons and antimesons~\cite{Braaten:2003he,Fleming:2007rp,Canham:2009zq,Mehen:2011yh}.
In this manuscript we extend the heavy hadron EFT formulated
in Ref.~\cite{Valderrama:2012jv} and put in use
in Refs.~\cite{Nieves:2012tt,HidalgoDuque:2012pq,Guo:2013sya}
to the case of the heavy baryon-antibaryon molecules.
As commented, these type of molecules might very well be discovered
in the next few years.
The purpose of this work is to explore the symmetry constrains and
the kind of EFT that is to be expected in these systems,
rather than to make concrete predictions of
the possible location of these states.
Yet we will speculate a bit about this later issue on the basis of
the relative strength of the long range pion exchange and
the saturation of the EFT low energy constants
by $\sigma$, $\rho$, $\omega$ and $\phi$ meson exchange.

The manuscript is structured as follows:
in Section \ref{sec:EFT} we make a brief introduction to the EFT formalism.
In Section \ref{sec:LO} we present the leading order EFT potential
for heavy baryon-antibaryon states, which consists of a series
of contact four-baryon vertices plus the time-honoured
one pion exchange potential.
In Section \ref{sec:perturbative} we explore the question of whether
pions are perturbative or not for this type of hadron molecule.
In Section \ref{sec:pc} we discuss the possible power countings to describe
molecular states.
In Section \ref{sec:pred} we speculate about which heavy baryon-antibaryon
molecules might be more probable.
Finally in Section \ref{sec:conclusions} we present our conclusions.
In Appendix \ref{app:ope} we present the complete derivation of
the one pion exchange potential, in Appendix \ref{app:OME} we
briefly explain the one eta and one kaon exchange potential
and in Appendix \ref{app:contact} we derive the heavy quark
symmetry constrains for the four-baryon contact vertices.

\section{Effective Field Theory for Heavy Baryon Molecules}
\label{sec:EFT}

Effective field theories (EFTs) are generic and systematic descriptions
of low energy processes.
They can be applied to physical systems in which there is a distinct
separation of scales, but where the underlying high energy theory
for that system is unknown or unsolvable.
Hadronic molecules are a good candidate for the EFT treatment:
the separation among the hadrons forming a hadronic molecule
is expected to be larger than the size of the hadrons.
When the hadrons are close to each other they overlap and
the ensuing description in terms of quantum chromodynamics
(QCD) is unsolvable.
But this is not the case when the hadrons are far away, in which case
their interactions can be described in terms of well-known physics
such as pion exchanges.
In the following lines we will present a brief introduction to
the application of the EFT framework to heavy baryon-antibaryon systems.

\subsection{The Effective Field Theory Expansion}

EFTs rely on the existence of a separation of scales, where a distincion is
made between low and high energy physics and their respetive characteristic
momentum scales $M_{\rm soft}$ and $M_{\rm hard}$, which are sometimes called
the soft (or light) and hard (or heavy) scales.
The separation of scales can be used to express physical quantities
at low energies as expansions in terms of the small parameter
$M_{\rm soft} / M_{\rm hard}$.
If we consider a system of heavy baryons for concreteness,
there are two possible EFT expansions depending on which
type of low energy symmetry we are considering:
\begin{itemize}
\item[(i)] heavy quark spin symmetry (HQSS) and
\item[(ii)] chiral symmetry.
\end{itemize}
For HQSS the soft and hard scales are $\Lambda_{\rm QCD} \sim 200\,{\rm MeV}$
and the mass of the heavy quark $m_Q$, which is either
$m_c \sim 1.5\,{\rm GeV}$ or $m_b \sim 4.5\,{\rm GeV}$. 
For chiral symmetry we call the soft and hard scales $Q$ and $M$, where
if there are no baryons we have $Q \sim m_{\pi} \sim q \sim 100-200\,{\rm MeV}$
(with $m_{\pi}$ the pion mass and $q$ the momenta of the pions)
and $M \sim 2\pi f_{\pi} \sim 1\,{\rm GeV}$.
If there are baryons $Q$ includes the soft momenta of the baryons,
while practical calculations in the two-baryon sector suggest
a more conservative value of $M \sim 0.5-1.0\,{\rm GeV}$
for the hard scale.
We advance that the scale $Q$ can contain more than the pion mass and
the momenta of the pions and baryons, as we will discuss
in Sect.~\ref{subsec:bound-pc}.

Heavy baryons are non-relativistic at the soft scales of either of the two
previous symmetries. This implies that their interactions can be described
in terms of an effective potential $V_{\rm EFT}$, which admits
the double expansion
\begin{eqnarray}
  V_{\rm EFT} = \sum_{\mu, \nu} \hat{V}^{(\mu, \nu)}\,
  {\left(\frac{\Lambda_{\rm QCD}}{m_Q}\right)}^{\mu}\,
  {\left(\frac{Q}{M}\right)}^{\nu} \, ,
  \label{eq:V-double-expansion}
\end{eqnarray}
where the indexes $\mu$ and $\nu$ indicate the order in the heavy quark and
chiral expansion, respectively, with $\mu \geq 0$ and $\nu \geq -1$
(this second point will be explained in Sect.~\ref{subsec:bound-pc}).
The HQSS expansion converges remarkably faster than the chiral expansion,
owing to the sizes of the soft and hard scales involved
in each of these expansions.
For this reason from now on we will work in the $m_Q \to \infty$ limit
and ignore any HQSS breaking effect.
With this in mind, the expansion of the EFT potential simplifies to
\begin{eqnarray}
  V_{\rm EFT} &=& \sum_{\nu = \nu_0}^{\nu_{\rm max}} V^{(\nu)} +
  \mathcal{O}{\left(\frac{Q}{M}\right)}^{\nu_{\max} +1} 
  \, ,
  \label{eq:V-chiral-expansion}
\end{eqnarray}
which converges for $Q < M$ and where we have simplified the notation
with respect to Eq.~(\ref{eq:V-double-expansion}).
The expansion begins at $\nu = \nu_0 \geq -1$ and we truncate
it at $\nu = \nu_{\rm max}$,, where the truncation error gives
the uncertainty of a calculation.
The lowest order $\nu = \nu_0$ is referred to as the leading order (LO).
For a two heavy baryon system the scale $Q$ includes the external momenta of
the hadrons, the pion mass and the binding momentum of
a potential bound state if there is any.
The rules by which we decide the order of each contribution
are called {\it power counting}.

The degrees of freedom of the EFT we are constructing are the heavy baryons
and the pion fields (or the pseudo Nambu-Goldstone boson fields
if we consider SU(3) chiral symmetry).
This choice of light degrees of freedom actually implies
that the EFT potential can be decomposed
into two different contributions
\begin{eqnarray}
  V_{\rm EFT} = V_{C} + V_{F} \, ,
\end{eqnarray}
where $V_C$ and $V_F$ are the contact-range and finite-range potentials.
While $V_C$ only involves direct interactions between the heavy baryon fields
(thus its contact-range nature), $V_F$ involves the exchange of pions and has
a finite-range determined by the inverse of the mass of the pion.
We can expand $V_C$ and $V_F$ according to power counting
\begin{eqnarray}
  V_{C} &=& \sum_{\nu = \nu_{0}(C)}^{\nu_{\rm max}} V_C^{(\nu)} +
  \mathcal{O}{\left(\frac{Q}{M}\right)}^{\nu_{\max} +1} \, , \\
  V_{F} &=& \sum_{\nu = \nu_{0}(F)}^{\nu_{\rm max}} V_F^{(\nu)} +
  \mathcal{O}{\left(\frac{Q}{M}\right)}^{\nu_{\max} +1} \, ,
\end{eqnarray}
where the power counting of $V_C$ and $V_F$ can differ:
we introduce $\nu_0(C)$ and $\nu_0(F)$ to indicate that
the expansion may begin at different orders.

For concreteness we will temporarily consider that (chiral) power counting is
given by naive dimensional analysis (NDA).
We warn that NDA is incompatible with the existence of bound states,
but we will address this problem later.
Within NDA the power counting of a contribution to the potential is determined
by the powers of the heavy baryon momenta, pion momenta and pion masses
included in a particular contribution.
In NDA the order of the LO contribution to the potential is $\nu = 0$.
For the contact-range potential this LO contribution is
a momentum- and energy-independent interaction
\begin{eqnarray}
  \langle \vec{p}' | V_{C}^{(0)} | \vec{p} \rangle = C^{(0)} \, ,
  \label{eq:V_C_LO}
\end{eqnarray}
with $C^{(0)}$ a coupling, which will depend on the quantum numbers of
the two-body system under consideration and will involve
spin and isospin operators.
The LO piece of the finite-range potential is given by one pion exchange (OPE),
which we write (again schematically) as
\begin{eqnarray}
  \langle \vec{p}' | V_{F}^{\rm LO} | \vec{p} \rangle = 
  F^{(0)}\,\vec{I}_1 \cdot \vec{I}_2\,
  \frac{\vec{a}_1 \cdot \vec{q}\,\vec{a}_2 \cdot \vec{q}}
  {q^2 + m_{\pi}^2}
  \, ,
  \label{eq:V_F_LO}
\end{eqnarray}
with $\vec{I}_1$ and $\vec{I}_2$ the isospin operators and
$\vec{a}_1$ and $\vec{a}_2$ the spin operators of the heavy baryons.
The couplings $C^{(0)}$ and $F^{(0)}$ have dimensions of $[{\rm energy}]^{-2}$,
which means that in NDA their size is given by~\footnote{
  Modulo numerical factors, which for non-relativistic scattering
  will be multiples of $4 \pi$. As we are focusing on the scaling
  only, we do not include these factors explicitly.
}
\begin{eqnarray}
  C^{(0)} \sim \frac{1}{M^2} \quad , \quad F^{(0)} \sim \frac{1}{M^2} \, ,
  \label{eq:NDA}
\end{eqnarray}
for which we have taken into account that the only scale
from which we can construct the couplings is $M$
(otherwise the counting of the LO potential will change and
we will not be talking about NDA).

Other possible contributions to $V_C$ and $V_F$ appear at
higher order in the EFT expansion.
The subleading terms in $V_C$ are derivative contact-range interactions,
i.e. they involve positive powers of the external heavy baryon momenta.
The subleading terms in $V_F$ include multi-pion exchanges,
by which it is meant irreducible diagrams involving
the exchange of two or more pions.
Here it is important to notice that iterations of the OPE potential indeed
involve the exchange of two or more pions, but these diagrams
are not irreducible: reducible multi-pion exchanges (iterated OPE)
will be lower order than irreducible multi-pion exchanges.

We will not consider the subleading terms in this work: subleading
interactions, in particular the contact-range ones, involve new
free parameters which require additional data to be determined.
These additional data are not expected to be experimentally available
in the near future.

\subsection{Bound States and Power Counting}
\label{subsec:bound-pc}

Power counting is not unique. NDA is simply the most obvious choice
for building a power counting, but not the only one.
In particular the existence of bound states requires modifications to
the power counting~\cite{Kaplan:1998tg,Kaplan:1998we,vanKolck:1998bw,Birse:1998dk},
as we will illustrate below.
Bound states are {solutions} of a dynamical equation, such as Schr\"odinger or
Lippmann-Schwinger, and require non-perturbative physics.
We can see this from the Lippmann-Schwinger equation as applied to bound states
\begin{eqnarray}
| \Psi_B \rangle = G_0 V | \Psi_B \rangle \, ,
\end{eqnarray}
where $| \Psi_B \rangle$ is the bound state wave function
and $G_0 = {1}/{(E - H_0)} $ the resolvent operator,
with $H_0$ the free hamiltonian.
When $G_0$ appears in loops it is counted as $Q$
\begin{eqnarray}
  \int\,\frac{d^3 \vec{l}}{(2 \pi)^3}\,\frac{1}{E - \frac{l^2}{2 \mu}} \sim \
  \mu\,Q \, ,
\end{eqnarray}
where $\mu$ refers to the reduced mass.
Generating a bound state requires the iteration of the potential,
which in terms of power counting is only consistent if
\begin{eqnarray}
\mathcal{O}(V) = \mathcal{O}(V G_0 V) = \mathcal{O}(V G_0 V G_0 V) = \dots \, ,
\end{eqnarray}
from which $V \sim Q^{-1}$ is required.
To explain why the potential is counted this way we have to revisit
the estimations of the size of the couplings $C^{(0)}$ and $F^{(0)}$
contained in Eq.~(\ref{eq:NDA}).
If any of these two couplings contains a light scale
\begin{eqnarray}
  C^{(-1)} \sim \frac{1}{M Q} \quad \mbox{and/or}
  \quad F^{(-1)} \sim \frac{1}{M Q} \, ,
\end{eqnarray}
the LO potential will be promoted from $Q^0$ to $Q^{-1}$,
allowing for the existence of bound states~\footnote{We mention in passing
that the promotion can also be understood in terms of the {\it anomalous}
dimension of the coupling $C_0$, i.e. to its scaling with respect
to the cut-off~\cite{Valderrama:2014vra,Valderrama:2016koj}.}.
The light momentum scale that appears in $C^{(-1)}$ can be identified
with the inverse scattering length of the two-body system~\cite{Kaplan:1998tg,Kaplan:1998we,vanKolck:1998bw,Birse:1998dk},
while the light scale in $F^{(-1)}$ is related
with the strength of the OPE potential~\cite{Birse:2005um}.
We stress that it is enough to promote one of the two couplings $C^{(0)}$
and $F^{(0)}$ from $Q^0$ to $Q^{-1}$, where for a more detailed discussion
we refer to Sect.~\ref{sec:pc}. 

\subsection{Coupled Channels}

Now we consider the power counting of coupled channel effects.
Heavy baryons can come in HQSS multiplets which are degenerate
in the heavy quark limit, e.g. the $\Sigma_c$ and the $\Sigma_c^*$
in the charm sector or the $\Sigma_b$ and the $\Sigma_b^*$
in the bottom sector.
If we take the $\Sigma_c$ and the $\Sigma_c^*$ heavy baryons as an example,
the two heavy baryon system can have transitions of
the type $\Sigma_c \Sigma_c \to \Sigma_c \Sigma_c^*$,
$\Sigma_c \Sigma_c \to \Sigma_c^* \Sigma_c^*$, etc.
In EFT these transitions have a characteristic momentum scale
\begin{eqnarray}
  \Lambda_{\rm CC} = \sqrt{2\mu \Delta_{\rm CC}} \, ,
\end{eqnarray}
with $\mu$ the reduced mass of the system and $\Delta_{\rm CC}$
the energy difference of the transition.
Coupled channel effects can be argued to be suppressed by a factor of
\begin{eqnarray}
  {\left( \frac{Q}{\Lambda_{\rm CC}} \right)}^2 \, ,
\end{eqnarray}
with $\Lambda_{\rm CC}$ the coupled channel scale.
For the $\Sigma_c \Sigma_c$ family of systems we have that
$\Lambda_{\rm CC} = 400 / 564\,{\rm MeV}$ depending on the transition,
while for the $\Sigma_b \Sigma_b$ case we have
$\Lambda_{\rm CC} = 350 / 495\,{\rm MeV}$.
This scale is softer than $M$, but not much softer:
we can effectively ignore the coupled channels at the price of
reducing the range of applicability of the EFT.

That is, there are two choices for constructing the EFT in this case:
(i) consider the coupled channel effects to be subleading (at the price
of reducing the convergence radius of the EFT),
(ii) include them at leading order.
Here we will opt for the first option, owing to its simplicity.
Be it as it may, most of the results of this manuscript
can be easily extended to the coupled channel case.

\subsection{Kaon/Eta Exchanges and SU(3) Symmetry}
\label{subsec:su3}

If we want to preserve SU(3) flavour symmetry, the exchange of kaons and
eta mesons should be treated on equal footing as the exchange of pions,
at least in principle.
But the mass of the kaon and the eta meson is of the order of $0.5\,{\rm GeV}$,
which is comparable to the hard scale $M$.
We have two choices: (i) ignore kaon and eta exchanges or, (ii) include
them explicitly.

In the first option, the contribution from kaon and eta exchange
are implicitly included in the contact-range potential.
There is a disadvantage, though: the contact-range potential breaks
SU(3)-flavour symmetry in this case.
We expect the size of this breaking to be parametrically small
for heavy baryon-baryon and heavy baryon-antibaryon systems of
the type $\Sigma_Q \Sigma_Q$, $\Sigma_Q \bar{\Sigma}_Q$,
$\Xi_Q' \Xi_Q'$, $\Xi_Q' \bar{\Xi}_Q'$, etc.,
i.e. systems containing only one
species of baryon.
Besides the pion, this type of system only exchange eta mesons,
where their coupling to the heavy baryons is considerably
weaker than that of the pions.
Regarding the kaons, they are relevant for heavy baryon-baryon systems
that involve different species: $\Sigma_Q \Xi_Q'$,
$\Xi_Q' \Omega_Q$, etc.
But if we consider instead heavy baryon-antibaryon system
the exchange of a single kaon implies a transition between
two-baryon states with different thresholds, which
involves coupled channel effects.
For example in the $\Xi_c' \bar{\Sigma}_c \to \Omega_c \bar{\Xi_c}'$
transition mediated by the exchange of a kaon/antikaon, 
the energy gap is $\Delta_{\rm CC} = 240\,{\rm MeV}$ and
the coupled channel scale is $\Lambda_{\rm CC} = 790\,{\rm MeV}$,
which is certainly hard.

From these reasons we expect the exchange of kaon and eta mesons
to have a small impact on the description of
heavy baryon-antibaryon systems in general.
This suggests that ignoring explicit kaon and eta exchange,
which amounts to include them implicitly in the contact-range couplings,
is not likely to generate a sizable breakdown of
SU(3)-flavour in the contact-range potential.
The implicit inclusion of eta and kaon effects in the contact-range couplings
will change their values from the ones expected from exact SU(3)-flavour
symmetry, but probably this change will be numerically smaller
than the usual $20\%$ uncertainty associated with
SU(3)-flavour symmetry relations.
This point is supported by an analysis of the strength of the eta and kaon
exchanges in Appendix \ref{app:OME}, where we also present the kaon
and eta exchange potentials in case one wants to include
them explicitly in the EFT.

\section{The Leading Order Potential}
\label{sec:LO}

In this section we write down the heavy baryon-antibaryon potential
at LO within the EFT expansion.
This potential can contain a contact- and a finite-range piece
\begin{eqnarray}
V^{(0)}_{\rm EFT} = V^{(0)}_{\rm C} + V^{(0)}_{\rm F} \, ,
\end{eqnarray}
where we are assuming NDA for the purpose of fixing the notation and
simplifying the discussion.
If there are bound states the actual power counting of
the heavy baryon-antibaryon system will differ from NDA,
see Sect.~\ref{subsec:bound-pc}.
But we will address this problem later in Sects.~\ref{sec:perturbative}
and ~\ref{sec:pc}.

The LO contact-range potential is a momentum- and energy-independent
potential in momentum space (or a Dirac-delta in coordinate space).
The LO finite-range potential is the OPE potential~\footnote{Regarding
  the exchange of the other $SU(3)$ Nambu-Goldstone bosons,
  we refer to the discussion in Section \ref{subsec:su3}.}.
For the contact-range component we cannot determine if it is perturbative
or not {\it a priori} without resorting to experimental or
phenomenological input.
For the finite-range components, i.e. the pion exchanges,
the situation is different and we can in fact determine
if they are perturbative, see Sect.~\ref{sec:perturbative}.
The discussion about the possible power countings that arise depending
on which pieces of the EFT potential are perturbative and non-perturbative
will be presented later in Sect.~\ref{sec:pc}.

This section is organized as follows:
we begin by explaining the details of how the heavy baryons are organized
in superfields that are well-behaved according to HQSS
in Sect.~\ref{subsec:superfields}.
Next we will consider the C- and G-parity properties of
the heavy baryon-antibaryon system in Sect.~\ref{subsec:C-parity}.
After this, we will first introduce the general form of
the contact-range potential and the constrains imposed on it
by heavy-quark spin symmetry, SU(2)-isospin and SU(3)-flavour symmetry
in Sect.~\ref{subsec:contact}.
Last, we will present the general form of the OPE potential and
its partial wave projection in Sect.~\ref{subsec:ope}.
Owing to the scope of the discussion, the notation will be complex.
We overview the most used notation
in this section in Table \ref{tab:symbols}.

\begin{table*}
\begin{center}
\begin{tabular}{|c|l|}
\hline \hline
Symbol & Meaning \\
\hline
$B_{\bar 3}$ & Antitriplet heavy baryon field \\
$B_{6}$ & Sextet heavy baryon field, ground state ($J=\tfrac{1}{2}$) \\
$B^*_{6}$ & Sextet heavy baryon field, excited state ($J=\tfrac{3}{2}$) \\
\hline
$M_{\bar 3}$ & Antitriplet heavy baryon mass \\
$M_{6}$ & Sextet heavy baryon mass, ground state ($J=\tfrac{1}{2}$) \\
$M_{6}^*$ & Sextet heavy baryon mass, excited state ($J=\tfrac{3}{2}$) \\
\hline
$T$ & Antitriplet heavy baryon superfield \\
$S$ & Sextet heavy baryon superfield \\
\hline
$C$ & (i) C-parity \\
& (ii) Coupling of the momentum and energy independent contact-interaction \\
\hline
$G$ & G-parity \\
\hline
$A^{(M)}_{S_L}$ & Antitriplet-antitriplet (A), antitriplet-sextet (B) and sextet-sextet (C) contact interactions \\
$B^{(M)}_{S_L D}$, $B^{(M)}_{S_L D}$ & M refers to the SU(3)-flavour representation, $S_L$ to total light-quark spin \\ 
$C^{(M)}_{S_L}$ & $D$ and $E$ to whether it is a direct or exchange term \\
\hline
$\vec{I}_i$ & Generic isospin operator for vertex $i=1,2$ of the two-body potential \\
$\vec{t}_i$ & Isosin-$0$ to isospin-$1$ transition matrices \\
$\vec{\tau}_i$ & Isospin-$\tfrac{1}{2}$ Pauli matrices \\
$\vec{T}_i$ & Isospin-$1$ matrices \\
\hline
$\vec{a}_i$ & Generic spin operator for vertex $i=1,2$ of the two-body potential \\
$\vec{\sigma}_i$ & Spin-$\tfrac{1}{2}$ Pauli matrices \\
$\vec{S}_i$, $\vec{S}_i^+$ & Spin-$\tfrac{1}{2}$ to spin-$\tfrac{3}{2}$ transition matrices \\
$\vec{\Sigma}_i$ & Spin-$\tfrac{3}{2}$ angular momentum matrices \\
\hline
$C_{12}$ & Spin-spin operator ($\vec{a}_1 \cdot \vec{a}_2$) \\
$S_{12}$ & Tensor operator
($3 \vec{a}_1 \cdot \hat{r}\,\vec{a}_2 \cdot \hat{r} - \vec{a}_1 \cdot \vec{a}_2$) \\
\hline
$^{2S+1}L_J$ & Spectroscopic notation for partial waves with $S$ the spin, \\
& $L$ the orbital angular momentum and $J$ the total angular momentum \\
\hline
${\bf C}_{12}$ & Matrix elements of the spin-spin operator in the partial wave
basis \\
${\bf S}_{12}$ & Matrix elements of the tensor operator in the partial wave
basis \\
\hline \hline
\end{tabular}
\end{center}
\caption{
  List of the most used symbols in Sect.~\ref{sec:LO}.
} \label{tab:symbols}
\end{table*}

\subsection{The Heavy Baryon Superfields}
\label{subsec:superfields}

Heavy baryons have the structure
\begin{eqnarray}
| Q (qq) \rangle \, ,
\end{eqnarray}
where $Q$ is the heavy quark and $qq$ the light quark pair,
which is in S-wave.
The light quarks can couple their spin to $S_L = 0, 1$.
If the light spin is $S_L = 0$, we have a $J^{P} = \frac{1}{2}^{+}$ heavy baryon
\begin{eqnarray}
B_{\bar 3} = | Q (qq)_{S_L = 0} \rangle \, .
\end{eqnarray}
This type of heavy baryon belongs to the $\bar{3}$ representation of the $SU(3)$
flavour group.
If the light spin is $S_L = 1$, we have instead a $J^{P} = \frac{1}{2}^{+}$
or a $J^{P} = \frac{3}{2}^{+}$ baryon
\begin{eqnarray}
B_6 &=& | Q (qq)_{S_L = 1} \rangle  \Big|_{J=1/2} \, , \\
B_6^* &=& | Q (qq)_{S_L = 1} \rangle \Big|_{J=3/2} \, ,
\end{eqnarray}
which belong to the $6$ representation of $SU(3)$.

If the heavy quark within the heavy baryons is a charm quark, $Q = c$,
the flavour components of the $B_{\bar 3}$ field are
\begin{eqnarray}
B_{\bar 3} =
\begin{pmatrix}
\phantom{+}\Xi_c^0 \\
-\Xi_c^+ \\
\phantom{+}\Lambda_c^{+}
\end{pmatrix} \, ,
\label{eq:B3c-su3}
\end{eqnarray}
where we follow the convention of Cho~\cite{Cho:1992cf}.
For the $B_6$ field the flavour components are
\begin{eqnarray}
B_6 =
\begin{pmatrix}
\Sigma_c^{++} & \frac{1}{\sqrt{2}}\,\Sigma_c^{+} &
\frac{1}{\sqrt{2}}\,\Xi_c^{+'} \\
\frac{1}{\sqrt{2}}\,\Sigma_c^{+} & \Sigma_c^{0} & \frac{1}{\sqrt{2}}\,\Xi_c^{0'} \\
\frac{1}{\sqrt{2}}\,\Xi_c^{+'}  & \frac{1}{\sqrt{2}}\,\Xi_c^{0'} & \Omega_c^0
\end{pmatrix} \, .
\label{eq:B6c-su3}
\end{eqnarray}
For the $B_6^*$ baryons we have exactly the same components as for $B_6$,
but with a star to indicate that they are spin-3/2 baryons.
Depending on the case, it can be practical to simply consider
the $SU(2)$-isospin structure rather than the complete $SU(3)$-flavour one.

The fields $B_{\bar 3}$, $B_6$ and $B_6^*$ can be organized into
the superfields $T$ and $S$, which have good transformation properties
under rotation of the heavy quark.
For non-relativistic heavy baryons we write the superfields
as~\cite{Cho:1992cf}
\begin{eqnarray}
T &=& B_{\bar 3} \quad , \quad
\vec{S} = \frac{1}{\sqrt{3}}\,\vec{\sigma}\,B_6 + \vec{B}_6^* \, ,
\end{eqnarray}
where the letters $T$ and $S$ stands for (anti-)triplet and sextet.
The definition of the non-relativistic superfield $T$ is redundant
--- it acts merely as a second name for $B_{\bar 3}$ ---
but we include it for completeness.
Notice that we have written the spin-3/2 heavy baryon field as a vector:
$\vec{B}_{6}^*$.
The reason is that this is a Rarita-Schwinger field, where the spin-3/2 nature
of this field is taken into account by {\it coupling} a spatial vector
with a Dirac spinor and then projecting to the spin-3/2 channel
with the condition $\vec{\sigma} \cdot \vec{B}_6^* = 0$.
Under rotations of the heavy quark spin the superfields behave as
\begin{eqnarray}
  T \to e^{\frac{i}{2} \vec{\epsilon} \cdot \vec{\sigma}_Q}\,T
  \quad , \quad
  \vec{S} \to e^{\frac{i}{2} \vec{\epsilon} \cdot \vec{\sigma}_Q}\,\vec{S} \, .
\end{eqnarray}
For a more complete account of the heavy baryon fields and superfields
we refer to Appendix~\ref{app:ope}.

\subsection{C- and G-Parity}
\label{subsec:C-parity}

We are considering heavy baryon-antibaryon states.
If a state is electrically neutral and does not have strangeness,
then C-parity will be a well-defined quantum number.
If the heavy baryon and antibaryon have identical $S_L$ and $J^P$,
i.e. if they have the structure
\begin{eqnarray}
  | B_{\bar 3} \bar{B}_{\bar 3} \rangle \, , \, | B_6 \bar{B}_6 \rangle \, , \,
  | B_6^* \bar{B}_6^* \rangle \, ,
\end{eqnarray}
then the C-parity of the system is
\begin{eqnarray}
  C = (-1)^{L+S} \, ,
\end{eqnarray}
with $L$ and $S$ the orbital angular momentum and spin~\footnote{ 
This comes from multiplying the intrinsic C-parity of
a fermion-antifermion system with the symmetry factors of
exchanging the particles, i.e. $C = (-1) \times (-1)^L \times (-1)^{S+1}$.}.
Examples of this type of heavy baryon-antibaryon system are
$\Lambda_c^{+} \Lambda_c^{-}$,
$\Sigma_c^{0} \bar{\Sigma}_c^{0}$
and $\Sigma_c^{*+} {\Sigma}_c^{*-}$.

If the light-quark spin $S_L$ and spin-parity $J^P$ of the heavy baryon
and antibaryon are not identical,
we first have to choose a C-parity convention, for instance:
\begin{eqnarray}
  C | B_{\bar 3} \rangle &=& + | \bar{B}_{\bar 3} \rangle \, , \\
  C | B_6 \rangle &=& + | \bar{B}_6 \rangle \, , \\
  C | B_6^* \rangle &=& - | \bar{B}_6^* \rangle \, ,
\end{eqnarray}
where there is a relative minus sign for the C-parity
transformation of the spin-$\frac{3}{2}$ fields
with respect to the spin-$\frac{1}{2}$ fields.
With this convention we define the states
\begin{eqnarray}
| B_{\bar 3} \bar{B}_6 (\eta) \rangle &=& \frac{1}{\sqrt{2}}\,\left[
| B_{\bar 3} \bar{B}_6 \rangle + \eta\, | B_6 \bar{B}_{\bar 3} \rangle \right] \, ,
\label{eq:C-parity-basis-a} \\
| B_{\bar 3} \bar{B}_6^* (\eta) \rangle &=& \frac{1}{\sqrt{2}}\,\left[
| B_{\bar 3} \bar{B}_6^* \rangle + \eta\, | B_6^* \bar{B}_{\bar 3} \rangle \right] \, ,
\label{eq:C-parity-basis-b} \\
| B_6 \bar{B}_6^* (\eta) \rangle &=& \frac{1}{\sqrt{2}}\,\left[
| B_6 \bar{B}_6^* \rangle + \eta\, | B_6^* \bar{B}_6 \rangle \right] \, ,
\label{eq:C-parity-basis-c}
\end{eqnarray}
where $\eta = \pm 1$, for which the C-parity is
\begin{eqnarray}
  C = \eta\,(-1)^{L+S} \, ,
\end{eqnarray}
where $L$ ($S$) is the total orbital angular momentum (spin)
of the heavy baryon-antibaryon pair~\footnote{
  The C-parity is the product of the intrinsic C-parity
  and the symmetry factor of exchanging the particles,
  which now includes a contribution from $\eta$, i.e.
  $C = (+\eta) \times (-1) \times (-1)^L \times (-1)^{S+1}$
  for $B_{\bar 3} B_6$ and
  $C = (-\eta) \times (-1) \times (-1)^L \times (-1)^{S}$
  for $B_{\bar 3} B_6^*$ / $B_6 B_6^*$.
}.
Examples of this type of molecule include $\Lambda_c^{+} \Sigma_c^{-}$,
$\Xi_c^{'0} \bar{\Xi}_c^{0}$ and $\Xi_c^{'+} {\Xi}_c^{*-}$.

For a heavy baryon-antibaryon state that is not electrically neutral
but has no strangeness and belongs to the same $SU(2)$ isospin
representation as a neutral state, C-parity is not a well-defined
quantum number but there exists an extension
that includes isospin.
This extension is G-parity~\cite{Lee:1956sw},
which can be defined as follows
\begin{eqnarray}
G = C\,e^{i \pi I_2} \, ,
\end{eqnarray}
that is, a C-parity transformation combined with a rotation
in isospin space~\footnote{Notice that the G-parity transformation is sometimes
  defined as $G = C\,e^{-i \pi I_2}$. with a minus sign. For particles
  with integer isospin this is equivalent to the definition
  with a plus sign, $G = C\,e^{+i \pi I_2}$. For particles
  with half-integer isospin each of these conventions
  generate anti-particle states that differ by a sign.
  This has no observable consequence, as it amounts to
  a global redefinition of the amplitudes by a phase.}.
For {a electrically charged state,} the G-parity is well-defined and
its eigenvalues are
\begin{eqnarray}
  G = C\,(-1)^{I} \, ,
\end{eqnarray}
where $I$ is the isospin of the electrically charged state and $C$ is
the C-parity of the electrically neutral component of
the isospin multiplet.
For example if we consider $\Sigma_c^{++} \bar{\Sigma}_c^{0}$,
its isospin is $I=2$ and the neutral component of its isospin multiplet
is a linear combination of $\Sigma_c^{0} \bar{\Sigma}_c^{0}$,
$\Sigma_c^{+} {\Sigma}_c^{-}$ and $\Sigma_c^{++} {\Sigma}_c^{--}$:
the G-parity of $\Sigma_c^{++} \bar{\Sigma}_c^{0}$ is then
$G = (-1)^I\,C = (-1)^{L+S}$.

\subsection{The Contact-Range Potential}
\label{subsec:contact}

The LO contact-range potential takes the generic form
\begin{eqnarray}
\langle p' | V^{(0)}_C | p \rangle = C \, ,
\end{eqnarray}
with $C$ a coupling constant and where $p$ ($p'$) is the center-of-mass
momentum of the incoming (outgoing) heavy baryon-antibaryon pair.
In principle there should be one independent coupling $C$ for each $J^{PC}$
quantum number and type of heavy baryon-antibaryon molecule.
But the contact-range potential is constrained by HQSS and
$SU(3)$-flavour symmetry, which greatly reduces
the number of possible couplings.
We first consider the HQSS structure of the contact-range potential
and then the $SU(3)$ flavour one.

\subsubsection{HQSS structure}

The application of HQSS to the heavy baryon-antibaryon system implies that
the contact-range coupling does not depend on the heavy quark spin,
only on the light quark spin.
This means that the coupling $C$ can be decomposed
in terms of light-quark components
\begin{eqnarray}
  C = \sum d_{S_L} C_{S_L} \, ,
\end{eqnarray}
where $S_L$ is the total light-quark spin of the heavy baryon-antibaryon system.
The $d_{S_L}$'s are coefficients that depend on the heavy- and light-quark
decomposition of the specific heavy baryon-antibaryon molecule,
see Appendix \ref{app:contact} for details.

The contact-range couplings of the $T\bar{T}$, $S\bar{T}$ and $S\bar{S}$
molecules are independent and we will use a different notation
for each case: $A$, $B$ and $C$ respectively.
For the $T\bar{T}$ system we write the contact-range potential as
\begin{eqnarray}
  \langle T \bar{T} | V_C^{(0)}\, | T \bar{T} \rangle &=& A_{0} \, ,
\end{eqnarray}
where we are already taking into account that
the total light quark spin is always $S_L = 0$
(we also ignore the coefficient because there is actually no decomposition).
For the $S\bar{T}$ system we write
\begin{eqnarray}
  \langle S \bar{T} | V_C^{(0)}\, | S \bar{T} \rangle &=&
  d_{1D} B_{1D} \, , \\
  \langle S \bar{T} | V_C^{(0)}\, | T \bar{S} \rangle &=&
  d_{1E} B_{1E} \, ,
\end{eqnarray}
where the total light spin is always $S_L = 1$, but where we have to make
the distinction between a diagonal and non-diagonal potential.
For the $S\bar{S}$ system we have
\begin{eqnarray}
  \langle S \bar{S} | V_C^{(0)}\, | S \bar{S} \rangle &=&
  d_0 C_0 + d_1 C_1 + d_2 C_2
  \, ,
\end{eqnarray}
where the total light spin is $S_L = 0, 1, 2$.
We list the contact-range potential for heavy baryon-antibaryon molecules
with well-defined C-parity in Table \ref{tab:contact-hqqs}, which also
applies by extension to the molecules with well-defined G-parity.

\begin{table}
\begin{center}
\begin{tabular}{|c|c|c|}
\hline \hline
System & $J^P$/$J^{PC}$ & $V_C$ (HQSS) \\
\hline
$B_{\bar 3} \bar{B}_{\bar 3}$ & $0^{-+}$ & $A_0$ \\
$B_{\bar 3} \bar{B}_{\bar 3}$ & $1^{--}$ & $A_0$ \\
\hline
$B_{\bar 3} \bar{B}_6$ & $0^{-+}$ & $B_{\rm  1D} - B_{\rm 1E}$ \\
$B_{\bar 3} \bar{B}_6$ & $0^{--}$ & $B_{\rm 1D} + B_{\rm 1E}$ \\
$B_{\bar 3} \bar{B}_6$ & $1^{-+}$ & $B_{\rm 1D} - \frac{1}{3}\,B_{\rm 1E}$ \\
$B_{\bar 3} \bar{B}_6$ & $1^{--}$ & $B_{\rm 1D} + \frac{1}{3}\,B_{\rm 1E}$ \\
\hline
$B_{\bar 3} \bar{B}_6^*$ & $1^{-+}$ & $B_{\rm 1D} + \frac{1}{3}\,B_{\rm 1E}$ \\
$B_{\bar 3} \bar{B}_6^*$ & $1^{--}$ & $B_{\rm 1D} - \frac{1}{3}\,B_{\rm 1E}$ \\
$B_{\bar 3} \bar{B}_6^*$ & $2^{-+}$ & $B_{\rm 1D} - B_{\rm 1E}$ \\
$B_{\bar 3} \bar{B}_6^*$ & $2^{--}$ & $B_{\rm 1D} + B_{\rm 1E}$ \\
\hline
$B_6 \bar{B}_6$ & $0^{-+}$ & $\frac{1}{3}\,C_0 + \frac{2}{3}\,C_1$ \\
$B_6 \bar{B}_6$ & $1^{--}$ & $\frac{1}{27}\,C_0 + \frac{6}{27}\,C_1 + \frac{20}{27}\,C_2$ \\
\hline
$B_6 \bar{B}_6^*$ & $1^{-+}$ & $C_1$ \\
$B_6 \bar{B}_6^*$ & $1^{--}$ &
$\frac{16}{27}\,C_0 + \frac{6}{27}\,C_1 + \frac{5}{27}\,C_2$ \\
$B_6 \bar{B}_6^*$ & $2^{-+}$ & $\frac{1}{3}\,C_1 + \frac{2}{3}\,C_2$ \\
$B_6 \bar{B}_6^*$ & $2^{--}$ & $C_2$ \\
\hline
$B_6^* \bar{B}_6^*$ & $0^{-+}$ & $\frac{2}{3}\,C_0 + \frac{1}{3}\,C_1$ \\
$B_6^* \bar{B}_6^*$ & $1^{--}$ &
$\frac{10}{27}\,C_0 + \frac{15}{27}\,C_1 + \frac{2}{27}\,C_2$ \\
$B_6^* \bar{B}_6^*$ & $2^{-+}$ & $\frac{2}{3}\,C_1 + \frac{1}{3}\,C_2$ \\
$B_6^* \bar{B}_6^*$ & $3^{--}$ & $C_2$ \\
\hline \hline
\end{tabular}
\end{center}
\caption{
  HQSS decomposition of the contact-range couplings
  for heavy baryon-antibaryon molecule
  with well-defined C-parity.
  $A$, $B$ and $C$ refers to the coupling of the $T\bar{T}$, $S\bar{T}$
  and $S\bar{S}$ system respectively, with the subscript indicating
  the light-quark spin decomposition.
  For the cases where C-parity is not well-define,
  we refer to the explanations in the main text.
} \label{tab:contact-hqqs}
\end{table}

If the molecules do not have well-defined C- or G-parity
(i.e. molecules with strangeness),
the form of the contact-range potential depends on the particular case.
For the family of molecules
\begin{eqnarray}
  | B_{\bar 3} \bar{B}_{\bar 3} \rangle \, , \, | B_6 \bar{B}_6 \rangle \, , \,
  | B_6^* \bar{B}_6^* \rangle \, , 
\end{eqnarray}
(e.g. $\Lambda_c \bar{\Xi}_c$, $\Sigma_c \bar{\Xi}_c^{'}$,
$\Sigma_c^{*} \bar{\Xi}_c^{*}$)
the contact-range couplings
are exactly as shown in Table \ref{tab:contact-hqqs}
for the case in which C-parity is well-defined.
For the molecules involving different types of heavy hadrons
the contact-range {potentials are} defined in coupled channels.
If we consider the bases
\begin{eqnarray}
\mathcal{B}_1 &=& \{ | B_{\bar 3} \bar{B}_6 \rangle ,
| B_6 \bar{B}_{\bar 3} \rangle \} \, ,  \\
\mathcal{B}_2 &=& \{ | B_{\bar 3} \bar{B}_6^* \rangle ,
| B_6^* \bar{B}_{\bar 3} \rangle \} \, , \\
\mathcal{B}_3 &=& \{ | B_6 \bar{B}_6^* \rangle ,
| B_6^* \bar{B}_6 \rangle \} \, ,
\end{eqnarray}
we end up with the following contact-range potentials
\begin{eqnarray}
V_{\mathcal{B}_1}(0^{-}) &=&
\begin{pmatrix}
B_{1D} & -B_{1E} \\
-B_{1E} & B_{1D} \\
\end{pmatrix} \, , \\
V_{\mathcal{B}_1}(1^{-}) &=&
\begin{pmatrix}
B_{1D} & \frac{1}{3}\,B_{1E} \\
\frac{1}{3}\,B_{1E} & B_{1D} \\
\end{pmatrix} \, ,
\end{eqnarray}
\begin{eqnarray}
V_{\mathcal{B}_2}(1^{-}) &=&
\begin{pmatrix}
B_{1D} & -\frac{1}{3}\,B_{1E} \\
-\frac{1}{3}\,B_{1E} & B_{1D} \\
\end{pmatrix} \, ,  \\
V_{\mathcal{B}_2}(2^{-}) &=&
\begin{pmatrix}
B_{1D} & - B_{1E} \\
-B_{1E} & B_{1D} \\
\end{pmatrix} \, .
\end{eqnarray}
\begin{eqnarray}
V_{\mathcal{B}_3}(1^{-}) &=&
\begin{pmatrix}
\frac{16\,C_0 +33\,C_1 + 5\,C_2}{54} &
\frac{16\,C_0 - 21\,C_1 + 5\,C_2}{54} \\
\frac{16\,C_0 - 21\,C_1 + 5\,C_2}{54} &
\frac{16\,C_0 +33\,C_1 + 5\,C_2}{54}
\end{pmatrix} \, , \\
V_{\mathcal{B}_3}(2^{-}) &=&
\begin{pmatrix}
\frac{1}{6}\,C_1 + \frac{5}{6}\,C_2 &
\frac{1}{6}\,C_1 - \frac{1}{6}\,C_2 \\
\frac{1}{6}\,C_1 - \frac{1}{6}\,C_2 &
\frac{1}{6}\,C_1 + \frac{5}{6}\,C_2
\end{pmatrix} \, .
\end{eqnarray}
Examples of bases $1$, $2$ and $3$ are the
$\Xi_c \bar{\Sigma}_c$-$\Xi_c' \bar{\Lambda}_c$,
$\Xi_c \bar{\Sigma}_c^*$-$\Xi_c^* \bar{\Lambda}_c$
and $\Xi_c' \bar{\Sigma}_c^*$-$\Xi_c^* \bar{\Sigma}_c$ systems.

\subsubsection{SU(3)-flavour structure}

Besides HQSS, heavy baryon-antibaryon systems also
have SU(3)-flavour symmetry.
In SU(3)-flavour the $T$ and $S$ heavy baryons belong to the antitriplet and
{sextet} representation ($\bar{3}$ and $6$), respectively.
For the $T\bar{T}$ the HQSS coupling is further divided into the SU(3)-flavour
representations $\bar{3} \otimes  3 = 1 \oplus 8$, a singlet and an octet.
That is, there are two independent SU(3)-flavour contact interactions
\begin{eqnarray}
A_0 &\to& \{ A_0^{(1)}, A_0^{(8)} \} \, .
\end{eqnarray}
For the $S\bar{T}$ case we have $6 \otimes 3 = 8 \oplus 10$:
\begin{eqnarray}
  B_{1D} &\to& \{ B_{1D}^{(8)}, B_{1D}^{(10)} \} \, , \\
  B_{1E} &\to& \{ B_{1E}^{(8)}, B_{1E}^{(10)} \} \, .
\end{eqnarray}
Finally for the $S\bar{S}$ case we have
$6 \otimes \bar{6} = 1 \oplus 8 \oplus 27$:
\begin{eqnarray}
C_{S_L} &\to& \{
C_{S_L}^{(1)}, C_{S_L}^{(8)} , C_{S_L}^{(27)} \} \, .
\end{eqnarray}
The decomposition for a specific molecule can be consulted
in Table \ref{tab:contact-su3}, which have been obtained
from the SU(3) Clebsch-Gordan coefficients for $\bar{3} \otimes 3$,
$6 \otimes 3$ and $6 \otimes \bar{6}$ of Ref.~\cite{Kaeding:1995vq}.
Notice that we are not explicitly considering the SU(2)-isospin structure
as it is a subgroup of SU(3)-flavour.

Finally,
we remind that the SU(3)-flavour structure of the contact-range potential
can be broken if the finite-range potential is not SU(3)-flavour symmetric.
Whether this happens depends on two factors.
The first is the particular power counting we are using and the order
we are considering within the EFT expansion, e.g. if the contact-range
interaction is leading, but the exchange of pions, kaons and etas
is subleading, the violations of SU(3)-flavour symmetry
if we ignore kaon and eta exchanges will be subleading.
The second factor is that kaon and eta exchanges are parametrically small,
as was explained in Sect.~\ref{subsec:su3} and where a more detailed
derivation can be found in Appendix \ref{app:OME}.

\begin{table}
\begin{center}
\begin{tabular}{|c|c|c|c|}
\hline \hline
System & Type & Isospin & $V_C$ \\
\hline
$\Lambda_c \bar{\Lambda}_c$ & $T\bar{T}$ & 0 &
$\frac{1}{3}\,A^{(1)} + \frac{2}{3}\,A^{(8)}$ \\
$\Xi_c \bar{\Xi}_c$ & $T\bar{T}$ & 0 &
$\frac{2}{3}\,A^{(1)} + \frac{1}{3}\,A^{(8)}$ \\
$\Xi_c \bar{\Xi}_c$ & $T\bar{T}$ & 1 & $A^{(8)}$ \\
\hline
$\Xi_c' \bar{\Xi}_c$ & $S\bar{T}$ & 0 & $B^{(8)}$ \\
$\Xi_c' \bar{\Xi}_c$ & $S\bar{T}$ & 1 &
$\frac{1}{3}\,B^{(8)} + \frac{2}{3}\,B^{(10)}$ \\
$\Sigma_c \bar{\Lambda}_c$ & $S\bar{T}$ & 1 &
$\frac{2}{3}\,B^{(8)} + \frac{1}{3}\,B^{(10)}$ \\
\hline
$\Omega_c \bar{\Omega}_c$ & $S\bar{S}$ & 0 &
$\frac{1}{6}\,C^{(1)} + \frac{8}{15}\,C^{(8)} + \frac{3}{10}\,C^{(27)}$ \\
$\Xi_c' \bar{\Xi}_c'$ & $S\bar{S}$ & 0 &
$\frac{1}{3}\,C^{(1)} + \frac{1}{15}\,C^{(8)} + \frac{3}{5}\,C^{(27)}$ \\
$\Sigma_c \bar{\Sigma}_c$ & $S\bar{S}$ & 0 &
$\frac{1}{2}\,C^{(1)} + \frac{2}{5}\,C^{(8)} + \frac{1}{10}\,C^{(27)}$ \\
$\Xi_c' \bar{\Xi}_c'$ & $S\bar{S}$ & 1 &
$\frac{1}{5}\,C^{(8)} + \frac{4}{5}\,C^{(27)}$ \\
$\Sigma_c \bar{\Sigma}_c$ & $S\bar{S}$ & 1 &
$\frac{4}{5}\,C^{(8)} + \frac{1}{5}\,C^{(27)}$ \\
$\Sigma_c \bar{\Sigma}_c$ & $S\bar{S}$ & 2 &
$C^{(27)}$ \\
\hline \hline
\end{tabular}
\end{center}
\caption{
  SU(3)-flavour decomposition of the contact-range couplings
  depending on the type heavy baryon-antibaryon molecule.
  Notice that the HQSS structure of the couplings is independent of
  the SU(3) one, which is why we do not show the light-spin indices
  for the couplings.
  For the $S$ heavy baryons we show the decomposition
  for the lightest member of the HQSS multiplet only.
} \label{tab:contact-su3}
\end{table}

\subsection{The One Pion Exchange Potential}
\label{subsec:ope}

\begin{table}
\begin{center}
\begin{tabular}{|c|c|c|c|c|}
\hline \hline
Vertex & $R_i$ & $\bar{R}_i$ & $\vec{I}_i$ & $\vec{a}_i$ \\
\hline
$\Lambda_c \to \Sigma_c$
& $\sqrt{\frac{2}{{3}}}$ & -$\sqrt{\frac{2}{{3}}}$ 
& $\vec{t}_i$ & $\vec{\sigma}_i$ \\
$\Lambda_c \to \Sigma_c^*$
& $\sqrt{2}$ & $\sqrt{2}$
& $\vec{t}_i$ & $\vec{S}_i^{\dagger}$ \\
\hline
$\Xi_c \to \Xi_c'$
& $\sqrt{\frac{2}{{3}}}$ & -$\sqrt{\frac{2}{{3}}}$ 
& $\frac{1}{2}\,\vec{\tau_i}$ & $\vec{\sigma}_i$ \\
$\Xi_c \to \Xi_c^*$
& $\sqrt{2}$ & $\sqrt{2}$
& $\frac{1}{2}\,\vec{\tau_i}$ & $\vec{S}_i^{\dagger}$ \\
\hline
$\Sigma_c \to \Sigma_c$
& $\frac{{2}}{3}$ & -$\frac{{2}}{3}$
& $\vec{T}_i$ & $\vec{\sigma}_i$ \\
$\Sigma_c^* \to \Sigma_c$
& $\frac{1}{\sqrt{3}}$ & $\frac{1}{\sqrt{3}}$
& $\vec{T}_i$ & $\vec{S}_i$ \\
$\Sigma_c \to \Sigma_c^*$
& $\frac{1}{\sqrt{3}}$ & $\frac{1}{\sqrt{3}}$
& $\vec{T}_i$ & $\vec{S}_i^{\dagger}$ \\
$\Sigma_c^* \to \Sigma_c^*$
& $\frac{{2}}{3}$ & -$\frac{{2}}{3}$
& $\vec{T}_i$ & $\vec{\Sigma}_i$ \\
\hline
$\Xi_c' \to \Xi_c'$
& $\frac{{2}}{3}$ & -$\frac{{2}}{3}$
& $\frac{1}{2}\,\vec{\tau}_i$ & $\vec{\sigma}_i$ \\
$\Xi_c^* \to \Xi_c' $
& $\frac{1}{\sqrt{3}}$ & $\frac{1}{\sqrt{3}}$
& $\frac{1}{2}\,\vec{\tau}_i$ & $\vec{S}_i$ \\
$\Xi_c' \to \Xi_c^* $
& $\frac{1}{\sqrt{3}}$ & $\frac{1}{\sqrt{3}}$
& $\frac{1}{2}\,\vec{\tau}_i$ & $\vec{S}_i^{\dagger}$ \\
$\Xi_c^* \to \Xi_c^* $
& $\frac{{2}}{3}$ & -$\frac{{2}}{3}$
& $\frac{1}{2}\,\vec{\tau}_i$ & $\vec{\Sigma}_i$ \\
\hline \hline
\end{tabular}
\end{center}
\caption{
Numerical, isospin and spin factor associated with each vertex
in the $\langle S \bar{T} | V | T \bar{S} \rangle$ and
$\langle S \bar{S} | V | S \bar{S} \rangle$ heavy
baryon-antibaryon potential. The arrows are used to indicate
the final baryon state in the vertex.
The symbols $\vec{\tau}_i$ and $\vec{T}_i$ represent
the Pauli matrices (in isospin space) and the isospin $I=1$ matrices
respectively, while $t_i$ is a special isospin matrix for connecting
the isoscalar $\Lambda_c$ with the isovector $\Sigma_c$ and the pion.
The symbols $\vec{\sigma}_i$ and $\vec{\Sigma}_i$ are the Pauli matrices
and the spin $S = 3/2$ matrices, while $\vec{S}_i$ is a
$2 \times 4$ matrix that is used for the transitions
involving a spin $1/2$ and spin $3/2$ baryon.
Notice that this table can also be used to compute
the heavy baryon-baryon potential.
} \label{tab:vertices}
\end{table}

\begin{figure}[htb]
\begin{center}
\includegraphics[height=4.5cm]{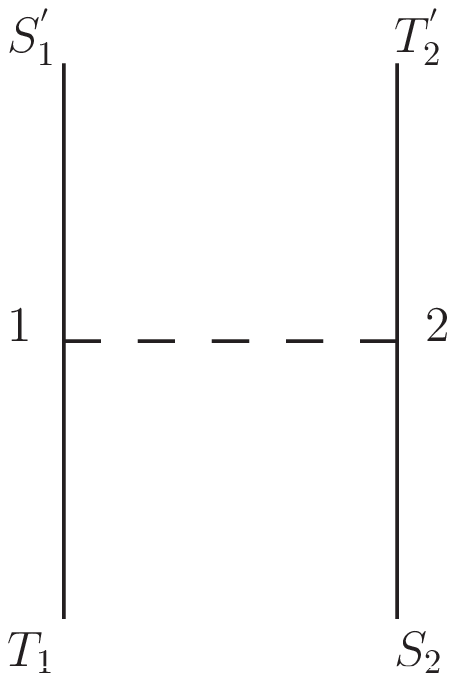}
\includegraphics[height=4.5cm]{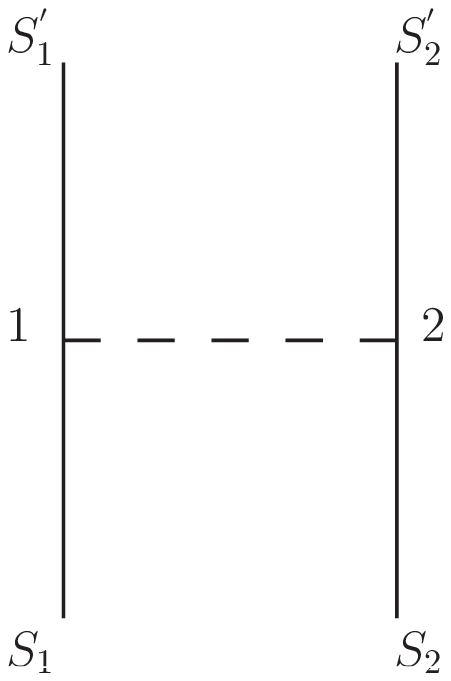}
\end{center}
\caption{
Diagrams for OPE potential between two heavy hadrons. In the left we show
the $T S$ potential and in the right the $S S$ potential, where $T$ ($S$)
is the heavy baryon with light spin $S_L = 0$ ($S_L = 1$).
}
\label{ope-diag}
\end{figure}

The OPE potential in momentum space reads
\begin{eqnarray}
  \langle T_1 \bar{T}_2 | V_{\rm F}^{(0)} | T_1' \bar{T}_2' \rangle &=& 0 \, ,
  \label{eq:V_TT_q} \\
\langle T_1 \bar{S}_2 | V_{\rm F}^{(0)} | S_1' \bar{T}_2' \rangle &=&
-R_1\,\bar{R}_2\,\frac{g_3^2}{2 f_{\pi}^2} \vec{I}_1 \cdot \vec{I}_2
\frac{\vec{a}_1 \cdot \vec{q}\, \vec{a}_2 \cdot \vec{q}}{q^2 + \mu_{\pi}^2} \, ,
\nonumber \\
\label{eq:V_ST_q} \\
\langle S_1 \bar{S}_2 | V_{\rm F}^{(0)} | S_1' \bar{S}_2' \rangle &=&
-R_1\,\bar{R}_2\,\frac{g_2^2}{2 f_{\pi}^2} \vec{I}_1 \cdot \vec{I}_2
\frac{\vec{a}_1 \cdot \vec{q}\, \vec{a}_2 \cdot \vec{q}}{q^2 + \mu_{\pi}^2} \, ,
\nonumber \\
\label{eq:V_SS_q}
\end{eqnarray}
where we have chosen the specific notation above to cover all
the possible combinations.
The subscripts $1$ and $2$ are used to denote the vertices $1$ and $2$
in the diagrams of Fig.~\ref{ope-diag}.
In the equation above $R_1$ and $\bar{R}_2$ are numerical factors
which {depend} on the transition we are considering,
see Table \ref{tab:vertices}
(the bar indicates an antibaryon to antibaryon transition).
$\vec{I}_1$ and $\vec{I}_2$ are isospin matrices,
while $\vec{a}_1$ and $\vec{a}_2$ spin matrices.
For the couplings we have that $g_2$ is the axial coupling for the $S_L = 1$
heavy baryon, $g_3$ the coupling involved in $T \to S \pi$ transitions
and $f_{\pi}$ the pion decay constant.
$\mu_{\pi}$ is the effective pion mass for the vertices involved
in the particular channel considered.
Finally we notice that OPE vanishes for the $T\bar{T}$ molecules,
which can be described solely in terms of contact-interactions at lowest order.

Regarding the isospin structure of the OPE potential, we have
that $\vec{I}_1$ and $\vec{I}_2$ are the isospin
matrices corresponding to vertex $1$ and $2$.
If we have a heavy baryon with isospin $1/2$ at vertex $i$ we can simply
make the substitution $\vec{I}_i = \frac{\vec{\tau}_i}{2}$,
where $\vec{\tau}$ are the Pauli matrices.
If the heavy baryon at vertex $i$ has isospin $1$ we use
the $J=1$ angular momentum matrices in isospin space,
for which we use the notation $\vec{T}_i$, i.e.
we make the substitution $\vec{I}_i = \vec{T}_i$.
The exact isospin factor for each type of vertex
can be consulted in Table \ref{tab:vertices}.

Regarding the spin structure, we note that the spin operators $a_1$ and $a_2$
depend on which is the initial and final spin of the heavy baryons
at vertex $1$ and $2$.
If the initial and final heavy baryons at vertex $1$($2$) have spin $1/2$
we have $\vec{a}_1 = \vec{\sigma}_1$ ($\vec{a}_2 = \vec{\sigma}_2$).
If the initial and final heavy baryons at vertex $1$($2$) have spin $3/2$
we have $\vec{a}_1 = \vec{\Sigma}_1$ ($\vec{a}_2 = \vec{\Sigma}_2$),
where $\vec{\Sigma}$ are the $S = 3/2$ spin matrices.
If the initial and final heavy baryons at vertex $1$($2$) switch from spin
1/2 to spin 3/2 (or viceversa), then $\vec{a}_1 = \vec{S}_1$
($\vec{a}_2 = \vec{S}_2$), where $\vec{S}$ are special 2x4 spin
matrices that describe the transition from a different initial
to final spin (see their definition in Appendix \ref{app:ope}).
As with isospin, the exact spin matrix to use in each type of vertex
can be checked in Table \ref{tab:vertices}.

Regarding the axial couplings $g_2$ and $g_3$,
we notice that the value of $g_3$ can be extracted
from the $\Sigma_c^{++} \to \Lambda_c^+ \pi^{+}$ decay
measured by Belle in Ref.~\cite{Lee:2014htd},
yielding $g_3 = 0.973^{+0.019}_{-0.042}$~\cite{Cheng:2015naa}
(notice that the previous reference originally uses the convention of
Yan et al.~\cite{Yan:1992gz} to define the axial couplings,
instead of the one by Cho~\cite{Cho:1992cf,Cho:1992gg} that we employ here
and we have consequently adapted the numbers of Ref.~\cite{Cheng:2015naa}
to our convention).
In contrast $g_2$ is experimentally unavailable,
but on the basis of quark model relations one can estimate it to be
$g_2 = -\sqrt{2} g_3 =  -1.38$.
If we consider the values of $g_2$ and $g_3$ from the lattice QCD
calculation of Ref.~\cite{Detmold:2012ge}, we obtain instead
$g_2 = -0.84 \pm 0.20$ and $g_3 = 0.71 \pm 0.13$,
where it is important to mention that they are calculated
in the $m_Q = \infty$ limit
(notice again that our convention for $g_2$ differs
from the definition used in Ref.~\cite{Detmold:2012ge}
by a sign, which has been taken into account).
Had we applied the quark model relations to the lattice QCD value of $g_3$,
we would have obtained $g_2 = -1.00$, which is considerably larger
than the lattice QCD determination but yet within its error bar.

For the effective pion mass, we have that if the particles in the vertex $1$
and $2$ have the same mass, then $\mu_{\pi} = m_{\pi}$.
On the other hand if they have different masses (e.g. $S_1 = B$, $S_1' = B'$)
and the mass splitting is given by $\Delta$, then we have that
$\mu_{\pi}^2 = m_{\pi}^2 - \Delta^2$ (a relation that assumes
heavy, non-relativistic baryons).

Finally we notice that we can also compute the heavy baryon-baryon potential
potential by making the change
\begin{eqnarray}
R_1 \bar{R}_2 \to R_1 R_2 \, ,
\end{eqnarray}
in Eqs.~(\ref{eq:V_ST_q}) and (\ref{eq:V_SS_q}) and consulting
the proper values in Table~\ref{tab:vertices}.

The most explicit way to construct the potential for one particular
channel is to make use of Table~\ref{tab:vertices}, where all the
factors are listed.
For instance if we are considering the $\Xi_c^* \bar{\Sigma}_c \to
\Xi_c' \bar{\Sigma}_c^*$ potential, we can see that it contains
a $\Xi_c^* \to \Xi_c'$ transition in vertex $1$ and a
$\bar{\Sigma}_c \to \bar{\Sigma}_c^*$ transition in vertex $2$.
If we use Table~\ref{tab:vertices} we find
$R_1 = 1/\sqrt{3}$, $I_1 = \vec{\tau}_1 / 2$, $\vec{a}_1 = \vec{S}_1$
for vertex 1 and $\bar{R}_2 = 1/\sqrt{3}$, $I_2 = \vec{T}_1$,
$\vec{a}_2 = \vec{S}_2^{\dagger}$ for vertex 2.
Putting the pieces together the potential reads
\begin{eqnarray}
\langle \, \Xi_c^* \bar{\Sigma}_c | V_{\rm F}^{(0)} | \Xi_c' \bar{\Sigma}^*_c \rangle &=&
- \frac{1}{3}\,\frac{g_3^2}{2 f_{\pi}^2} \frac{\vec{\tau}_1 \cdot \vec{T}_2}{2}
\frac{\vec{S}_1 \cdot \vec{q}\, \vec{S}_2^{\dagger}
\cdot \vec{q}}{q^2 + \mu_{\pi}^2} \, ,
\end{eqnarray}
where $\mu^2_{\pi} = m_{\pi}^2 - \Delta^2$,
with $\Delta \simeq (m(\Xi_c^*) - m(\Xi_c')) \simeq (m(\Sigma_c^*) - m(\Sigma_c))
\simeq  70\,{\rm MeV}$
and $\mu_{\pi} \sim 120\,{\rm MeV}$.
The other cases can be obtained analogously.

\subsubsection{The OPE Potential in Coordinate Space}

If we Fourier-transform the potential into coordinate space we obtain
\begin{eqnarray}
\langle T_1 \bar{S}_2 | & V_{\rm F}^{(0)}(\vec{r}) & | S_1' \bar{T}_2' \rangle =
\nonumber \\ &&
- R_1\,\bar{R}_2\,\vec{I}_1 \cdot \vec{I}_2
\frac{g_3^2}{6\,f_{\pi}^2} \, C_{12}
\delta^3(\vec{r}) \nonumber \\
&&  +R_1\,\bar{R}_2\,\vec{I}_1 \cdot \vec{I}_2 \,
\Big[ C_{12} \, W_C(r) \nonumber \\
&& + S_{12}(\hat{r}) \, W_T(r)
\Big] \, ,  \\
\langle S_1 \bar{S}_2 | & V_{\rm F}^{(0)}(\vec{r}) & | S_1' \bar{S}_2' \rangle =
\nonumber \\ &&
- R_1\,\bar{R}_2\,\vec{I}_1 \cdot \vec{I}_2
\frac{g_2^2}{6\,f_{\pi}^2} \, C_{12}
\delta^3(\vec{r}) \nonumber \\
&&  +R_1\,\bar{R}_2\,\vec{I}_1 \cdot \vec{I}_2 \,
\Big[ C_{12} \, W_C(r) \nonumber \\
&& + S_{12}(\hat{r}) \, W_T(r)
\Big] \, , \label{eq:VOPE}
\end{eqnarray}
where $C_{12}$ and $S_{12}$ are the spin-spin and tensor operators, defined as
\begin{eqnarray}
\label{eq:C12}
C_{12} &=& \vec{a}_1 \cdot \vec{a}_2 \, , \\
\label{eq:S12}
S_{12}(\hat{r}) &=&
3 \vec{a_1} \cdot \hat{r} \, \vec{a}_2 \cdot \hat{r} -
\vec{a}_1 \cdot \vec{a}_2 \, .
\end{eqnarray}
The OPE potential contains a Dirac-delta contribution which can be
reabsorbed into the contact-range potential if one wishes too.
The spin-spin and tensor pieces of the potential $W_C$ and $W_T$
can be written as
\begin{eqnarray}
\label{eq:WC}
W_C &=& \frac{g_i^2 \mu_{\pi}^3}{24 \pi f_{\pi}^2}\,\frac{e^{-\mu_{\pi} r}}{\mu_{\pi} r} \, , \\
\label{eq:WT}
W_T &=& \frac{g_i^2 \mu_{\pi}^3}{24 \pi f_{\pi}^2}\,\frac{e^{-\mu_{\pi} r}}{\mu_{\pi} r}
\,\left( 1 + \frac{3}{\mu_{\pi} r} + \frac{3}{(\mu_{\pi} r)^2} \right) \, ,
\end{eqnarray}
where $g_i = g_2$ or $g_3$ depending on the case and
$\mu_{\pi}$ is the effective pion mass for the channel under consideration.
The spin-spin piece of the OPE potential is often referred
to as {\it central} OPE, a naming convention that often
appears in nuclear physics for historical reasons and
which permeates the notation $W_C$ and $W_T$.
{\it Central} is used in opposition to {\it tensor} to convey the idea
that the {\it central} piece carries no orbital angular momentum
(while the tensor piece carries two units of orbital
angular momentum).
The term central OPE is indeed convenient and we will use it
in what follows (instead of the more accurate spin-spin OPE).
We notice that the OPE potential also contains a contact-range contribution,
which is mostly harmless: it can be reabsorbed into the EFT contact-range
contribution to the potential by a redefinition of the couplings.
Hence it can be simply ignored.

\subsubsection{Partial Wave Projection of the OPE Potential}

We consider now the projection of the coordinate space potential
into the partial wave basis.
For that we work with baryon-antibaryon states with well-defined total
angular momentum and parity $J^P$.
If the total strangeness of the baryon-antibaryon state is zero,
we will consider states with well-defined C-parity
(for neutral systems) or G-parity (if the system is not electrically neutral).
Besides we will only consider states that contain an S-wave,
as they are the more likely to form a bound state.
If we use the spectroscopic notation $^{2S+1}L_J$ to denote
the partial waves, we have the following combinations
\begin{eqnarray}
| B_Q \bar{B}_6 (0^{-}) \rangle &=& \{ ^1S_0 \} \, , \\
| B_Q \bar{B}_6 (1^{-}) \rangle &=& \{ ^3S_1, {}^3D_1 \} \, , \\
\nonumber \\
| B_Q \bar{B}_6^* (1^{-}) \rangle &=&
\{ ^3S_1 , {}^3D_1, {}^5D_1 \} \, , \\
| B_Q \bar{B}_6^* (2^{-}) \rangle &=&
\{ ^3D_2, {}^5S_2 , {}^5D_2 , {}^5G_2 \}
\label{eq:basis-2}
\, , \\
\nonumber \\
| B_6^* \bar{B}_6^* (0^{-}) \rangle &=& \{ ^1S_0 , {}^5D_0 \}
\, , \\
| B_6^* \bar{B}_6^* (1^{-}) \rangle &=& \{ ^3S_1 , {}^3D_1 , {}^7D_1 ,{}^7G_1 \}
\, , \\
| B_6^* \bar{B}_6^* (2^{-}) \rangle &=& \{ ^1D_2 , {}^5S_2 , {}^5D_2, {}^5G_2 \}
\, , \\
| B_6^* \bar{B}_6^* (3^{-}) \rangle &=&
\{ ^3D_3 , {}^3G_3 , {}^7S_3 ,{}^7D_3 ,{}^7G_3, {}^7I_3 \}
\, , 
\label{eq:basis-3}
\end{eqnarray}
with $B_Q = B_{\bar 3}$ or $B_6$.
The calculation of the matrix elements is in general straightforward,
where we refer to Appendix~\ref{app:ope} for the details.
The result of these calculations is that the $C_{12}$ and $S_{12}(\hat{r})$
operators can be expressed as matrices, which we denote
with the ${\bf C}_{12}$ and ${\bf S}_{12}$ notation.
With this in mind for $r > 0$ we write the OPE potential  as
\begin{eqnarray}
  V_{\rm F}^{(0)}(r) = R_1 \bar{R}_2\,\vec{I}_1 \cdot \vec{I}_2\,
  \left[
    {\bf C}_{12} W_C(r) + {\bf S}_{12} W_T(r) \right] \, , \nonumber \\
\end{eqnarray}
where the dimension of the matrices is set by the number of partial waves.
The explicit matrices that apply in each case can be consulted
in Appendix~\ref{app:ope}, where it is also explained
how they are calculated.

\section{How to Count the One Pion Exchange Potential}
\label{sec:perturbative}
 
The ${\rm LO}$ EFT heavy baryon-antibaryon potential can in principle
contain a contact- and a finite-range piece, where the latter is
the well-known OPE potential.
While there is no {\it a priori} way to determine
if the contact-range potential is perturbative,
this is not the case for the OPE potential where there exist a series of
theoretical developments to evaluate its strength.
In this section we will check how these ideas apply
to the central and tensor pieces of the OPE potential.
Before starting the discussion it is important to stress
that we make a very explicit distinction between iterated OPE
(or reducible multi-pion exchange) and irreducible multi-pion exchanges.
The former is merely the outcome of iterating the EFT potential
within the Schr\"odinger or Lippmann-Schwinger equations
while the latter is a genuine contribution to the EFT potential,
though a subleading one: the lowest order two pion exchange irreducible
diagrams enter at order $Q^2$ in the chiral expansion.

\subsection{The Central Potential}

The perturbative nature of the central piece of OPE can be determined
from the comparison of tree-level versus once-iterated central OPE,
i.e. $V$ and $V G_0 V$ in operator form.
This type of comparison was made in Ref.~\cite{Fleming:1999ee}
in the context of nucleon-nucleon scattering~\footnote{Recently a more
sophisticated method for determining the perturbativeness of OPE has
been developed in Ref.~\cite{PavonValderrama:2016lqn}
for peripheral waves with $L \geq 1$.
Unfortunately it has not been extended yet to S-waves.}.
Here we are merely adapting it to the particular case of
the heavy baryon-antibaryon system.
The ratio of iterated vs tree-level OPE
can be expressed as a ratio of scales
\begin{eqnarray}
\frac{\langle p | V G_0 V  | p \rangle}{\langle p | V | p \rangle} =
\frac{Q}{\Lambda_C} \, ,
\end{eqnarray}
where $Q$ is a light scale (either the external momentum $p$ or the pion mass
$m_{\pi}$) and $\Lambda_C$ is a scale that characterizes central OPE.
The evaluation of this ratio at $p = 0$ leaves the pion mass as the only
light scale left, in which case we obtain the following value
for the central scale
\begin{eqnarray}
\Lambda_C =
\frac{1}{| \sigma \, \tau |}\,
\frac{24 \pi f_{\pi}^2}{\mu |R_1 \bar{R}_2|\, g_i^2}  \, ,
\label{eq:central-scale}
\end{eqnarray}
with $\mu$ the reduced mass of the system,
$\sigma$ and $\tau$ the evaluation of the spin and isospin operators
corresponding to the particular case under consideration and
where $R_1$ and $\bar{R}_2$ can be consulted on Table~\ref{tab:vertices}.
For the charm and bottom sectors the value is respectively
\begin{eqnarray}
\Lambda_C (Q = c) &=&
\frac{1060 \, {\rm MeV}}{| \sigma \tau |\,| R_1 \bar{R}_2 |\,g_i^2} \, , \\
\Lambda_C (Q = b) &=&
\frac{450 \, {\rm MeV}}{| \sigma \tau |\,| R_1 \bar{R}_2 |\,g_i^2} \, ,
\end{eqnarray}
which depend on the value of the couplings $g_2$ and $g_3$.

The discussion about the values of the axial couplings,
and in particular $g_2$, is important because it can change
the value of $\Lambda_C$ by a large factor.
For the $T\bar{S}$ and $S\bar{S}$ molecules the value of $\Lambda_C$
in the charm sector is
\begin{eqnarray}
\Lambda_C^{T\bar{S}}(Q=c) &\sim&
\frac{1120^{+100}_{-40} \, {\rm MeV}}{| \sigma \tau |\,| R_1 \bar{R}_2 |} \, ,
\\
\Lambda_C^{S\bar{S}}(Q=c) &\sim&
\frac{(560-1500) \, {\rm MeV}}{| \sigma \tau |\,| R_1 \bar{R}_2 |} \, ,
\end{eqnarray}
where $\Lambda_C^{S\bar{S}}$ can change almost by a a factor of $3$
owing to the uncertainty of $g_2$ (notice that instead of a number
with an error, we have simply indicated a range of possible values).
In the bottom sector it is instead more advisable to use the lattice QCD
determination for $g_2$ and $g_3$, leading to
\begin{eqnarray}
\Lambda_C^{T\bar{S}}(Q=b) &\sim&
\frac{900^{+440}_{-260}\, {\rm MeV}}{| \sigma \tau |\,| R_1 \bar{R}_2 |} \, , \\
\Lambda_C^{S\bar{S}}(Q=b) &\sim&
\frac{660^{+440}_{-220}\, {\rm MeV}}{| \sigma \tau |\,| R_1 \bar{R}_2 |} \, .
\end{eqnarray}
The previous values have to be combined with
the $| \sigma \tau |\,| R_1 \bar{R}_2 |$ factor.
The maximum value of this factor happens
for the channels with lowest spin and isospin.
In Table \ref{tab:Lambda-C} we list the scale $\Lambda_C$ for a
few representative heavy baryon-antibaryon molecules.
In general $\Lambda_C \sim 0.5\,{\rm GeV}$ (if not harder) in most cases,
which means that we expect central OPE to be perturbative.
The exceptions are the isoscalar $0^{-+}$ $\Sigma_Q \bar{\Sigma}_Q$ and
$0^{-+}$, $1^{--}$ $\Sigma_Q^* \bar{\Sigma}_Q^*$ molecules,
at least in the bottom sector.
In the charm sector the scale $\Lambda_C$ varies considerably
as a consequence of the uncertainty of the axial coupling $g_2$.
In particular if the absolute value of the axial coupling $|g_2|$ is on
the high end, i.e. the value $g_2 = -1.38$ deduced from the quark model,
central OPE will be important for certain molecules
in the hidden charm sector.
In the bottom sector the situation is more clear: central OPE will
be non-perturbative for the aforementioned $\Sigma_Q \bar{\Sigma}_Q$ and
$\Sigma_Q^* \bar{\Sigma}_Q^*$ molecules.
Finally for having a comparison with a well-known state, we mention that
the central scale for OPE in the two-nucleon system is
$\Lambda_C \simeq 590\,{\rm MeV}$.

\begin{table*}
\begin{center}
\begin{tabular}{|c|c|c|c|c|c|c|c|}
\hline \hline
Channel & $I$ & Sign & $\Lambda_C$ &
Channel & $I$ & Sign & $\Lambda_C$ \\
\hline
$\Xi_c \bar{\Xi}_c' (0^{-\pm})$ & $0$ & $\mp$ & $720^{+70}_{-30}$ &
$\Xi_b \bar{\Xi}_b' (0^{-\pm})$ & $0$ & $\mp$ & $580^{+290}_{-170}$
\\
$\Lambda_c \bar{\Sigma}_c (0^{-\pm})$ & $1$ & $\pm$ & $580^{+50}_{-20}$ &
$\Lambda_b \bar{\Sigma}_b (0^{-\pm})$ & $1$ & $\pm$ & $450^{+220}_{-130}$
\\
\hline
$\Xi_c' \bar{\Xi}_c' (0^{-+})$
& $0$ & $-$ & $530-1420$ &
$\Xi_b' \bar{\Xi}_b' (0^{-+})$
& $0$ & $-$ & $620^{+450}_{-210}$ \\
$\Sigma_c \bar{\Sigma}_c (0^{-+})$
& $0$ & $-$ & $210-560$ &
$\Sigma_b \bar{\Sigma}_b (0^{-+})$
& $0$ & $-$ & $240^{+170}_{-80}$ \\
\hline
$\Xi_c^* \bar{\Xi}_c' (1^{-+})$
& $0$ & $-$ & $580-1560$ &
$\Xi_b^* \bar{\Xi}_b' (1^{-+})$
& $0$ & $-$ & $680^{+490}_{-230}$  \\
$\Sigma_c^* \bar{\Sigma}_c (1^{-+})$
& $0$ & $-$ & $230-610$ &
$\Sigma_b^* \bar{\Sigma}_b (1^{-+})$
& $0$ & $-$ & $260^{+190}_{-90}$  \\
$\Sigma_c^* \bar{\Sigma}_c (1^{--})$
& $0$ & $-$ & $280-750$&
$\Sigma_b^* \bar{\Sigma}_b (1^{--})$
& $0$ & $-$ & $320^{+230}_{-110}$ \\
$\Sigma_c^* \bar{\Sigma}_c (2^{--})$
& $0$ & $+$ & $280-750$ &
$\Sigma_b^* \bar{\Sigma}_b (2^{--})$
& $0$ & $+$ & $320^{+230}_{-110}$ \\
\hline
$\Xi_c^* \bar{\Xi}_c^* (0^{-+})$
& $0$ & $-$ & $410-1100$ &
$\Xi_b^* \bar{\Xi}_b^* (0^{-+})$
& $0$ & $-$ & $490^{+350}_{-170}$  \\
$\Sigma_c^* \bar{\Sigma}_c^* (0^{-+})$
& $0$ & $-$ & $160-440$ &
$\Sigma_b^* \bar{\Sigma}_b^* (0^{-+})$
& $0$ & $-$ & $190^{+140}_{-70}$  \\
$\Sigma_c^* \bar{\Sigma}_c^* (1^{--})$
& $0$ & $-$ & $220-590$ &
$\Sigma_b^* \bar{\Sigma}_b^* (1^{--})$
& $0$ & $-$ & $260^{+190}_{-90}$  \\
$\Sigma_c^* \bar{\Sigma}_c^* (3^{--})$
& $0$ & $+$ & $270-740$ &
$\Sigma_b^* \bar{\Sigma}_b^* (3^{--})$
& $0$ & $+$ &  $320^{+230}_{-110}$ \\
\hline \hline
\end{tabular}
\end{center}
\caption{
  {
    The central scale $\Lambda_C$ (in units of ${\rm MeV}$)
    for a few selected heavy baryon-antibaryon molecules.
  For momenta above this scale, $p > \Lambda_C$,
  the central force becomes non-perturbative.
  The scale $\Lambda_C$ is inversely proportional to $\sigma \tau$,
  reaching its minimum for the channels with the lowest total spin
  and isospin and growing quickly for other configurations.
  The selection includes the heavy baryon-antibaryon states for which
  $\Lambda_C < 0.5\,{\rm GeV}$, most of which are of the
  $\Sigma_Q^{(*)} \bar{\Sigma}_Q^{(*)}$ type.
  In addition we include the $\Xi_Q^{('/*)} \bar{\Xi}_Q^{('/*)}$ molecule
  for which the central force is strongest.
  and the antitriplet-sextet molecules.
  The uncertainty in the antitriplet-sextet case comes from the axial coupling,
  which is taken to be $g_3 = 0.973^{+0.019}_{-0.042}$ in the charm
  sector~\cite{Cheng:2015naa} and $g_3 = 0.71 \pm 0.13$
  in the bottom one~\cite{Detmold:2012ge}.
  For the sextet-sextet case in the charm sector the axial coupling is not
  well determined and we take it in the range $|g_2| \sim 0.84-1.38$,
  while in the bottom sector
  we have $g_2 = -0.84 \pm 0.20$~\cite{Detmold:2012ge}.
  For comparison in the deuteron channel of the two-nucleon system
  the central scale is $\Lambda_C \simeq 590\,{\rm MeV}$.
  }
} \label{tab:Lambda-C}
\end{table*}

\subsection{The Tensor Potential}

The tensor piece of the OPE potential requires a more involved analysis.
A direct comparison of $V$ and $V G_0 V$ is not possible.
The reason is that the iteration of the tensor piece of OPE diverges,
see for instance Refs.~\cite{Valderrama:2012jv,PavonValderrama:2016lqn}
for a detailed explanation.
Thus we must resort to a method that does not involve the direct evaluation
of iterated tensor OPE.

The type of power-law behaviour of the tensor OPE potential is analogous
to the behaviour of a few physical systems studied in atomic physics.
The potential between two dipoles is of the $1/r^3$ type,
just like the tensor force.
The failure of standard perturbation theory for these systems is well-known
in atomic physics, where alternative techniques have been developed to
deal with this type of potentials.
The work of Cavagnero~\cite{Cavagnero:1994zz} explains that the divergences of
perturbation theory in these type of systems is similar to the role of
{\it secular} perturbations in classical mechanics, i.e.
a type of perturbation that is small at short time scales
but ends up diverging at large time scales.
The solution is to redefine (or, loosely speaking, {\it renormalize}~\footnote{
  We simply adopt the terminology in use in the field of atomic physics
  for these redefinitions in secular perturbation theory,
  though it does not exactly corresponds
  with the standard meaning of renormalization.})
some quantity in order to obtain a finite result again.
For the perturbative series of the $1/r^3$ potentials the quantity
we renormalize is the angular momentum.
The zero-th order term in the perturbative expansion
of the wave function is now
\begin{eqnarray}
  \Psi_l^{(0)}(r ; k) = \frac{J_{\nu(k a_3)}(k r)}{\sqrt{r}} \, ,
  \label{eq:wf-zeroth}
\end{eqnarray}
instead of the standard $\frac{J_{l + 1/2}(k r)}{\sqrt{r}}$,
where $J_{\nu}(z)$ refers to the Bessel function of order $\nu$.
In the expression above $\nu$ is the {\it renormalized} angular momentum,
which happens to be a function of the momentum $k$ and a
length scale $a_3$ that is related to the strength of the potential
(it will be defined later).
The secular series is built not only by adding higher order terms
but also by making $\nu (x)$ to depend on $\kappa = k a_3$.
If we switch off the potential and take $k a_3 = 0$,
we have $\nu(0) = l + \frac{1}{2}$ and we recover the free wave function.
For small enough values of $\kappa$ we expect $\nu(\kappa)$ to be expansible
in powers of $x$, i.e. to be perturbative.
By reexpanding the secular series and the renormalized angular momentum
we can recover the original perturbative series.
However the interesting feature of the series above is that we can
determine the values of $\kappa$ for which $\nu (\kappa)$ is analytic.
When $\nu(\kappa)$ is not analytic, it does not admit a power series
in $\kappa$ anymore.
This in turn means that there is no way to rearrange the secular series
into the standard perturbative series, leading to its failure.

For the $1/r^3$ potential, which is equivalent to the tensor force
for distances $m_{\pi} r < 1$, the secular series has been studied
in detail by Gao~\cite{Gao:1999zz} for the uncoupled channel case.
Birse~\cite{Birse:2005um} extended the previous techniques
for the coupled channel case and particularized them
for the nucleon-nucleon system.
In a previous publication by one of the authors~\cite{Valderrama:2012jv}
the analysis of Birse was applied to the heavy meson-antimeson system.
In this work we extend it to the heavy baryon-antibaryon system.

We will consider the tensor force in the limit $m_{\pi} \to 0$,
for which the OPE potential can be written as
\begin{eqnarray}
2\mu \, {\bf V}(r) =
\frac{a_3}{r^3} \, {\bf S}_j \, ,
\label{eq:ope-chiral}
\end{eqnarray}
where the potential is a matrix in the coupled channel space
and ${\bf S}_j$ is the tensor operator (in matrix form,
as written in Sect.~\ref{subsec:ope}), with
$j$ referring to the total angular momentum.
We have that $\mu$
is the reduced mass of the heavy baryon-antibaryon
system and $a_3$ is the length scale that determines the strength of the tensor
piece of the potential.
The potential in this limit is amenable to the secular perturbative series
developped in Refs.~\cite{Gao:1999zz,Birse:2005um,Valderrama:2012jv}.
The corrections stemming from the finite value of $m_{\pi}$ where considered
in Ref.~\cite{Valderrama:2012jv} and will be discussed later on
in this section.

\subsection{The Renormalized Angular Momentum}

\begin{table}
\begin{center}
\begin{tabular}{|c|c|}
\hline \hline
Channel & $\kappa_c$ \\
\hline
$B_3\bar{B}_6(0^-)$/$B_6\bar{B}_6(0^-)$ & - \\
$B_3\bar{B}_6(1^-)$/$B_6\bar{B}_6(1^-)$ & 0.6835  \\
\hline
$B_3\bar{B}_6^*(1^{-})$ & 1.412 \\
$B_3\bar{B}_6^*(2^{-})$ & 1.934 \\
\hline
$B_6\bar{B}_6^*(1^{-+})$ & 0.8533 \\
$B_6\bar{B}_6^*(1^{--})$ & 0.7264 \\
$B_6\bar{B}_6^*(2^{-+})$ & 0.5784 \\
$B_6\bar{B}_6^*(2^{--})$ & 0.6998 \\
\hline
$B_6^*\bar{B}_6^*(0^{-+})$ & 0.6378 \\
$B_6^*\bar{B}_6^*(1^{--})$ & 0.6674 \\
$B_6^*\bar{B}_6^*(2^{-+})$ & 0.6833 \\
$B_6^*\bar{B}_6^*(3^{--})$ & 0.5922 \\
\hline \hline
\end{tabular}
\end{center}
\caption{
  Reduced critical momenta $\kappa_c$ for the different S-wave
  heavy baryon-antibaryon systems $B' \bar{B}$.
} \label{tab:kappa-c}
\end{table}

Now we explain how the secular perturbation theory looks like and most
importantly, how to calculate the renormalized angular momenta $\nu$.
We begin by writing the reduced Schr\"odinger equation in coupled channels
(the uncoupled channel can be consulted in Ref.~\cite{Valderrama:2012jv})
for the particular case of a pure $1/r^3$ potential
\begin{eqnarray}
\label{eq:schroe_r3_c}
- {\bf u}_{k,j}'' + \left[ {\bf S}_j\,\frac{a_3}{r^3}
  + \frac{{\bf L}_j^2}{r^2} \right]\,{\bf u}_{k,j}(r) =
k^2 \, {\bf u}_{k,j}(r) \, ,
\end{eqnarray}
where we are considering $N$ angular momentum channels.
Notice that we have taken the chiral limit $m_{\pi} \to 0$, which means
that only the tensor piece of the OPE potential survives,
see Eq.~(\ref{eq:ope-chiral}).
In the equation above ${\bf S}_j$ is the tensor operator matrix,
while ${\bf L}_j$ is a diagonal matrix representing the angular
momentum operator $\vec{L}^2$
\begin{eqnarray}
{\bf L}_j^2 = {\rm diag}(l_1(l_1 + 1), \dots, l_N(l_N + 1)) \, .
\end{eqnarray}
The reduced wave function ${\bf u}_{k,j}$ is an $N \times N$ matrix,
where column $j$ represents a solution that behaves as a free wave
with angular momentum $l_j$ when we take $a_3 \to 0$.
The solution of the Schr\"odinger equation is a linear combination of
the functions ${\bm \xi}$ and ${\bm \eta}$:
\begin{eqnarray}
{\bf u}_{k,j}(r) = \sum_{\{l_j\}}\,
\left[ \alpha_{l_j} {\bm \xi}_{l_j}(r; k) + \beta_{l_j} {\bm \eta}_{l_j}(r; k)
\right] \, ,
\end{eqnarray}
where we sum over the possible values of the angular momenta
and with $\bm \xi$ and $\bm \eta$ $N$-component vectors
that can be written as sums of Bessel functions:
\begin{eqnarray}
{\bm \xi}_{l_j}(r; k) &=&
\sum_{m=-\infty}^{\infty} {\bm b}_m(\nu_{l_j})\,
\sqrt{r}\,J_{m + \nu_{l_j}}(k r) \, , \label{eq:xi-exp} \\
{\bm \eta}_{l_j}(r; k) &=&
\sum_{m=-\infty}^{\infty} (-1)^m {\bm b}_{-m}(\nu_{l_j})\,
\sqrt{r}\,J_{-m - \nu_{l_j}}(k r) \, , \nonumber \\ \label{eq:eta-exp}
\end{eqnarray}
where the $\nu_{l_j}$'s are functions of $\kappa = k a_3$,
i.e. $\nu_{l_j} = \nu_{l_j}(\kappa)$.
The $\nu_{l_j}$'s are the {\it renormalized} angular momenta
that we previously introduced in Eq.~(\ref{eq:wf-zeroth}).
In turn the  expansion of the reduced wave functions $\bm \xi$ and $\bm \eta$
in Eqs.~(\ref{eq:xi-exp}) and (\ref{eq:eta-exp}) is simply the extension
of Eq.~(\ref{eq:wf-zeroth}) to arbitrary orders~\footnote{We note
  that Eq.~(\ref{eq:wf-zeroth}) is written in terms of the standard
  wave function, while Eqs.~(\ref{eq:xi-exp}) and (\ref{eq:eta-exp})
  use the reduced wave function instead.
}.
We have $N$ different solutions for $\bm \xi$ and $\bm \eta$ that
we have labeled with the subscript ${l_j}$ to indicate
that for $\kappa = 0$ they behave as a free wave of
angular momentum $l_j$.
The recursive relation from which one can compute
${\bm b}_{m}(\nu_{l_j})$ can be consulted in Ref.~\cite{Birse:2005um}
but are of no concern if we are only interested in the $\nu_{l_i}$'s.
For $\kappa = 0$ only the ${\bm b}_{m}(\nu_{l_j})$
coefficient for $m = 0$ survives.

The renormalized angular momenta $\nu_{l_i} = \nu_{l_i}(\kappa)$
(with $\kappa = k a_3$) can be calculated as follows.
First we define the following ${N} \times {N}$ matrix
\begin{eqnarray}
{\bf F}_j(\nu, \kappa) \equiv {\bf f}_j(\nu) - \frac{\kappa^2}{\nu}\,
\left[ {\bf R}_1(\nu) - {\bf R}_1(-\nu)\right] \, ,
\end{eqnarray}
which depends on two other matrices, ${\bf f}_j(\nu)$ and ${\bf R}_1(\nu)$;
${\bf f}_j(\nu)$ is a diagonal matrix defined as
\begin{eqnarray}
\frac{{\bf f}_j(\nu)}{2 \, \nu} = {\rm diag}(
\nu^2 - (l_1 + \frac{1}{2})^2, \dots,
\nu^2 - (l_N + \frac{1}{2})^2) \, , \nonumber \\
\end{eqnarray}
while ${\bf R}_1(\nu)$ is given by the recursive relation
\begin{eqnarray}
{\bf R}_n(\nu) &=&
{\left[ {\bf f}_j(n+\nu) - \kappa^2\,{\bf S}_j\,{\bf R}_{n+1}(\nu)\,{\bf S}_j
    \right]}^{-1 } \, , \nonumber \\
\label{eq:R-recursive}
\end{eqnarray}
which can be accurately solved with between 20 and 30
iterations~\cite{Birse:2005um} (that is, one takes ${\bf R}_N = 0$ for
large enough $N$, e.g. $20-30$, and solves the recursion relation backwards).
Once we have the matrix $F_j$, we obtain $\nu_{l_i} = \nu_{l_i}(\kappa)$
by finding the zeros of
\begin{eqnarray}
\label{eq:characteristic-c}
\det{\left( {\bf F}_j(\nu, \kappa) \right)} = 0 \, ,
\end{eqnarray}
This equation admits $N$ solutions, one for each value of the angular momentum.
For $\kappa = 0$ these solutions behave as
\begin{eqnarray}
\nu_{l_i}(\kappa = 0) = l_i + \frac{1}{2} \, ,
\end{eqnarray}
with $i = 1, \dots, N$.
As $\kappa$ increases $\nu_{l_i}(\kappa)$ moves slowly downwards.
Once we reach $\nu_{l_i} = l_i$ at the critical value $\kappa = \kappa_c$,
we have that $\nu_{l_i}$ splits into the complex conjugate solutions
$\nu_{l_i}(\kappa) = l_i \pm i \rho_{l_i}(\kappa)$.
This is a {non-analyticity} which marks the point above
which $\nu_{l_i}(\kappa)$ cannot be expressed
as a perturbative series.
This in turn defines $\kappa_c$, the critical value of $\kappa$
for which there is a $\nu_{l_i}(\kappa)$ that becomes
non-analytic in $\kappa$.
Usually the first $\nu_{l_j}$ to split is the one that corresponds
to the smallest angular momentum and also the one that determines
the breakdown of the perturbative series.

For computing the critical momenta we need first the matrix elements of
the tensor operator in the channel under consideration.
For the $B_{\bar 3} \bar{B}_{\bar 3}$, $B_6 \bar{B}_6$ and $B_6^* \bar{B}_6^*$
molecules this is trivial: first we take the tensor force matrix
${\bf S}_j$, which can be found in Eq.~(\ref{eq:S12_1}) and
Eqs.~(\ref{eq:S12_ss_0}-\ref{eq:S12_ss_3}) of Appendix \ref{app:ope},
then plug this matrix into Eq.~(\ref{eq:R-recursive}), from which
finally we solve Eq.~(\ref{eq:characteristic-c})
to obtain $\kappa_c$.
The $B_3 \bar{B}_6^*$ molecules the procedure is the same,
with the tensor force matrices defined
in Eqs.(\ref{eq:S12_E_1C}) and (\ref{eq:S12_E_2C}).
The $B_6 \bar{B}_6^*$ molecules are the most complicated because they contain
a direct and exchange tensor operator, which mediate
the $B_6 \bar{B}_6^* \to B_6 \bar{B}_6^*$ and
$B_6 \bar{B}_6^* \to B_6^* \bar{B}_6$ potential,
respectively.
In addition the effective pion masses are different for the direct and
exchange tensor operators.
Here we ignore this effect: in the present calculation
we are making the approximation that HQSS is exact and
therefore there is no mass splitting between
the $B_6$ and $B_6^*$ heavy baryons.
However the length scale $a_3$ is different for the direct and exchange
operators, i.e.
\begin{eqnarray}
2\mu \, {\bf V}_{B_6\bar{B}_6^*}(r) =
\frac{a_3^D}{r^3} \, {\bf S}^D_j
+ \frac{a_3^E}{r^3} \, {\bf S}^E_j \, .
\end{eqnarray}
It happens that both scales are proportional to each other
\begin{eqnarray}
a_3^E = -\frac{3}{4}\,a_3^D \, ,
\end{eqnarray}
as can be checked by inspecting the coupled-channel form of
the potential in Eq.~(\ref{eq:V_SS}) from Appendix~\ref{app:ope}.
Thus in the $B_6\bar{B}_6^*$ system we will be computing the critical values
of the matrix
\begin{eqnarray}
  {\bf S}_j = {\bf S}_j^D - \frac{3}{4}\,{\bf S}_j^E \, .
  \label{eq:SDE}
\end{eqnarray}

\subsection{Critical Momenta}

The critical $\kappa_c$ for which the convergence criterion fails
are listed in Table \ref{tab:kappa-c} for the different possible
heavy baryon-antibaryon states that contain an S-wave.
The previous values have been obtained under the assumption
that the effective pion mass can be taken to be zero.
The effect of finite pion mass was considered in Ref.~\cite{Valderrama:2012jv},
where it was found that it increases the range of momenta where
the tensor part of OPE is perturbative by the following factor
\begin{eqnarray}
  \kappa_c(m_{\pi}) = \kappa_c \, e^{+m_{\pi} R_c} \, ,
  \label{eq:kappa-mpi}
\end{eqnarray}
where $R_c$ is the radius below which we do not expect
the OPE potential to be valid.
The value of this radius is rather ambiguous.
In Ref.~\cite{Valderrama:2012jv} the estimation $R_c = 0.5-0.8\,{\rm fm}$
was proposed, yielding
\begin{eqnarray}
  \kappa_c(m_{\pi}) \simeq 1.5\,\kappa_c \, .
\end{eqnarray}
Higher values might be more appropriate indeed,
but here we will stick to this value.

To obtain the critical momenta we multiply $\kappa_c(m_{\pi})$ by
the relation $k_c = \kappa_c(m_{\pi}) / | a_3 |$,
where $a_3$ is the tensor length scale.
If we match the $m_{\pi} r \to 0$ limit of the OPE potential
to the $1/r^3$ form we have used to derive $\kappa_c$,
we find that
\begin{eqnarray}
| a_3 | = | R_1 \bar{R}_2\,\tau |
\,\frac{\mu g_i^2}{4 \pi f_{\pi}^2} \,
\, ,
\end{eqnarray}
where $\tau = \vec{T}_1 \cdot \vec{T}_2$ and $g_i = g_2$, $g_3$ depending
on whether we are considering the $S \bar{S}$ or the $T \bar{S}$ potential.
The factors $R_1$ and $\bar{R}_2$ and the proper isospin operator
to use can be checked in Table~\ref{tab:vertices}.
For the $B_6\bar{B}_6^*$
case we will use the factors corresponding to the direct
channels, i.e. $B_6 \to B_6$ and $B_6^* \to B_6^*$,
in agreement with the convention that we have used in Eq.~(\ref{eq:SDE})
for writing their tensor matrices.
From the previous, we can define the {\it tensor scale} as
\begin{eqnarray}
\Lambda_T(m_{\pi}) = \frac{\kappa(m_{\pi})}{| a_3 |} =
\frac{\kappa(m_{\pi})}{| R_1 \bar{R}_2\tau |}\,
\frac{4 \pi f_{\pi}^2}{\mu g_i^2} \, ,
\label{eq:tensor-scale}
\end{eqnarray}
which is useful because it allows a direct comparison with the central scale
$\Lambda_C$ that we defined in Eq.~(\ref{eq:central-scale}).
If we particularize for the $T \bar{S}$ and $S\bar{S}$ in the charm sector
\begin{eqnarray}
\Lambda_T^{T\bar{S}}(m_{\pi}, Q=c) &\simeq&
\kappa_c(m_{\pi})\,
\frac{186^{+18}_{-8} \, {\rm MeV}}{| R_1 \bar{R}_2\,\tau |}
\, ,  \\
\Lambda_3^{S\bar{S}}(m_{\pi}, Q=c) &\simeq&
\kappa_c(m_{\pi})\,
\frac{90-250 \, {\rm MeV}}{| R_1 \bar{R}_2\,\tau |}
\, ,
\end{eqnarray}
while for the bottom sector we obtain
\begin{eqnarray}
  \Lambda_T^{T\bar{S}}(m_{\pi}, Q=b) &\simeq&
  \kappa_c(m_{\pi})\,
\frac{140^{+80}_{-40} \, {\rm MeV}}{| R_1 \bar{R}_2\,\tau |}
\, ,  \\
\Lambda_T^{S\bar{S}}(m_{\pi}, Q=b) &\simeq&
\kappa_c(m_{\pi})\,
\frac{100^{+80}_{-40} \, {\rm MeV}}{| R_1 \bar{R}_2\,\tau |}
\, ,
\end{eqnarray}
where we have used the lattice QCD values of
$g_2$ and $g_3$~\cite{Detmold:2012ge}.
These scales look rather soft at first sight
but the factors $R_1$, $\bar{R}_2$ and $\tau$
will increase the values of $\Lambda_T$ considerably in most cases.
A few representative values of $\Lambda_T$ are compiled
in Table \ref{tab:Lambda-T} both for the chiral limit
and the physical pion mass.
In general we find that tensor OPE is considerably stronger than central OPE,
For the $\Xi_Q \bar{\Xi}_Q'$ and $\Lambda_Q \bar{\Sigma}_Q$ molecules
we find that $\Lambda_T$ is markedly softer than $\Lambda_C$,
and in the bottom sector the tensor force probably requires
a non-perturbative treatment.
For the iso-scalar $\Sigma_Q \bar{\Sigma}_Q$, $\Sigma_Q^* \bar{\Sigma}_Q$,
$\Sigma_Q^* \bar{\Sigma}_Q^*$ molecules the tensor scale $\Lambda_T$ is
moderately soft, particularly in the bottom sector.
We notice that the same comments are also valid in the two-nucleon system,
in which $\Lambda_T(0) = 66\,{\rm MeV}$ in the chiral limit~\cite{Birse:2005um}
and $\Lambda(m_{\pi}) = 99\,{\rm MeV}$ for the physical pion mass.

\begin{table*}
\begin{center}
\begin{tabular}{|c|c|c|c|c|c|c|c|}
\hline \hline
Channel & $I$ & $\Lambda_T(0)$ & $\Lambda_T(m_{\pi})$ &
Channel & $I$ & $\Lambda_T(0)$ & $\Lambda_T(m_{\pi})$ \\
\hline
$\Xi_c \bar{\Xi}_c' (1^{-\pm})$ & $0$  & $246^{+23}_{-10}$ & $368^{+34}_{-14}$ &
$\Xi_b \bar{\Xi}_b' (1^{-\pm})$ & $0$  & $200^{+100}_{-60}$ & $300^{+150}_{-90}$
\\
$\Lambda_c \bar{\Sigma}_c (1^{-\pm})$ & $1$ & $196^{+18}_{-7}$ & $295^{+27}_{-11}$ &
$\Lambda_b \bar{\Sigma}_b (1^{-\pm})$ & $1$ & $150^{+80}_{-40}$ & $230^{+110}_{-70}$
\\
\hline
$\Xi_c' \bar{\Xi}_c' (1^{--})$
& $0$ & $180-490$ & $270-740$ &
$\Xi_b' \bar{\Xi}_b' (1^{--})$
& $0$ & $210^{+160}_{-70}$ & $320^{+230}_{-110}$ \\
$\Sigma_c \bar{\Sigma}_c (1^{--})$
& $0$ & $70-190$ & $110-290$ &
$\Sigma_b \bar{\Sigma}_b (1^{--})$
& $0$ & $80^{+60}_{-30}$ & $120^{+90}_{-40}$ \\
\hline
$\Xi_c^* \bar{\Xi}_c' (2^{-+})$
& $0$ & $150-410$ & $230-610$ &
$\Xi_b^* \bar{\Xi}_b' (2^{-+})$
& $0$ & $180^{+130}_{-60}$ & $270^{+200}_{-90}$ \\
$\Sigma_c^* \bar{\Sigma}_c (1^{-+})$
& $0$ & $90-240$ & $130-350$ &
$\Sigma_b^* \bar{\Sigma}_b (1^{-+})$
& $0$ & $100^{+80}_{-30}$ & $150^{+110}_{-50}$   \\
$\Sigma_c^* \bar{\Sigma}_c (1^{--})$
& $0$ & $70-200$ & $110-300$ &
$\Sigma_b^* \bar{\Sigma}_b (1^{--})$
& $0$ & $90^{+60}_{-30}$ & $130^{+90}_{-40}$   \\
$\Sigma_c^* \bar{\Sigma}_c (2^{-+})$
& $0$ & $60-160$ & $90-240$ &
$\Sigma_b^* \bar{\Sigma}_b (2^{-+})$
& $0$ & $70^{+50}_{-20}$ & $100^{+80}_{-30}$   \\
$\Sigma_c^* \bar{\Sigma}_c (2^{--})$
& $0$ & $70-190$ & $110-290$ &
$\Sigma_b^* \bar{\Sigma}_b (2^{--})$
& $0$ & $80^{+60}_{-30}$ & $130^{+90}_{-50}$   \\
\hline
$\Xi_c^* \bar{\Xi}_c' (3^{--})$
& $0$ & $160-420$ & $230-630$ &
$\Xi_b^* \bar{\Xi}_b' (3^{--})$
& $0$ & $190^{+130}_{-70}$ & $280^{+200}_{-100}$ \\
$\Sigma_c^* \bar{\Sigma}_c^* (0^{-+})$
& $0$ & $60-170$ & $100-260$ &
$\Sigma_b^* \bar{\Sigma}_b^* (0^{-+})$
& $0$ & $80^{+50}_{-30}$ & $110^{+80}_{-40}$  \\
$\Sigma_c^* \bar{\Sigma}_c^* (1^{--})$
& $0$ & $70-180$ & $100-270$ &
$\Sigma_b^* \bar{\Sigma}_b^* (1^{--})$
& $0$ & $80^{+60}_{-30}$ & $120^{+90}_{-40}$  \\
$\Sigma_c^* \bar{\Sigma}_c^* (2^{-+})$
& $0$ & $70-190$ & $100-280$ &
$\Sigma_b^* \bar{\Sigma}_b^* (2^{-+})$
& $0$ & $80^{+60}_{-30}$ & $120^{+90}_{-40}$  \\
$\Sigma_c^* \bar{\Sigma}_c^* (3^{--})$
& $0$ & $60-160$ & $90-240$ &
$\Sigma_b^* \bar{\Sigma}_b^* (3^{--})$
& $0$ & $70^{+50}_{-20}$ & $100^{+80}_{-40}$  \\
\hline \hline
\end{tabular}
\end{center}
\caption{ The tensor scale $\Lambda_T$ (in units of ${\rm MeV}$)
  for a series of heavy baryon-antibaryon molecules.
  For momenta above this scale, $p > \Lambda_T$,
  the tensor force becomes non-perturbative.
  The tensor scale depends on the pion mass:
  $\Lambda_T(0)$ and $\Lambda_T(m_{\pi})$ represent
  the value in the chiral limit and physical pion mass, respectively.
  The values of $\Lambda_T$ are shown for the antitriplet-sextet
  and sextet-sextet molecules. In the latter case we concentrate
  on the isoscalar $\Sigma_Q^{(*)} \bar{\Sigma}_Q^{(*)}$ molecules,
  for which pion exchanges are stronger (for the isovector case
  the value of $\Lambda_T$ is twice that of the isoscalar case).
  For the $\Xi_Q^{('/*)} \bar{\Xi}_Q^{('/*)}$ molecules we only show
  the channel in which the tensor force is strongest.
  For comparison purposes, in the deuteron channel of the two-nucleon system
  the tensor scale is $\Lambda_T(0) = 66\,{\rm MeV}$ and
  $\Lambda_T(m_{\pi}) = 99\,{\rm MeV}$, respectively.
} \label{tab:Lambda-T}
\end{table*}

\section{Power Counting for Heavy Baryon Molecules}
\label{sec:pc}

In this section we discuss the different possible power {counting rules}
for the heavy baryon-antibaryon states.
We are interested in the case where there are bound states.
This excludes NDA, for which
\begin{eqnarray}
V_C^{\rm LO}(\vec{q}) \sim Q^0 \quad , \quad V_F^{\rm LO} \sim Q^0 \, ,
\end{eqnarray}
as this counting leads to purely perturbative
heavy baryon-antibaryon interactions.
The existence of bound states requires that at least one of the components
of the potential is promoted to $Q^{-1}$, see the discussion
in Sect.~\ref{subsec:bound-pc} for details.
There are different choices depending on which piece of
the interaction is promoted.
We will consider three scenarios:
\begin{itemize}
\item[(a)] promotion of the contact terms,
\item[(b)] promotion of central OPE and,
\item[(c)] promotion of tensor OPE.
\end{itemize}
Each scenario represents a different binding mechanism:
\begin{itemize}
\item[(a)] short-range,
\item[(b)] long-range and
\item[(c)] a combination of both,
\end{itemize}
where we notice that (c) is not obvious but a consequence of
the technicalities of power counting, as we will explain.
We present an overview of these scenarios in Table \ref{tab:pc-map}.
But we stress that the discussion here will be theoretical:
in the absence of experimental data it is not particularly useful
to consider the subleading orders of the EFT expansion.
The exploration in this section provides information about
the theoretical uncertainties that are to be expected
from a LO calculation in each scenario,
\begin{itemize}
\item[(a)] $({Q}/{M})$,
\item[(b)] $({Q}/{M})$,
\item[(c)] ${( {Q}/{M} )}^{\frac{5}{2}}$,
\end{itemize}
where we will explain in detail how we obtain these uncertainties and
also what is the general form of the first subleading corrections,
which can actually be consulted in Table \ref{tab:pc-map}.
For a more in depth discussion of the power counting of heavy meson-antimeson
in particular and two-body systems in general we refer
the reader to Refs.~\cite{Valderrama:2012jv,Valderrama:2016koj}.

\begin{table*}
\begin{center}
\begin{tabular}{|c|c|c|c|c|c|}
\hline \hline
Power Counting & $Q$ & ${\rm LO}$ & $V^{\rm LO}$  & ${\rm NLO}$ & $V^{\rm NLO}$ \\ 
\hline
NDA &
$p$, $m_{\pi}$ & $Q^0$ & $C$, $V_{\rm OPE}$ &
$Q^2$ & $D\,(p^2 + p'^2)$, $V_{\rm TPE}$ \\
\hline
(a) & $p$, $m_{\pi}$, $\sqrt{2 \mu B_2}$
& $Q^{-1}$ & $C$ & $Q^0$ & $D\,(p^2 + p'^2)$, $V_{\rm OPE}$ \\
\hline
(b) & $p$, $m_{\pi}$, $\Lambda_C$
& $Q^{-1}$ & $V_{\rm OPE(C)}$ & $Q^0$ & $C$, $V_{\rm OPE(T)}$ \\
\hline
(c) & $p$, $m_{\pi}$, $\Lambda_T$
& $Q^{-1}$ & $C$, $V_{\rm OPE}$ & $Q^{3/2}$ & $D\,(p^2 + p'^2)$ \\
\hline \hline
\end{tabular}
\end{center}
\caption{
  Possible power countings for the heavy baryon-antibaryon system: NDA
  (naive dimensional analysis) and the three scenarios (a), (b), (c)
  that we consider in Sect.~\ref{sec:pc}.
  $Q$ refers to the soft scales in each power counting, which include
  the momenta $p$ of the heavy baryons and pions, the mass $m_{\pi}$
  of the pions, the binding momentum $\sqrt{2 \mu B_2}$ of a heavy
  baryon-antibaryon bound state (with $\mu$ the reduced mass and
  $B_2$ the binding energy) and the central and tensor scales
  $\Lambda_C$ and $\Lambda_T$ defined in Eqs.~(\ref{eq:central-scale})
  and (\ref{eq:tensor-scale}) and listed in Tables \ref{tab:Lambda-C}
  and \ref{tab:Lambda-T} for a series of heavy baryon-antibaryon systems.
  The $\rm LO$ and $\rm NLO$ columns
  indicate the counting $Q^{\nu}$ of the leading order
  and first subleading correction.
  In the $V^{\rm LO}$ and $V^{\rm NLO}$ columns
  we write what are the contributions to the EFT potential in each case.
  $C$ and $D\,(p^2 + p'^2)$ refer to a contact-range potential without
  derivatives and with two derivatives of the heavy baryon field, respectively.
  $V_{\rm OPE}$ is the OPE potential, $V_{\rm OPE(C)}$ its central piece and
  $V_{\rm OPE(T)}$ its tensor piece,
  Finally $V_{\rm TPE}$ refers to two pion exchange potential (i.e. irreducible
  diagrams containing two pions), which we have not considered in this work.
} \label{tab:pc-map}
\end{table*}

\subsection{Counting with Perturbative Pions}

The first possibility --- scenario (a) --- is that the binding mechanism
for heavy baryon-antibaryon molecules is of a short-range nature.
Within the EFT language this amounts to the promotion of
the contact-range potential from $Q^{0}$ to $Q^{-1}$.
Within this power counting the leading order
(${\rm LO} \equiv Q^{-1}$ in this case) potential
will be composed of contact terms, while the next-to-leading order potential
(${\rm NLO} \equiv Q^{0}$) will contain the OPE potential
plus a few additional contact interactions
\begin{eqnarray}
V^{\rm LO} &=& V_C^{(-1)} \, , \\
V^{\rm NLO} &=& V_C^{(0)} + V_{\rm OPE} \, .
\end{eqnarray}
We do not have to promote all the possible contact interactions
that we obtain from the heavy-light spin decomposition:
in general a subset of it will be enough.

There is one important detail with this counting.
If we consider the S-wave contact-range interactions in EFT,
they admit the momentum expansion
\begin{eqnarray}
\langle p' | V_C | p \rangle = C + D ({p'}^2 + p^2) + \dots \, .
\end{eqnarray}
where the dots denote couplings involving more derivatives of the baryon fields.
Here we use $C$ and $D$ as a generic notation for the couplings
of a contact-range potential with no derivatives ($C$) or
with two derivatives ($D$).
The naive expectation for the scaling of the $C$ and $D$ couplings is
\begin{eqnarray}
C \sim \frac{1}{M^2} \quad , \quad D \sim \frac{1}{M^{4}} \, .
\end{eqnarray}
But if we promote the coupling $C$ to LO,
the coupling $D$ must also be
promoted~\cite{Kaplan:1998tg,Kaplan:1998we,vanKolck:1998bw}
\begin{eqnarray}
C \sim \frac{1}{M Q} \quad , \quad D \sim \frac{1}{M^{2} Q^2} \, .
\end{eqnarray}
As a consequence the ordering of the contact-range potential will be
\begin{eqnarray}
\label{eq:VCm1}
\langle p' | V_C^{\rm LO} | p \rangle &=& C \, , \\
\label{eq:VC0}
\langle p' | V_C^{\rm NLO} | p \rangle &=& D ({p'}^2 + p^2) \, .
\end{eqnarray}
That is, the ${\rm NLO}$ potential will contain a contact-range interaction
with two-derivatives on the baryon fields.
As a consequence if we promote a particular $C$ coupling to $Q^{-1}$,
the corresponding derivative coupling with $D$ will be promoted to $Q^0$.
We notice that in this work we have not explicitly considered
a contact-range potential with derivatives.
The take-home message is that in this scenario the theoretical uncertainty
of the calculations is $Q / M$ because the first correction to a ${\rm LO}$
calculation is suppressed by one order in the EFT expansion.

\subsection{Counting with Non-Perturbative Central OPE}

The second possibility is that the binding mechanism
depends also on the attraction provided by central OPE.
We can distinguish two cases:
(i) the binding depends on central OPE alone, i.e. scenario (b), and
(ii) the binding depends on the interplay of the contact terms
and central OPE, i.e. scenario (a+b).

In the first case --- scenario (a) --- we have a relatively
simple power counting in which
\begin{eqnarray}
V^{\rm LO} &=& V_{\rm OPE(C)} \, ,
\end{eqnarray}
where by $\rm OPE(C)$ it is meant the central piece of OPE.
The ${\rm NLO}$ potential contains tensor OPE and the contact interactions
\begin{eqnarray}
V^{\rm NLO} &=&  V_{\rm OPE(T)} + V_{\rm C}^{(0)} \, ,
\end{eqnarray}
where
\begin{eqnarray}
\langle p' | V_C^{(0)} | p \rangle &=& C_0 \, .
\end{eqnarray}
Contacts with $2 n$ derivatives on the baryon fields will enter
at order $Q^{2n}$.
In this scenario the relative uncertainty of a ${\rm LO}$ calculation
is $Q / M$ because the first correction to the EFT potential
enters at $NLO$.

The second case --- scenario (a+b) --- is identical to the power counting
of scenario (a) except for the fact that we include OPE in the ${\rm LO}$:
\begin{eqnarray}
V^{\rm LO} &=& V_C^{\rm LO} + V_{\rm OPE(C)} \, , \\
V^{\rm NLO} &=& V_C^{\rm NLO} + V_{\rm OPE(T)} \, ,
\end{eqnarray}
where $V_{\rm OPE(T)}$ is the tensor piece of OPE, while $V_C^{\rm LO}$
and $V_C^{\rm NLO}$ are the contact-range potentials
of Eqs.~(\ref{eq:VCm1}) and (\ref{eq:VC0}).
The uncertainty of the ${\rm LO}$ calculation is $Q / M$.

\subsection{Counting with Non-Perturbative Tensor OPE}
\label{subsec:pc-tensor}

The third possibility --- scenario (c) --- arises when tensor OPE
is non-perturbative. This is the most involved of the three power
countings considered.
Tensor OPE is a singular potential, which means that it diverges as fast as
(or faster than) $1/r^2$ for $r \to 0$.
Singular potentials in general lead to non-trivial consequences
in EFT~\cite{Beane:2000wh,PavonValderrama:2005gu,PavonValderrama:2005wv,PavonValderrama:2005uj,Valderrama:2008kj,PavonValderrama:2010fb}.
The tensor force is not only singular, but also attractive
for the case at hand: for an S-wave heavy baryon-antibaryon
state that mixes with a D-wave, there is always a configuration
for which the tensor force is attractive~\footnote{
  This can be seen by inspecting the
  partial wave projection of the tensor operator $S_{12}$,
  which can be consulted in Appendix \ref{app:ope}.
  It happens that for this matrices there is always
  at least one positive and one negative eigenvalue.}.
For attractive singular potentials short-range physics is enhanced:
the non-perturbative treatment of attractive singular potentials
requires the inclusion of a contact-range interaction
at ${\rm LO}$~\cite{PavonValderrama:2005wv,PavonValderrama:2005uj}.

The application of these ideas for a heavy baryon-antibaryon S-wave molecule
implies that a non-perturbative tensor force requires
a non-perturbative contact potential.
As a consequence, the ${\rm LO}$ potential will be
\begin{eqnarray}
V^{\rm LO} &=& V_{\rm OPE} + V_C^{-1} \, ,
\end{eqnarray}
with $V_C^{(-1)}$ the lowest order contact-range potential~\footnote{
We have simply included the full OPE potential in ${\rm LO}$ because
the addition of central OPE does not further modify
the power counting induced by tensor OPE.}.
The counting of the contacts will be modified
as follows~\cite{Birse:2005um,Valderrama:2016koj}
\begin{eqnarray}
C \sim \frac{1}{M Q} \quad , \quad
D \sim \frac{1}{M^{7/2} Q^{1/2}} \, .
\end{eqnarray}
or equivalently we can write
\begin{eqnarray}
\langle p' | V_C^{(-1)} | p \rangle &=& C \, , \\
\langle p' | V_C^{(3/2)} | p \rangle &=& D ({p'}^2 + p^2) \, ,
\end{eqnarray}
where the contacts with derivatives get promoted by half an order.
Thus the theoretical uncertainty of a ${\rm LO}$ calculation is $(Q / M)^{5/2}$
(the first subleading correction, a derivative contact interaction,
enters $Q^{5/2}$ orders after $\rm LO$), which is considerably
better than for the other scenarios.

The previous analysis is a simplification though:
tensor forces mix channels with different orbital angular {momentum},
which might lead to complications in certain cases (in particular
the power counting of contact-interactions mixing partial waves).
We have not addressed these problems here: they depend on the particular system
under consideration and the aim of the present discussion is
to provide an overview of power counting rather
than a detailed account.

\section{Predicting Heavy Baryon Molecules}
\label{sec:pred}

In this section we investigate the question of
whether we can predict heavy baryon molecules.
The answer to this question depends on which is the binding mechanism.
If the binding mechanism is of a short-range nature,
the prediction of bound states will rely on phenomenology.
Within the EFT framework this is illustrated by the fact
that the contact-range couplings are free parameters.
If there is no preexisting experimental information
about the heavy baryon-antibaryon system,
we will have to determine the contact-range couplings
by matching to a phenomenological model.
Conversely if the binding mechanism is of a long-range nature,
the prediction of bound states is possible within EFT.
Examples are the $\Lambda_{c1} \bar{\Sigma}_c$~\cite{Geng:2017jzr} and
$D D_{s0}^*$ / $D^* D_{s1}^*$~\cite{SanchezSanchez:2017xtl} systems,
which interact via a long-range Yukawa potential
that is strong enough to bind.
This is not the standard situation {though} and more often than not
we will need phenomenological input.

At this point it is interesting to notice the relation between power counting
and the predictability of heavy baryon-antibaryon molecules.
In Sect.~\ref{sec:pc} we proposed three power counting scenarios:
(a), (b) and (c).
Scenario (a) corresponds to a short-range binding mechanisms, which requires
phenomenological input. Scenario (b) corresponds to a long-range binding
mechanism, which allows for EFT predictions. Finally scenario (c)
is a mixture of short- and long-range binding, which
in a few cases will lead to predictions.
The heavy baryon-antibaryon system belongs to scenario (a) or (c)
depending on the particular state and quantum numbers considered.

Theoretical studies of hadronic molecules have attributed
the binding mechanism to either short- or long-range causes.
In the pioneering work of Voloshin and Okun~\cite{Voloshin:1976ap}
it is the exchange of light mesons ($\pi$, $\sigma$, $\rho$ and $\omega$)
which generates heavy hadron molecules, i.e. a mixture of short- and
long-range physics.
Early speculations~\cite{Tornqvist:1991ks,Manohar:1992nd,Ericson:1993wy,Tornqvist:1993ng}
often predicted binding from the OPE potential (long-range physics)
alone. It is notable to mention that Ericson and Karl~\cite{Ericson:1993wy}
indicated that hadronic molecules should be possible in the charm sector and
that T\"ornqvist~\cite{Tornqvist:1993ng} predicted the existence of a
isoscalar $1^{++}$ $D^*\bar{D}$ bound state.
The experimental discovery of the $X(3872)$ a decade after~\cite{Choi:2003ue}
suggests that these theoretical speculations were on the right track.
At this point we find it interesting to notice
that a molecular $X(3872)$ also arises naturally
from short-range physics~\cite{Gamermann:2007fi}.
Before the discovery of the $P_c(4450)$ by the LHCb~\cite{Aaij:2015tga},
which is suspected (but not confirmed) to be a $\bar{D}^* \Sigma_c$
molecule~\cite{Roca:2015dva,Xiao:2015fia,Burns:2015dwa,Geng:2017hxc},
there were theoretical predictions of its existence too.
The work or Refs.~\cite{Wu:2010jy,Xiao:2013yca} used contact-range interactions
derived from vector meson exchange saturation to make quantitative predictions
of an $I=\tfrac{1}{2}$, $J^P = \tfrac{3}{2}^-$ $\bar{D}^* \Sigma_c$ molecule
(among others).
Meanwhile the work of Ref.~\cite{Karliner:2015ina} used the OPE potential
instead to make qualitative predictions about probable hadronic molecules,
including a possible $I=\tfrac{1}{2}$, $J^P = \tfrac{3}{2}^-$
$\bar{D}^* \Sigma_c$ molecule.
Finally EFT and EFT-inspired works explain the properties of shallow molecular
states solely on the basis of short-range
interactions~\cite{Braaten:2003he,Voloshin:2005rt,Mehen:2011yh},
without making explicit assumptions about the binding mechanism.

In this section we will examine the short- and long-range binding
mechanisms for heavy baryon-antibaryon molecules.
The most obvious short-range mechanism is the saturation of
the EFT contact-range couplings from scalar and vector meson exchange,
while the most important long-range mechanisms is the OPE potential.
Now we will explain these binding mechanisms in detail.

\subsection{Short-Range Binding Dynamics}
\label{subsec:saturation}

First we explore the short-range dynamics,
in particular scalar and vector meson exchange.
For taking this effect into account we saturate the contact-range
couplings of the EFT with the exchange of a meson with mass $m_S$
of the order of the hard scale of the EFT ($m_S \sim M$),
see Ref.~\cite{Epelbaum:2001fm} for a detailed exposition of this idea.
For this we expand the exchange potential $V_S$ for momenta
${\vec{q}\,}^2 \ll m_S$ and match it with the expansion of
the EFT contact-range potential
\begin{eqnarray}
  V_S(\vec{q}) &=& V_0 + V_2 \, {\vec{q}\,}^2 + \dots \, , \\
  V_C(\vec{q}) &=& C + D \, {{\vec{q}\,}^2} + \dots \, ,
\end{eqnarray}
from which we arrive to
\begin{eqnarray}
  C(\Lambda \sim m_S) \sim V_0 = V_S(|\vec{q}\,| = 0) \, , \label{eq:C-sat}
\end{eqnarray}
where $\Lambda$ is the cutoff. Notice that we take $\Lambda \sim m_S$:
this is because the saturation hypothesis is only expected to work
if the cutoff is of the order of the mass of
the exchange meson~\cite{Epelbaum:2001fm}.
For a Yukawa-like meson exchange potential 
\begin{eqnarray}
  V_S(\vec{q}) = \frac{g_S^2}{\vec{q}^2 + m_S^2} \, ,
\end{eqnarray}
the saturated contact-range coupling is proportional to
\begin{eqnarray}
  C \propto  \frac{g_S^2}{m_S^2} \, ,
\end{eqnarray}
where the proportionality constant will depend on the details of
the regularization process. 
This argument is independent of the nature of the exchanged meson,
it only matters that the mass of this meson is of the order of
the hard scale.

Next we calculate the scalar and vector meson exchange contribution to
the saturation of the EFT coupling.
We begin by considering scalar meson exchange.
The sigma meson exchange potential is
\begin{eqnarray}
  V_{\sigma}(\vec{q}) = - \frac{g_{\sigma}^2}{\vec{q}^2 + m_{\sigma}^2} \, ,
\end{eqnarray}
with $g_{\sigma}$ the sigma coupling.
We can determine $g_{\sigma}$ from the quark model,
in which $g_{\sigma}$ is simply proportional to the number of
$u$ and $d$ quarks in the hadron.
Here we take the sigma-nucleon-nucleon coupling as {input},
which in the non-linear sigma model~\cite{GellMann:1960np}
is $g_{\sigma NN} = \sqrt{2}\, M_N / f_{\pi} \sim 10.2$,
where $M_N$ is the nucleon mass and $f_{\pi}$ the pion decay constant.
From this we have $g_{\sigma} = g_{\sigma NN} / 3 \sim 3.4$
for $\Xi_Q$, $\Xi_Q'$ and $\Xi_Q^*$
and $g_{\sigma} = 2 g_{\sigma NN} / 3 \sim 6.8$
for $\Lambda_Q$, $\Sigma_Q$ and $\Sigma_Q^*$.
The contributions of the scalar meson to the saturation of
the contact-range couplings are listed in Table \ref{tab:saturation}.

We continue with the vector meson exchange potential, for which
the starting point is the heavy baryon - vector meson Lagrangian
for the $SSV$ and $TTV$ vertices (where $V$ represents the vector meson).
If we consider interactions with no derivatives, which allow for saturation
of the lowest order EFT couplings, we can write the following Lagrangians
\begin{eqnarray}
  \mathcal{L}_{TTV} &=& \lambda_T\,
\epsilon_{ikl}\,\epsilon_{jkm}\,\bar{T}_Q^l  v_{\mu} (V^{\mu})^{i}_{j} T_{Q m} \, , \\
\mathcal{L}_{SSV} &=& \lambda_S\,
\bar{S}_{Q \nu i k} v_{\mu} (V^{\mu})^{i}_{j} S^{\nu j k}_Q \, ,
\end{eqnarray}
where the latin indices indicate the sum over the SU(3) components~\footnote{
Notice that we are not considering $TSV$ vertices because they involve
derivatives and do not saturate the LO contact-range couplings.}.
The vector meson nonet field is given by
\begin{eqnarray}
V =
\begin{pmatrix}
\frac{\rho^0}{\sqrt{2}} + \frac{\omega^0}{\sqrt{2}} & \rho^{+} & K^{*+} \\
\rho^{-} & -\frac{\rho^0}{\sqrt{2}} + \frac{\omega^0}{\sqrt{2}} & K^{*0} \\
K^{*-} & \bar{K}^{*0} & \phi
\end{pmatrix} \, ,
\end{eqnarray}
where the Lorentz index $\mu$ is implicitly understood.
The vector meson exchange contribution to the potential can be worked out
along the lines of Appendix \ref{app:ope}. A few representative
examples are
\begin{eqnarray}
\langle \Xi_Q \bar{\Xi}_Q | V | \Xi_Q \bar{\Xi}_Q \rangle
&=&  \frac{\lambda_T^2}{\vec{q}^2 + m_V^2}\,
\frac{\left( \vec{\tau}_1 \cdot \vec{\tau}_2 - 3\right)}{2}
\, , \label{eq:V-rho-1} \\
\langle \Xi_Q' \bar{\Xi}_Q | V | \Xi_Q' \bar{\Xi}_Q \rangle
&=&  \frac{\lambda_S\,\lambda_T}{\vec{q}^2 + m_V^2}\,
\frac{\left( \vec{\tau}_1 \cdot \vec{\tau}_2 - 3\right)}{4}
\, , \label{eq:V-rho-2} \\
\langle \Xi_Q' \bar{\Xi}_Q' | V | \Xi_Q' \bar{\Xi}_Q' \rangle
&=&  \frac{\lambda_S^2}{\vec{q}^2 + m_V^2}\,
\frac{\left( \vec{\tau}_1 \cdot \vec{\tau}_2 - 3\right)}{8} \, ,
 \label{eq:V-rho-3} \\
\langle \Sigma_Q \bar{\Sigma}_Q | V | \Sigma_Q \bar{\Sigma}_Q \rangle
&=&
\frac{\lambda_S^2 }{\vec{q}^2 + m_V^2}\,
\frac{\left( \vec{T}_1 \cdot \vec{T}_2 - 1 \right)}{2}\, \label{eq:V-rho-4}
 \, ,
\end{eqnarray}
which have been calculated in the SU(3) limit. We have taken
$m_{\rho} = m_{\omega} = m_{\phi} = m_V$ with $m_V$ the vector meson mass.
Notice that we do not have to write explicitly the potential for
the sextet spin-$3/2$ heavy baryons: the vector meson potentials
for the $\Sigma_Q^*$, $\Xi_Q^*$ and $\Omega_Q^*$ are identical
to the ones for $\Sigma_Q$, $\Xi_Q'$ and $\Omega_Q$.
The couplings $\lambda_S$ and $\lambda_T$ are not arbitrary:
they can be determined  from the universality of
the $\rho$ coupling constant~\cite{Sakurai:1960ju}.
If we consider the $\rho$-meson exchange potential
between two isospin $1/2$ baryons
\begin{eqnarray}
  V_{\rho}(\vec{q}) = \frac{g_{\rho}^2}{\vec{q}^2 + m_{\rho}^2}\,
  \vec{\tau}_1 \cdot \vec{\tau}_2 \, ,
\end{eqnarray}
with $m_{\rho} = 770\,{\rm MeV}$, the universality of
the $\rho$ coupling implies that $g_{\rho} = m_{\rho} / 2 f_{\pi} \simeq 2.9$.
If we match to the potentials in Eqs.~(\ref{eq:V-rho-1}) to (\ref{eq:V-rho-4}),
we find
\begin{eqnarray}
  \lambda_T = \sqrt{2}\,g_{\rho} \quad \mbox{and} \quad
  \lambda_S = 2\sqrt{2}\,g_{\rho} \, .
\end{eqnarray}
The saturation of the EFT contact couplings by the vector mesons
is easy to obtain and can be consulted in Table \ref{tab:saturation}.
We mention that is is possible to consider the contributions
of the different vector mesons separately
\begin{eqnarray}
  C_V = C_{\rho} + C_{\omega} + C_{\phi} \, .
\end{eqnarray}
This form is interesting because it makes it easy to deduce
the strength of the heavy baryon-baryon short-range interaction
from the heavy baryon-antibaryon one.
This merely involves changing the sign of the contributions from
the negative G-parity mesons, the $\omega$ and the $\phi$,
yielding $C_V' = C_{\rho} - C_{\omega} - C_{\phi}$.
Though the vector meson saturation of the heavy baryon-baryon system
is not listed here, they can be obtained from Table \ref{tab:saturation}
where the $C_{\rho}$, $C_{\omega}$ and $C_{\phi}$ contributions are listed.

Finally we add the contribution to the EFT contact couplings
from scalar and vector meson exchange saturation, that is
\begin{eqnarray}
  C(\Lambda \sim m_V, m_{\sigma}) \sim C_S + C_V \, ,
\end{eqnarray}
where $C_S$ and $C_V$ are the scalar and vector meson {contributions}.
At this point it is interesting to compare saturation
in the heavy baryon-antibaryon system with the heavy
meson-antimeson and heavy meson-antibaryon cases.
The $X(3872)$ and $P_c(4450)$ are $D^*\bar{D}$ and $\bar{D}^* {\Sigma}_c$
molecular candidates for which we can apply the saturation argument as well,
as can be seen in Table \ref{tab:saturation}.
If we compare the saturated contact-range couplings of
the $X(3872)$ and $P_c(4450)$ with the ones
for the heavy baryon-antibaryon system,
we can identify the most promising molecular candidates.
Heavy baryon-antibaryon systems for which the short-range interaction
is expected to be more attractive than the $X(3872)$ include
\begin{eqnarray}
  \Lambda_Q \bar{\Lambda}_Q \, , \, \Xi_Q \bar{\Xi}_Q (I=0)
  \, , \, \Xi_Q' \bar{\Xi}_Q (I=0) \, , \, \nonumber \\
  \Sigma_Q \bar{\Lambda}_Q \, , \, \Xi_Q' \bar{\Xi}_Q' (I=0)
  \, , \, \Sigma_Q \bar{\Sigma}_Q (I=0,1) \, , \nonumber \\
  \label{eq:list-Xc}
\end{eqnarray}
to which we have to add the molecules containing the excited sextet baryons,
i.e. the molecules we obtain from the substitutions
$\Xi_Q \to \Xi_Q^*$ and $\Sigma_Q \to \Sigma_Q^*$.
The systems for which there is more short-range attraction
than for the $P_c(4450)$ include
\begin{eqnarray}
  \Lambda_Q \bar{\Lambda}_Q \, , \,
  \Sigma_Q \bar{\Lambda}_Q \, , \,
  \Sigma_Q \bar{\Sigma}_Q (I=0,1) \, ,
  \label{eq:list-Pc}
\end{eqnarray}
where we notice that they are a subset of Eq.~(\ref{eq:list-Xc}).
The obvious conclusion is that the heavy baryon-antibaryon pairs listed
in Eq.~(\ref{eq:list-Pc}) are the strongest candidates to bind.
The particular case of $\Lambda_c \bar{\Lambda}_c$ has been recently studied
in Ref.~\cite{Chen:2017vai}, leading to binding in agreement
with our conclusions.

If we consider the heavy baryon-baryon system instead,
the contribution of the $\omega$ and $\phi$ mesons is repulsive
and in general there is less attraction that in the heavy
baryon-antibaryon case.
Yet for the following heavy baryon-baryon system
\begin{eqnarray}
  \Sigma_Q {\Sigma}_Q (I=0) \, ,
  \label{eq:BB-list-Pc}
\end{eqnarray}
there is more short-range attraction than in the $X(3872)$ and the $P_c(4450)$.
Further candidates for binding can be inferred from a comparison
with the deuteron, for which the short-range interaction is repulsive.
In Table \ref{tab:saturation} we see that for the nucleon-nucleon system
there is a strong short-range repulsion from the exchange of
the $\omega$ meson but also a strong attraction coming
from the exchange of the $\sigma$ meson.
The existence of the deuteron indicates that attraction wins in this case.
This is not surprising if we notice that $m_{\sigma} < m_{\omega}$ and
$(g_{\sigma NN} / 3) >  g_{\rho}$, which suggests that $\sigma$ meson
saturation overcomes $\omega$ meson saturation
($|C_S| > |C_{\omega}|$).
Here it is worth noticing
that binding in non-relativistic systems depends on the reduced
potential, the product of the potential by twice the reduced mass of
the system. This in turn implies that for the following systems
\begin{eqnarray}
  \Lambda_Q {\Lambda}_Q \, , \,
  \Sigma_Q {\Lambda}_Q \, , \,
  \Sigma_Q {\Sigma}_Q (I=0,1) \, ,
  \label{eq:list-NN}
\end{eqnarray}
the net effect of the short-range attraction from the $\sigma$ meson
will be larger than in the two-nucleon system, i.e.
\begin{eqnarray}
  2 \mu | C_S | > 2 \mu_{\rm NN} | C_S^{\rm NN} | \, ,
\end{eqnarray}
with $\mu$ and $\mu_{\rm NN}$ the reduced masses of the systems listed
in Eq.~(\ref{eq:list-NN}) and the two-nucleon system respectively, and
$C_S$ and $C_S^{\rm NN}$ their $\sigma$-saturated couplings.
But we warn that this argument is incomplete: common-wisdom in nuclear physics
attributes binding in the deuteron to the interplay of short- and
long-range physics, in particular the short-range repulsion from the
$\omega$ meson, the attraction from the $\sigma$ meson
and the tensor force from OPE.
This suggests that, with the exception of the isoscalar $\Sigma_c \Sigma_c$
system, the other molecular candidates listed in Eq.~(\ref{eq:list-NN})
require a more thorough theoretical exploration to determine
if there is binding.

The saturation argument probably provides incomplete information
about the ${\rm LO}$ contact-range couplings.
The saturated couplings are independent of the total light spin of
the heavy hadron-antihadron system.
This is compatible with HQSS --- it represents a subset of
the possible interactions that respect HQSS ---
but not necessarily with {experiments}.
If we review the heavy meson-antimeson system, to which the $X(3872)$
is suspected to belong, scalar and vector meson exchange saturation
predicts exactly the same potential for $D\bar{D}$, $D^*\bar{D}$,
$D\bar{D}^*$ and $D^*\bar{D}^*$ irrespectively of
the spin and C-parity quantum numbers.
That is, the saturation argument leads to the prediction of
six isoscalar heavy meson-antimeson molecules.
This is to be compared with only one obvious molecular candidate,
the $X(3872)$.
Analogously, the application of this argument to the heavy meson-antibaryon
molecules leads to the prediction of seven $\bar{D} {\Sigma}_c$,
$\bar{D} {\Sigma}_c^*$, $\bar{D}^* {\Sigma}_c$ and
$\bar{D}^* {\Sigma}_c^*$ molecules but only one experimental
candidate, the $P_c(4450)$.
This situation also happens in other theoretical approaches
that derive heavy hadron interactions
from vector meson saturation~\cite{Gamermann:2007fi,Wu:2010jy,Xiao:2013yca}.
The probable conclusion is that we are probably missing something
in the resonance saturation arguments we are using to derive
the ${\rm LO}$ couplings.
Be it as it may, for the set of molecules in Eq.~(\ref{eq:list-Pc})
the short-range attraction is expected to be remarkably stronger
than in the $X(3872)$ and $P_c(4450)$.

\begin{table}
\begin{center}
\begin{tabular}{|c|c|c|c||c|c|c|}
\hline \hline
System & Isospin & $C_S$ & $C_V$ & $C_{\rho}$ & $C_{\omega}$ & $C_{\phi}$ \\
\hline
$NN$ & $0$ & $-9$ & $+6$ & $-3$ & $+9$ & $\phantom{-}0$
\\
$NN$ & $1$ & $-9$ & $+10$ & $+1$ & $+9$ & $\phantom{-}0$
\\
\hline
$D \bar{D}$ & $0$ & $-1$ & $-4$ & $-3$ & $-1$ & $\phantom{-}0$ 
\\
$D \bar{D}$ & $1$ & $-1$ & $\phantom{-}0$ & $+1$ & $-1$ & $\phantom{-}0$  
\\
\hline
$\Sigma_c \bar{D}$ & $\frac{1}{2}$ & $-2$ & $-2$ & $-4$ & $+2$ & $\phantom{-}0$ 
\\
$\Sigma_c \bar{D}$ & $\frac{3}{2}$ & $-2$ & $+4$ & $+2$ & $+2$ & $\phantom{-}0$  
\\
\hline
$\Lambda_Q \bar{\Lambda}_Q$ & $0$ & $-4$ & $-4$ & $\phantom{-}0$  & $-4$ & $\phantom{-}0$ 
\\
$\Xi_Q \bar{\Xi}_Q$ & $0$ & $-1$ & $-6$ & $-3$ & $-1$ & $-2$ 
\\
$\Xi_Q \bar{\Xi}_Q$ & $1$ & $-1$ & $-2$ & $+1$ & $-1$ & $-2$ 
\\
\hline
$\Xi_Q' \bar{\Xi}_Q$ & $0$ & $-1$ & $-6$ & $-3$ & $-1$ & $-2$ 
\\
$\Xi_Q' \bar{\Xi}_Q$ & $1$ & $-1$ & $-2$ & $+1$ & $-1$ & $-2$ 
\\
$\Sigma_c \bar{\Lambda}_Q$ & $1$ & $-4$ & $-4$ & $\phantom{-}0$ & $-4$ & $\phantom{-}0$ 
\\
\hline
$\Omega_Q \bar{\Omega}_Q$ & $0$ & $\phantom{-}0$ & $-8$ & $\phantom{-}0$ & $\phantom{-}0$ & $-8$ 
\\
$\Xi_Q \bar{\Xi}_Q$ & $0$ & $-1$ & $-6$ & $-3$ & $-1$ & $-2$ 
\\
$\Xi_Q \bar{\Xi}_Q$ & $1$ & $-1$ & $-2$ & $+1$ & $-1$ & $-2$
\\
$\Sigma_Q \bar{\Sigma}_Q$ & $0$ & $-4$ & $-12$ & $-8$ & $-4$ & $\phantom{-}0$ 
\\
$\Sigma_Q \bar{\Sigma}_Q$ & $1$ & $-4$ & $-8$ & $-4$ & $-4$ & $\phantom{-}0$ 
\\
$\Sigma_Q \bar{\Sigma}_Q$ & $2$ & $-4$ & $\phantom{-}0$ & $+4$ & $-4$ & $\phantom{-}0$ 
\\
\hline
\hline \hline
\end{tabular}
\end{center}
\caption{
  Contact-range coupling from saturation of vector and scalar meson exchange
  in units proportional to $g_{\rho}^2 / m_V^2$ and $g_{\sigma q q} / m_{\sigma}^2$,
  with $g_{\rho} \simeq 2.9$ and $g_{\sigma q q} \simeq 3.4$.
  Scalar and vector meson exchange saturation of the spin-$\tfrac{3}{2}$
  sextet heavy baryons ($\Sigma^*_Q$, $\Xi^*_Q$, $\Omega^*_Q$) \
  is identical to their spin-$\tfrac{1}{2}$ partners
  ($\Sigma_Q$, $\Xi_Q$, $\Omega_Q$) and
  are not listed independently.
  The contributions to vector meson exchange are also listed separately
  as $C_V = C_{\rho} + C_{\omega} + C_{\phi}$.
  Though not listed, the saturation for the heavy baryon-baryon system
  can be obtained from the heavy baryon-antibaryon case by changing
  the sign of the $\omega$ and $\phi$ contributions.
  For comparison purposes we also include the $NN$ (two-nucleon),
  $D^* \bar{D}$ and $\bar{D}^* \Sigma_c$ systems, which are related
  to the deuteron, the $X(3872)$ and the $P_c(4450)$, respectively.
} \label{tab:saturation}
\end{table}

\subsection{Long-Range Binding Dynamics}

\begin{table*}
\begin{center}
\begin{tabular}{|c|c|c|c|c|c|c|c|}
\hline \hline
Channel &
$I=0$ & $I=1$ & $I=2$ &
Channel &
$I=0$ & $I=1$ & $I=2$ \\
\hline
$\Sigma_c \bar{\Sigma}_c (0^{-+})$ & $-$ & $-$ & $-$ &
$\Sigma_b \bar{\Sigma}_b (0^{-+})$
& $0^{+0.23}$ & $-$ & $-$ \\
$\Sigma_c \bar{\Sigma}_c (1^{--})$
& $0.42-0.94$ & $0.22-0.55$ & $0.19-0.48$ &
$\Sigma_b \bar{\Sigma}_b (1^{--})$
& $0.86^{+0.30}_{-0.31}$ & $0.49^{+0.21}_{-0.19}$ & $0.43^{+0.21}_{-0.18}$ \\
\hline
$\Sigma_c \bar{\Sigma}_c^* (1^{-+})$
& $0.39-1.00$ & $0.19-0.53$ & $0.15-0.36$ &
$\Sigma_b \bar{\Sigma}_b^* (1^{-+})$
& $0.88^{+0.38}_{-0.35}$ & $0.46^{+0.23}_{-0.20}$ & $0.32^{+0.13}_{-0.12}$ \\
$\Sigma_c \bar{\Sigma}_c^* (1^{--})$
& $0.44-1.11$ & $0.22-0.59$ & $0.18-0.44$ &
$\Sigma_b \bar{\Sigma}_b^* (1^{--})$
& $0.99^{+0.40}_{-0.39}$ & $0.52^{+0.26}_{-0.22}$ & $0.39^{+0.16}_{-0.15}$ \\
$\Sigma_c \bar{\Sigma}_c^* (2^{-+})$
& $0.46-1.01$ & $0.24-0.59$ & $0.25-0.64$ &
$\Sigma_b \bar{\Sigma}_b^* (2^{-+})$
& $0.91^{+0.31}_{-0.32}$ & $0.51^{+0.21}_{-0.21}$ & $0.56^{+0.24}_{-0.22}$ \\
$\Sigma_c \bar{\Sigma}_c^* (2^{--})$
& $0.39-0.85$ & $0.21-0.50$ & $0.20-0.55$ &
$\Sigma_b \bar{\Sigma}_b^* (2^{--})$
& $0.77^{+0.27}_{-0.27}$ & $0.45^{+0.18}_{-0.17}$ & $0.48^{+0.23}_{-0.21}$ \\
\hline
$\Sigma_c^* \bar{\Sigma}_c^* (0^{-+})$
& $0.56-1.36$ & $0.27-0.75$ & $0.18-0.41$ &
$\Sigma_b^* \bar{\Sigma}_b^* (0^{-+})$
& $1.20^{+0.46}_{-0.46}$ & $0.64^{+0.32}_{-0.27}$ & $0.36^{+0.14}_{-0.30}$ \\
$\Sigma_c^* \bar{\Sigma}_c^* (1^{--})$
& $0.52-1.24$ & $0.26-0.68$ & $0.18-0.42$ &
$\Sigma_b^* \bar{\Sigma}_b^* (1^{--})$
& $1.09^{+0.42}_{-0.41}$ & $0.59^{+0.29}_{-0.25}$ & $0.37^{+0.15}_{-0.14}$ \\
$\Sigma_c^* \bar{\Sigma}_c^* (2^{-+})$
& $0.44-1.02$ & $0.22-0.57$ & $0.20-0.50$ &
$\Sigma_b^* \bar{\Sigma}_b^* (2^{-+})$
& $0.91^{+0.34}_{-0.34}$ & $0.50^{+0.23}_{-0.20}$ & $0.43^{+0.18}_{-0.17}$ \\
$\Sigma_c^* \bar{\Sigma}_c^* (3^{--})$
& $0.45-0.91$ & $0.25-0.55$ & $0.28-0.69$ &
$\Sigma_b^* \bar{\Sigma}_b^* (3^{--})$
& $0.82^{+0.27}_{-0.28}$ & $0.48^{+0.20}_{-0.18}$ & $0.60^{+0.28}_{-0.25}$ \\
\hline \hline
\end{tabular}
\end{center}
\caption{
Critical radius $R_c$ for which OPE is able to bind certain
heavy baryon-antibaryon molecules, where $I$ refers to
the isospin of the system.
For the $\Sigma_Q \bar{\Sigma}_C (0^{-+})$, where there is no tensor force,
the notation $0^{+0.23}$ indicate that there is only binding
for $R_c \leq 0.23\,{\rm fm}$ if the coupling $g_2$ lies
on the high end of the lattice calculations.
For comparison purposes, the critical radius for the deuteron and the
$P_c^{+}(4450)$ pentaquark is $1.00\,{\rm fm}$ and
$0.30-0.49\,{\rm fm}$, respectively.
} \label{tab:R-c}
\end{table*}

The long range dynamics of the heavy baryon-antibaryon system is driven by OPE.
We assess the relative strength of the OPE potential
for each channel in the following way:
first we modify the OPE potential by including a cut-off
\begin{eqnarray}
  V_{\rm OPE}(r ; r_c) = V_{\rm OPE}(r)\,\theta(r - r_c) \, ,
\end{eqnarray}
where $r_c$ is the cut-off.
Then we calculate the largest $r_c$ for which OPE alone is able to bind
a molecule. We call this radius $r_c = R_c$ the {\it critical radius}.
Notice that we are in fact assuming that (i) OPE is valid from infinity
till the critical radius and (ii) there is no short-range physics.
If this critical radius turns out to be {\it large enough}
we will consider that the system is likely to bind.
By large enough we mean for instance that the critical radius is larger
than the size of the hadrons or the range of other contributions
to the hadron-hadron potential that have not been taken
into account (e.g. two-pion exchange).

It is important to notice that most heavy baryon-antibaryon molecules bind
if $r_c$ is sufficiently small because of the tensor force.
Thus the crucial factor is not whether there is a critical radius for which
the molecule binds, but whether the critical radius $R_c$ is reasonable or not.
The reasons why the tensor force is able to bind in most cases is because
for S-wave molecules it behaves as an attractive singular potential,
see the discussion in Sect.~\ref{subsec:pc-tensor}.
This is why it is important to consider whether the distance
at which OPE binds is reasonable or not.

We list the critical radii for the $\Sigma_Q^{(*)} \bar{\Sigma}_Q^{(*)}$
molecules in Table \ref{tab:R-c}.
We have chosen the $\Sigma_Q^{(*)} \bar{\Sigma}_Q^{(*)}$ system because
this is the case in which the OPE potential is expected to be stronger
owing to the higher isospin of the $\Sigma_Q$'s.
Besides, from scalar and {vector} meson exchange saturation
we expect a very strong short range attraction.
The isoscalar molecules are the ones showing more attraction and
higher critical radii, reaching in a few cases $1\,{\rm fm}$.
For the hidden charm molecules the uncertainty is really big
as the value of $g_2$ is not experimentally known.
To give a sense of scale we mention that for the deuteron
the critical radius is $1.00\,{\rm fm}$.
For the $P_c(4450)$ pentaquark-like state as a $\Sigma_c \bar{D}^*$ molecule,
the critical radius is $0.30-0.49\,{\rm fm}$ (where the uncertainty
is again a consequence of $g_2$).
In comparison for the heavy meson-antimeson molecular candidates
the radii are $0.30\,{\rm fm}$, $0.10\,{\rm fm}$ and $0.26\,{\rm fm}$
for the $X(3872)$, $Z_c(3900)$ and $Z_b(10610)$, respectively.
The rather small critical radii of the $X$, the $Z_c$ and the $Z_b$
suggest that these hadron molecules depend on the short-range
attraction (instead of OPE) to bind.
For the deuteron the critical radius is significantly larger,
indicating that OPE is an important component of the binding mechanism.
Last the situation for the $P_c(4450)^+$ pentaquark seems to be
in the middle.
From Table \ref{tab:R-c} it is apparent that
for the heavy baryon-antibaryon system OPE can provide
as much attraction as in the deuteron.
If we combine this observation with what we know about short-range physics
according to Table \ref{tab:saturation}, the conclusion is that
there will be a rich $\Sigma_Q^{(*)} \bar{\Sigma}_Q^{(*)}$
molecular spectrum, particularly
in the $I = 0, 1$ configurations.

\section{Conclusions}
\label{sec:conclusions}

In this work we have presented a general EFT framework
for the heavy baryon-antibaryon system.
EFTs exploit the existence of a separation of scales to express
the observable quantities of a low energy system as a power series.
In the case at hand the size of a hadron molecule is expected to be larger
than the hadrons forming it.
As a consequence this type of system is amenable to an EFT description.
Besides, heavy hadron molecules are constrained by chiral,
SU(2)-isospin, SU(3)-flavour and HQSS symmetries.
This degree of symmetry translates into a few interesting
regularities in their spectrum.

EFT explains the heavy baryon-antibaryon interaction
in terms of contact-range interactions and pion exchanges.
The relative importance of these two contributions
changes from system to system.
In general the ${\rm LO}$ EFT description involves four-baryon contact-range
interactions and pion exchanges (OPE), but this depends on the molecule.
Pion exchanges are expected to be particularly important in the isoscalar
$\Sigma_Q \bar{\Sigma}_Q$, $\Sigma_Q^* \bar{\Sigma}_Q$ and
$\Sigma_Q^* \bar{\Sigma}_Q^*$ molecules (but less important
for other configurations).
In contrast OPE vanishes in the $\Xi_Q \bar{\Xi}_Q$ and
$\Lambda_Q \bar{\Lambda}_Q$ molecules, which can be described
in terms of a contact-theory at ${\rm LO}$.
For the $\Xi_Q' \bar{\Xi}_Q'$, $\Xi_Q^* \bar{\Xi}_Q'$ and
$\Xi_Q^* \bar{\Xi}_Q^*$ molecules, particularly in the hidden
charm sector, OPE is probably a ${\rm NLO}$ effect.
We warn that the conclusions about the relevance of the OPE potential
are only well-established for the bottom sector. In the charm sector
the value of the $g_2$ axial coupling that appears in the
$\Sigma_c \to \Sigma_c \pi$ amplitude is not known
experimentally, and a determination either in a future experiment or
in the lattice will be welcomed.
Particle coupled channels, i.e. transitions in which a heavy baryon
changes from the ground to the excited state ($B_6 \to B_6^*$),
are subleading if the molecules are not too tightly bound,
i.e. for binding momenta $\gamma = \sqrt{M_{6}\,|E_B|} \leq 350-400\,{\rm MeV}$.
The previous findings regarding pion exchanges and coupled channels are
analogous to the ones in the heavy meson-antimeson
molecules~\cite{Valderrama:2012jv}.
It is also worth mentioning that right now
the ${\rm LO}$ EFT is more than enough for the description of
heavy hadron-antihadron molecules, where the scarcity of
experimental data makes it superfluous to calculate subleading orders.

The EFT potential is constrained by HQSS.
This is particularly evident for S-wave interactions, such
as the ${\rm LO}$ contact-range potential and central OPE.
Symmetries in the S-wave interaction are likely to translate
into symmetries in the spectrum.
For the $T\bar{T}$ case the ${\rm LO}$ EFT potential does not depend
on the total spin of the system:
\begin{eqnarray}
\langle T\bar{T} | V_s(0^{-}) | T\bar{T} \rangle
&=& \langle T\bar{T} | V_s(1^{-}) | T\bar{T} \rangle \, ,
\end{eqnarray}
where the subscript $s$ is used to indicate S-wave.
That is, the $T\bar{T}$ heavy baryon molecules are expected to come in pairs.
For the $T\bar{S}$/$S\bar{T}$ molecules we have the following two relations:
\begin{eqnarray}
\langle T\bar{S} | V_s(0^{-\pm}) | T\bar{S} \rangle
&=& \langle T\bar{S} | V_s(2^{-\pm}) | T\bar{S} \rangle \, , \\
\langle T\bar{B}_6 | V_s(1^{-\pm}) | T\bar{B}_6 \rangle
&=& \langle T\bar{B}_6^* | V_s(1^{-\mp}) | T\bar{B}_6^* \rangle \, ,
\end{eqnarray}
where in the second line we have explicitly indicated whether
the sextet heavy baryon is in the ground or excited state.
The conclusion is again that $T\bar{S}$ molecules appear in pairs.
For $S\bar{S}$ molecules the contacts have a far richer structure, with
only one obvious symmetry relation:
\begin{eqnarray}
\langle S\bar{S} | V_s(2^{--}) | S\bar{S} \rangle =
\langle S\bar{S} | V_s(3^{--}) | S\bar{S} \rangle \, .
\end{eqnarray}
Tensor OPE mixes partial waves and will induce deviations from the previous
relations, which will be moderate if the bound states are shallow.
At this point it is worth noticing the analogy
with the heavy meson-antimeson case,
where this type of twin structure also happens for
(i) the $1^{+-}$ $D^*\bar{D}$ and $1^{+-}$ $D^*\bar{D}^*$ 
and (ii) the $1^{++}$ $D^*\bar{D}$ and $2^{++}$ $D^*\bar{D}^*$ molecules.
The first of these relations explains why the $Z_c$'s and $Z_b$'s
resonances appear in pairs~\cite{Bondar:2011ev,Mehen:2011yh},
while the latter predicts that the $X(3872)$ should have a $2^{++}$ {partner},
the $X(4012)$~\cite{Valderrama:2012jv,Nieves:2012tt}.
In the heavy meson-antimeson system there are a series of dynamical effects
(besides the aforementioned tensor OPE) that might break these patterns,
which include decays into nearby channels~\cite{Albaladejo:2015dsa},
coupled channel dynamics~\cite{Baru:2016iwj},
the existence of nearby quarkonia~\cite{Cincioglu:2016fkm}
and annihilation~\cite{Dai:2017ont}.
Though they have not been studied in the heavy baryon-antibaryon case,
these effects could be relevant.

Finally there is the important question of whether
the existence of heavy baryon molecules can be predicted.
EFTs are generic frameworks that usually require preexisting experimental input
to make predictions.
The EFT potential is composed of a long-range and short-range piece.
The short-range piece {involves} unknown couplings, which have to be determined
from external information.
In the absence of experimental data, there is the possibility of using
phenomenological arguments to estimate the contact-range couplings.
If we assume the saturation of these couplings from
$\rho$-, $\omega$-, $\phi$- and $\sigma$-meson exchange,
the most probable candidates for a heavy baryon-antibaryon bound state
are the isoscalar $\Lambda_c \bar{\Lambda}_c$,
$\Sigma_c \bar{\Sigma}_c$, $\Sigma_c^* \bar{\Sigma}_c$
and $\Sigma_c^* \bar{\Sigma}_c^*$ molecules,
located at $4573$, $4906$, $4970$ and $5035\,{\rm MeV}$ and
the isovector $\Lambda_c \bar{\Sigma}_c$ and $\Lambda_c \bar{\Sigma}_c^*$
molecules at $4740$ and $4805\,{\rm MeV}$.
If we consider the heavy baryon-baryon system instead, saturation indicates
that the isoscalar, doubly-charmed $\Sigma_c {\Sigma}_c$, $\Sigma_c^* {\Sigma}_c$
and $\Sigma_c^* {\Sigma}_c^*$ molecules are good candidates for binding,
followed by their isovector counterparts, the isoscalar
$\Lambda_c \Lambda_c$ and isovector
$\Lambda_c {\Sigma}_c$ and $\Lambda_c {\Sigma}_c^*$ systems.
For the heavy baryon-antibaryon system we supplement the saturation argument
with an estimate of the relative strength of the OPE potential,
which we assess by calculating the radius for which OPE
would be able to bind the system by itself.
This second argument also points to isoscalar molecules
as the most likely to bind.
It might be possible to observe these heavy baryon-antibaryon molecules
in experiments such as LHCb and PANDA, which is expected to be
particularly suited for the precision study of hidden charm
exotic states~\cite{PANDA:2018zjt}, or alternatively
in the lattice.

\section*{Acknowledgments}

M.P.V thanks Johann Haidenbauer for discussions and the Institute de
Physique Nucl\'eaire d'Orsay, where part of this work was carried out,
for its hospitality.
We also thank Chu-Wen Xiao for his comments on the manuscript.
This work is partly supported by the National Natural Science Foundation
of China under Grants No. 11375024,  No.11522539, No. 11735003, 
the Fundamental Research Funds for the Central Universities
and and the Thousand Talents Plan for Young Professionals.

\appendix
\section{The One Pion Exchange Potential
in Heavy Hadron Chiral Perturbation Theory}
\label{app:ope}

The OPE potential is the $\rm LO$ piece of the finite-range EFT potential.
In this appendix we explain how to compute it.
The idea is to obtain non-relativistic amplitudes for processes involving
an incoming and outgoing heavy baryon and a pion,
which we write as
\begin{eqnarray}
\mathcal{A}(B \to B' \pi; \vec{q}) \, ,
\end{eqnarray}
where $B$($B'$) are the initial/final baryon and $\vec{q}$ the momentum of
the pion if outgoing (if incoming we change the momentum to $-\vec{q}$).
If written in a suitable normalization these amplitudes can be combined
to compute the OPE potential (or for that matter any one boson exchange
potential) as follows
\begin{eqnarray}
\langle B_1' B_2' | V | B_1 B_2 \rangle =
\frac{\mathcal{A}_1(\vec{q})\,\mathcal{A}_2(-\vec{q})}{\vec{q}^2 + \mu_{\pi}^2} \, ,
\label{eq:ope-derivation}
\end{eqnarray}
where $1$ and $2$ refer to the pion vertices $1$ and $2$ and $\mu_{\pi}$ is the
effective pion mass for this particular transition (which is not necessarily
the physical pion mass because $m_{B_1'} - m_{B_1} \neq 0$ and gives the pion
a non-vanishing zero-th component to its 4-momentum).
The amplitudes $\mathcal{A}_1$ and $\mathcal{A}_2$ may refer to baryons or
antibaryons indistinctively.
In the following lines we will explain how to do the derivation in detail.

\subsection{The Heavy Baryon Field}

Heavy baryons contain a heavy quark and two light quarks,
i.e. they have the structure $| Q \, qq \rangle$.
The total spin of the light quark pair is $S_L = 0, 1$.
If the light quark spin is $S_L = 0$, we have an antitriplet spin-1/2
heavy baryons, which we denote by the Dirac field
\begin{eqnarray}
  \Psi_{\bar 3 Q} \, .
\end{eqnarray}
If the light quark spin is $S_L = 1$, we have spin-1/2 and spin-3/2
heavy baryons which we denote by the Dirac and Rarita-Schwinger fields
\begin{eqnarray}
  \Psi_{6 Q} \quad , \quad \Psi_{6 Q \mu}^* \, .
\end{eqnarray}
Notice that the Rarita-Schwinger field contains a Lorenzt index:
it is the external product of a Minkowsky vector
and a Dirac spinor.
The product contains a spurious spin-$\tfrac{1}{2}$ component that
can be removed with the condition
\begin{eqnarray}
\gamma^{\mu} \Psi_{6 Q\mu}^{*} = 0 \, . \label{eq:rarita-condition}
\end{eqnarray}
Within heavy hadron EFT it is customary to use the fields
$T(v)$ and $S_{\mu}(v)$ instead, which have good
transformation properties under rotations of
the spin of the heavy quark (where $v$ refers
to the velocity of the heavy quark).
For the $\Psi_{\bar 3 Q}$ heavy baryon the definition is
\begin{eqnarray}
T_Q(v) &=& \frac{1 + \slashed v}{2}\,\Psi_{\bar 3 Q} \, , \\
\bar{T}_Q(v) &=& \bar{\Psi}_{\bar 3 Q}\,\frac{1 + \slashed v}{2} \, .
\end{eqnarray}
The superfield $T_Q(v)$ transforms as
\begin{eqnarray}
T_Q(v) \to e^{-i \vec{\epsilon} \cdot \vec{S}_v} T_Q(v) \, ,
\end{eqnarray}
where $\vec{S}_v$ is related to $SU(2)_{v}$,
the $SU(2)$ spin group of the heavy quark $Q$ moving at velocity $v$.
For the $\Psi_{6 Q}$ and $\Psi_{6 Q \mu}^*$ sextet heavy baryons the definition is
\begin{eqnarray}
S_{Q\mu}(v) &=&
\frac{1}{\sqrt{3}}\,(\gamma_{\mu} + v_{\mu})\,\gamma_5\,
\frac{1 + \slashed v}{2}\,\Psi_{6 Q} \nonumber \\
&+& \frac{1 + \slashed v}{2}\,\Psi_{6 Q \mu}^* \, ,
\\
\bar{S}_{Q\mu}(v) &=&
-\frac{1}{\sqrt{3}}\,\bar{\Psi}_{6 Q}\,\frac{1 + \slashed v}{2}\,\gamma_5\,
(\gamma_{\mu} + v_{\mu}) \nonumber \\
&+& \bar{\Psi}_{6 Q\mu}^*\,\frac{1 + \slashed v}{2}  \, ,
\end{eqnarray}
where $\gamma_{\mu}$, $\gamma_5$ are the Dirac matrices.
The $S_{Q\mu}(v)$ superfield contains a Lorentz index that comes
from the Rarita-Schwinger field $\Psi_{6 Q\mu}^*$.
It obeys a contrain analogous to Eq.~(\ref{eq:rarita-condition})
\begin{eqnarray}
v^{\mu} S_{Q \mu} (v) = 0 \quad \mbox{and} \quad {\slashed v}
S_{Q \mu} (v) = S_{Q \mu}(v) \, . \nonumber \\
\end{eqnarray}
The $S_{Q\mu}(v)$ superfield transforms as
\begin{eqnarray}
S_{Q\mu}(v) \to e^{-i \vec{\epsilon} \cdot \vec{S}_v} S_{Q\mu}(v) \, .
\end{eqnarray}
In general we take the velocity parameter to be $v = (1, \vec{0})$.

We are interested in heavy baryon-antibaryon molecules, i.e. we need the
antibaryon fields.
Here it is important to notice that
\begin{eqnarray}
\bar{T}_Q(v) &=& T_Q^{\dagger}(v) \, \gamma_0 \, , \\
\bar{S}_{Q\mu}(v) &=& S_{Q\mu}^{\dagger}(v) \,\gamma_0 \, ,
\end{eqnarray}
are the operators for creating heavy baryons, which are unrelated to
the antibaryon fields.
Heavy antibaryons require the definition of new $T_{\bar{Q}}(v)$,
$\bar{T}_{\bar{Q}}(v)$, $S_{\bar{Q}}(v)$ and $\bar{S}_{\bar{Q}}(v)$ fields.
We will not need to define them explicitly though. Instead
we will use C- and G-parity transformations to deduce
the interactions of the heavy antibaryons
with the pions.

The heavy baryon fields have SU(2)-isospin and SU(3)-flavour structure.
If we add SU(3)-flavour indices for the heavy baryons
with $Q = b$ and $S_L = 0$, we have
\begin{eqnarray}
\Psi_{\bar 3 b} =
\begin{pmatrix}
\Xi_b^{-} \\
-\Xi_b^{0} \\
\Lambda_b^{0}
\end{pmatrix}
\end{eqnarray}
while for $Q = b$, $S_L = 1$ we have
\begin{eqnarray}
\Psi_{\bar 6 b} =
\begin{pmatrix}
\Sigma_b^{+} & \frac{1}{\sqrt{2}}\,\Sigma_b^{0} &
\frac{1}{\sqrt{2}}\,\Xi_b^{0'} \\
\frac{1}{\sqrt{2}}\,\Sigma_b^{0} & \Sigma_b^{-} & \frac{1}{\sqrt{2}}\,\Xi_b^{-'} \\
\frac{1}{\sqrt{2}}\,\Xi_b^{0'}  & \frac{1}{\sqrt{2}}\,\Xi_b^{-'} & \Omega_b^-
\end{pmatrix} \, ,
\end{eqnarray}
where the corresponding expressions for the spin-$\tfrac{3}{2}$
heavy baryons $\Psi_{6 c}^*$ are identical.
For the $SU(3)$-flavour structure of the $Q = c$ charmed baryons
we refer to Eqs.~(\ref{eq:B3c-su3}) and (\ref{eq:B6c-su3}).
In the following we will mostly consider the $SU(2)$-isospin structure:
when we talk about the $T_Q$ we could either be referring to $\Lambda_Q$
(isoscalar) or $\Xi_Q$ (isospinor), while when we talk about $S_Q$ it could
either be $\Sigma_Q$ (isovector), $\Xi_Q'$ (isospinor) or $\Omega_Q$
(isoscalar).

Notice that we are interested in the OPE potential: the isoscalar $\Omega_Q$
cannot exchange a single pion and will not be further considered here.
The isoscalar $\Lambda_Q$ and the isospinor $\Xi_Q$ can only exchange pions
in vertices involving a $\Xi_Q'$ and a $\Sigma_Q^{(*)}$ respectively,
i.e. there is no $\Xi_Q \Xi_Q \pi$ or $\Lambda_Q \Lambda_Q \pi$
vertex but there are $\Xi_Q \Xi_Q^{(*)'} \pi$ and
$\Lambda_Q \Sigma_Q^{(*)} \pi$ vertices.

\subsection{The Heavy Baryon Chiral Lagrangian at $\rm LO$}

The interaction of heavy baryons and pions
can be written as~\cite{Cho:1992cf,Cho:1992gg}
\begin{eqnarray}
\mathcal{L}_{T T \pi} &=& 0 \, , \\
\mathcal{L}_{S T \pi} &=& g_3\,\Big[
\epsilon_{ijk} \bar{T}_Q^i\,(A^{\mu})^{j}_{l}\,S^{k l}_{Q \mu} + \nonumber \\
&& \quad \epsilon^{ijk} \bar{S}^{\mu}_{Q k l}\,(A^{\mu})^{l}_{j}\,T_{Q i} \Big]
\, , \\
\mathcal{L}_{S S \pi} &=& i \, g_2 \, \epsilon_{\mu \nu \sigma \lambda} \,
\bar{S}^{\mu}_{Q i k} \, v^{\nu} \, (A^{\sigma})^{i}_{j} \,(S^{\lambda})_Q^{jk} \, ,
\end{eqnarray}
where the latin indices $i$,$j$,$k$,$l$ indicate either the SU(2)-isospin
or the SU(3)-flavour components, $g_2$, $g_3$ are coupling constants
and $\epsilon_{\mu \nu \sigma \lambda}$ is the 4-dimensional
Levi-Civita symbol.
In the equation above $A^{\mu}$ is the pseudo Goldstone-boson field,
\begin{eqnarray}
A^{\mu} &=& \frac{i}{2}\,
\left(
\xi^{\dagger} \partial^\mu \xi - \xi \partial^{\mu} \xi^{\dagger}
\right) \, ,
\end{eqnarray}
where $\xi$ is defined as
\begin{eqnarray}
\xi &=& e^{\frac{i}{f_{\pi}}\,M} \, ,
\end{eqnarray}
with the matrix $M$
\begin{eqnarray}
M =
\begin{pmatrix}
\frac{\pi^0}{\sqrt{2}} + \frac{\eta}{\sqrt{6}} & \pi^{+} & K^{+} \\
\pi^{-} & - \frac{\pi^0}{\sqrt{2}} + \frac{\eta}{\sqrt{6}} &  K^{0} \\
K^{-} & \bar{K}^0 & - \sqrt{\frac{2}{3}}\,\eta
\end{pmatrix} \, ,
\end{eqnarray}
which entails that we are taking the normalization choice
$f_{\pi} \simeq 132\,{\rm MeV}$.

If we consider the $SU(2)$ subgroup of $SU(3)$, the $A^{\mu}$ field reduces to
the following expansion in the pion field
\begin{eqnarray}
A^{\mu}
&=& - \frac{1}{f_{\pi}}\,\partial^{\mu} {\pi} + \frac{1}{6 f_{\pi}^3}\,
\left[ {\pi}, [{\pi}, \partial^{\mu} \pi]
\right] + \dots \, ,
\end{eqnarray}
where $\pi$ refers to the $SU(2)$ submatrix in the equation above
(after removing the contribution from the $\eta$), i.e.
\begin{eqnarray}
\pi =
\begin{pmatrix}
\frac{\pi^0}{\sqrt{2}} & \pi^+ \\
\pi^{-} & -\frac{\pi^0}{\sqrt{2}}
\end{pmatrix} \, .
\end{eqnarray}

\subsection{The Non-Relativistic Limit}

The potential is well-defined in the non-relativistic limit,
where the heavy baryon fields reduce to
\begin{eqnarray}
\Psi_{\bar 3 Q} &\to& \sqrt{2 M_{\bar 3}}\,
\begin{pmatrix}
\chi_s' \\
0
\end{pmatrix} \, , \label{eq:B3-spinor} \\
\Psi_{6 Q} &\to& \sqrt{2 M_6}\,
\begin{pmatrix}
\chi_s \\
0
\end{pmatrix} \, , \label{eq:B6-spinor} \\
\Psi_{6 Q\mu}^* = \left\{ {\Psi}^*_{6 Q 0} , \vec{\Psi}^*_{6 Q} \right \}
&\to& \sqrt{2 M_{6}^*}\,
\left\{
\begin{pmatrix}
0 \\
0
\end{pmatrix} \, , \,
\begin{pmatrix}
\vec{\chi}_s \\
0
\end{pmatrix}
\right\}
\, ,  \nonumber \\ \label{eq:BB6-spinor}
\end{eqnarray}
where $\chi_s'$, $\chi_s$ are standard spinors,
while $\vec{\chi}_s = (\chi_{s1}, \chi_{s2}, \chi_{s3})$ 
is a vector in which each component is a spinor.
The vector $\vec{\chi}_s$ fulfills the condition
\begin{eqnarray}
\vec{\sigma} \cdot \vec{\chi}_s = 0 \, ,
\end{eqnarray}
i.e. the non-relativistic version of Eq.~(\ref{eq:rarita-condition}),
which ensures that $\Psi^*_{6 Q\mu}$ is a genuine spin-$\tfrac{3}{2}$ field.
We have that $M_{\bar 3}$ is the $S_L = 0$ heavy baryon mass,
while $M_6$, $M_{6}^*$ are the $S_L = 1$ heavy baryon masses.
In the heavy quark limit the $S_L = 1$ baryon masses
are identical:
\begin{eqnarray}
  M_6 = M_{6}^* \quad \mbox{for $m_Q \to \infty$.}
\end{eqnarray}
However this does not happen with the $M_{\bar 3}$ mass, which remains
different from $M_6$ and $M_{6}^*$ in the heavy quark limit.
Putting all the pieces together, in the non-relativistic limit
the $S_L =0$ heavy field reduces to
\begin{eqnarray}
\frac{1}{\sqrt{2 M_{\bar{3}}}}\,T_Q(v) &\to&
\begin{pmatrix}
B_{\bar 3} \\
0
\end{pmatrix} \, , \\
\frac{1}{\sqrt{2 M_{\bar 3}}}\,\bar{T}_Q(v) &\to&
\begin{pmatrix}
B_{\bar 3}^{\dagger} & 0
\end{pmatrix} \, ,
\end{eqnarray}
while the $S_L = 1$ heavy fields read
\begin{eqnarray}
\frac{1}{\sqrt{2 M_{6}}}\,\vec{S}_Q(v) &\to&
\begin{pmatrix}
\sqrt{\frac{1}{3}}\,\vec{\sigma}\,B_{6} + \vec{B}_{6}^* \\
0
\end{pmatrix} \, , \\
\frac{1}{\sqrt{2 M_{6}}}\,\bar{\vec{S}}_{Q}(v) &\to&
\begin{pmatrix}
\sqrt{\frac{1}{3}}\,\vec{\sigma}\,B_{6}^{\dagger} +
\vec{B}_{6}^{*\dagger} & 0
\end{pmatrix} \, ,
\end{eqnarray}
with $B_{\bar 3}$, $B_{6}$ and $B_6^*$ the non-relativistic heavy baryon fields.
The notation can be further simplified by noticing
that (i) there is no difference between the $\bar{T}_Q$ / $\bar{S}_Q$
and $T_Q^{\dagger}$ / $S_Q^{\dagger}$  fields in the non-relativistic limit,
(ii) by ignoring the antibaryon components and (iii) by absorbing
the normalization factors $\sqrt{2 M_{\bar 3}}$ and $\sqrt{2 M_{6}}$
in a field redefinition.
In this case we end up with
\begin{eqnarray}
T_Q &=& B_{\bar 3} \, , \\
T_Q^{\dagger} &=& B_{\bar 3}^{\dagger} \, , \\
\nonumber \\
\vec{S}_Q &=& \sqrt{\frac{1}{3}}\,\vec{\sigma}\,B_{6} +
\vec{B}_{6}^* \, , \\
\vec{S}_Q^{\dagger} &=& \sqrt{\frac{1}{3}}\,B_{6}^{\dagger}\,\vec{\sigma} +
\vec{B}_{6}^{*\dagger} \, .
\end{eqnarray}
The pion field $A_{\mu}$ reduces in the heavy baryon non-relativistic limit to
\begin{eqnarray}
\vec{A} = -\frac{1}{f_{\pi}} \vec{\nabla} \vec{\pi} +
\mathcal{O}(\frac{\pi^3}{f_{\pi}^3}) \, ,
\end{eqnarray}
where we ignore the zero-th component of $A_{\mu}$ because it couples to
the zero-th component of $S_{Q \mu}(v)$, which vanishes
in the non-relativistic limit.
From this we can rewrite the Lagrangian as
\begin{eqnarray}
\mathcal{L}_{S T \pi} &=& -g_3\,\Big[
\epsilon_{ijk} {T}_Q^{i\dagger}\,
(\vec{A})^{j}_{l} \cdot \vec{S}_Q^{k l} + \nonumber \\
&& \quad \epsilon^{ijk} \vec{S}^{\dagger}_{Q k l}\,\cdot\,
(\vec{A})^{l}_{j}\,T_{Qi} \Big]
\, , \\
\mathcal{L}_{SS\pi} &=& - i \, g_2 \, {\rm Tr}\left[ \vec{S}_Q^{\dagger} \cdot
\left( \vec{A} \times \vec{S}_Q \right)  \right] \, ,
\end{eqnarray}
where the trace is over isospin space.
Alternatively we can expand the Lagrangian in terms of the fields
{$B_{\bar{3}}$, $B_{6}$ and $B_{6}^*$}:
\begin{eqnarray}
\mathcal{L}_{ST\pi} &=&
 \frac{g_3}{\sqrt{3} f_{\pi}} \,B_{\bar 3}^{\dagger} \,
\vec{\sigma} \cdot \vec{\nabla} \pi \,B_{6}
\nonumber \\ &+&
 \frac{g_3}{\sqrt{3} f_{\pi}} \,B_{6}^{\dagger} \,
\vec{\sigma} \cdot \vec{\nabla} \pi \,B_{6}'
\nonumber \\ &+&
 \frac{g_3}{f_{\pi}}\,B_{\bar 3}^{\dagger}\,
\vec{\nabla} \pi \,\cdot\,\vec{B}^*_{6}
\nonumber \\ &+&
 \frac{g_3}{f_{\pi}} \,\vec{B}_{6}^{*\dagger} \,
\,\cdot \vec{\nabla} \pi \,B_{\bar 3}
\end{eqnarray}
\begin{eqnarray}
\mathcal{L}_{SS\pi} &=&
 i \, \frac{g_2}{3 f_{\pi}} \,
B_{6}^{\dagger}\,\vec{\sigma}\,\cdot\,
\left(
\vec{\nabla}\,\vec{\pi} \times \sigma
\right) \, B_{6} \nonumber \\ &+&
 i \, \frac{g_2}{f_{\pi}}\,
\vec{B}_{6}^{*\dagger}\,\cdot\,
\left(
\vec{\nabla}\,\vec{\pi} \times \vec{B}_{6}^*
\right) \nonumber \\ &+&
 i \, \frac{g_2}{\sqrt{3}\,f_{\pi}}\,
{B}_{6}^{\dagger}\,\vec{\sigma}\,\cdot\,
\left(
\vec{\nabla}\,\vec{\pi} \times \vec{B}_{6}^*
\right) \nonumber \\ &+&
 i \, \frac{g_2}{\sqrt{3}\,f_{\pi}}\,\vec{B}_{6}^{*\dagger}\,\cdot\,
\left(
\vec{\nabla}\,\vec{\pi} \times \vec{\sigma}
\right) B_{6}  \, , 
\end{eqnarray}
where  we have removed the isospin / flavour indices to make
the expressions shorter.

\subsection{The Spin and Isospin Factors}

Now we calculate the matrix elements of the different vertices,
which depend on a series of spin and isospin factors.
In the charm sector ($Q=c$) the relations between the isospin
and particle basis are
\begin{eqnarray}
B_{\bar{3}}(\Lambda_c) &=& \Lambda_c^{+} \, , \\
B_{\bar{3}}(\Xi_c) &=& \begin{pmatrix}
\Xi_c^{+'} \\
\Xi_c^{0'}
\end{pmatrix} \, , \\
B_{6}(\Xi_c') &=& \begin{pmatrix}
\Xi_c^{+'} \\
\Xi_c^{0'}
\end{pmatrix} \, , \\
B_{6}(\Sigma_c) &=& \begin{pmatrix}
\Sigma_c^{++} & \frac{1}{\sqrt{2}}\,\Sigma_c^{+} \\
\frac{1}{\sqrt{2}}\,\Sigma_c^{+} & \Sigma_c^0
\end{pmatrix} \, ,
\end{eqnarray}
for the $\Lambda_c$, $\Xi_c$, $\Xi_c'$ and $\Sigma_c$ baryons respectively.
The relations for the excited $\Xi_c^{'*}$ and $\Sigma_c^{*}$ sextet baryons
{are identical to those} of the $\Xi_c^{'}$ and $\Sigma_c$ baryons.
Alternatively if we consider isospin vectors we can write
\begin{eqnarray}
\Lambda_c^{+} &=&  | 0 0 \rangle_I  \, , \\
\left\{ \Xi_c^{+} , \Xi_c^{0} \right\} &=&
\left\{ {| \frac{1}{2}, \phantom{+}\frac{1}{2} \rangle}_I ,
{| \frac{1}{2}, -\frac{1}{2} \rangle}_I \right\} \, , \\
\left\{ \Xi_c^{+'} , \Xi_c^{0'} \right\} &=&
\left\{ {| \frac{1}{2}, \phantom{+}\frac{1}{2} \rangle}_I ,
{| \frac{1}{2}, -\frac{1}{2} \rangle}_I \right\} \, , \\
\left\{ \Sigma_c^{++}, \Sigma_c^{+}, \Sigma_c^{0} \right\} &=&
\left\{ {| 1 , \phantom{+}1 \rangle}_I  ,
{| 1 , \phantom{+}0 \rangle}_I  ,
{| 1 , -1 \rangle}_I \right\} \, . \nonumber \\
\end{eqnarray}
The isospin factors can be extracted by first expanding the isospin / flavour
indices in the particle basis and later reinterpreting the result in terms
of matrices in the isospin space.
We begin with the $S T \pi$ Lagrangian, for which
\begin{eqnarray}
\Lambda_c^{\dagger} \pi^a \Sigma_c = t^a \, , \\
\Xi_c^{\dagger} \pi^a \Xi_c' = \frac{\tau^a}{2} \, ,
\end{eqnarray}
where $\pi^a$ is the pion field in the Cartesian basis, $\tau^a$
are the Pauli matrices and $t^a$ are given by
\begin{eqnarray}
t^1 &=&
\begin{pmatrix}
\phantom{+} \frac{1}{\sqrt{2}} \\
\phantom{+} 0 \\
-\frac{1}{\sqrt{2}}
\end{pmatrix} \, , \\
t^2 &=&
\begin{pmatrix}
\frac{i}{\sqrt{2}} \\
0 \\
\frac{i}{\sqrt{2}}
\end{pmatrix} \, , \\
t^3 &=&
\begin{pmatrix}
0 \\
1 \\
0
\end{pmatrix} \, .
\end{eqnarray}
In the $S S \pi$ case we have
\begin{eqnarray}
\Xi_c^{'\dagger} \pi^a \Xi_c' &=& \frac{\tau^a}{2\,\sqrt{2}} \, , \\
\Sigma_c^{\dagger} \pi^a \Sigma_c &=& \frac{T^a}{\sqrt{2}} \, ,
\end{eqnarray}
where $T_a$ are the $J=1$ angular momentum matrices in isospin space.
The isospin factors for the $\Xi_c^*$ and $\Sigma_c^*$ baryons are
identical to those of the $\Xi_c$ and $\Sigma_c$ baryons.

Next we factor out the spin in terms of
angular momentum matrices or equivalent expressions.
For the $S T \pi$ vertices the factors are
\begin{eqnarray}
B_{\bar{3}}^{ \dagger}\, \vec{\sigma} \cdot \vec{q} \, B_{6} &=&
\vec{\sigma} \cdot \vec{q} \, , \\
B_{\bar{3}}^{ \dagger}\, \vec{q} \cdot \, \vec{B}^*_{6} &=&
\vec{S} \cdot \vec{q} \, ,
\end{eqnarray}
while for the $S S \pi$ vertices we have
\begin{eqnarray}
B_{6}^{\dagger}\,\vec{\sigma}\cdot
\left( \vec{q} \times \vec{\sigma} \right) \, B_{6}
&=& - i \, 2 \, \vec{\sigma} \cdot \vec{q} \, ,  \\
\vec{B}_{6}^{*\dagger}\,\cdot \left( \vec{q} \times
\vec{B}_{6}^* \right)
&=& -i \, \frac{2}{3} \vec{\Sigma} \cdot \vec{q} \, ,  \\
B_{6}^{\dagger}\,\vec{\sigma}\, \cdot
\left( \vec{q} \times \vec{B}_{6}^* \right)
&=& - i \, \vec{S} \cdot \vec{q} \, , \\
\vec{B}_{6}^{*\dagger} \cdot \left( \vec{q} \times \vec{\sigma}
\right) B_{6}
&=& - i \, \vec{S}^{\dagger} \cdot \vec{q} \, ,
\end{eqnarray}
where $\vec{\sigma}$ are the Pauli matrices,
$\vec{\Sigma}$ are the $J=\tfrac{3}{2}$ angular momentum matrices
and $\vec{S}$ are $2\times4$ matrices that connect
the spin-$\tfrac{1}{2}$ and spin-$\tfrac{3}{2}$ baryons.
These $\vec{S}$ matrices read
\begin{eqnarray}
S_1 &=&
\begin{pmatrix}
\frac{1}{\sqrt{2}} & 0 & - \frac{1}{\sqrt{6}} & 0 \\
0 & \frac{1}{\sqrt{6}} & 0 & -\frac{1}{\sqrt{2}}
\end{pmatrix}
\, , \\
S_2 &=&
\begin{pmatrix}
\frac{i}{\sqrt{2}} & 0 & \frac{i}{\sqrt{6}} & 0 \\
0 & \frac{i}{\sqrt{6}} & 0 & \frac{i}{\sqrt{2}}
\end{pmatrix}
\, , \\
S_3 &=&
\begin{pmatrix}
0 & -\sqrt{\frac{2}{3}} & 0 & 0 \\
0 & 0 & -\sqrt{\frac{2}{3}} & 0
\end{pmatrix}
\, ,
\end{eqnarray}
which are normalized as follows
\begin{eqnarray}
S_i S_j^{\dagger} = \frac{2\,\delta_{ij} - i \epsilon_{ijk} \sigma_k}{3} \, .
\end{eqnarray}
Now we define the non-relativistic amplitudes as
\begin{eqnarray}
\mathcal{A}(B \to B \pi^a) &=&
- i \langle B \pi^a | \mathcal{L} | B \rangle \, ,
\end{eqnarray}
with $B = B_{\bar 3}$, $B_6$, $B_6^*$.
For the transitions involving $\Lambda_c$ we have
\begin{eqnarray}
\mathcal{A}(\Lambda_c \to \Sigma_c \pi^a) &=&
\frac{g_3}{\sqrt{3} f_{\pi}}\,t^a\,\vec{\sigma} \cdot \vec{q} \, ,
\label{eq:amp-first} \\
\mathcal{A}(\Lambda_c \to \Sigma_c^* \pi^a) &=&
\frac{g_3}{f_{\pi}}\,t^a\,\vec{S}^{\dagger} \cdot \vec{q} \, , \\
\mathcal{A}(\Sigma_c^* \to \Lambda_c \pi^a) &=&
\frac{g_3}{f_{\pi}}\,t^a\,\vec{S} \cdot \vec{q} \, .
\end{eqnarray}
For the transitions involving $\Xi_c$ we have
\begin{eqnarray}
\mathcal{A}(\Xi_c' \to \Xi_c \pi^a) &=&
\frac{g_3}{\sqrt{3} f_{\pi}}\,\frac{\tau^a}{2}\,
\vec{\sigma} \cdot \vec{q} \, , \\
\mathcal{A}(\Xi_c' \to \Xi_c^* \pi^a) &=&
\frac{g_3}{f_{\pi}}\,\frac{\tau^a}{2}\,\vec{S}^{\dagger} \cdot \vec{q} \, , \\
\mathcal{A}(\Xi_c^* \to \Xi_c' \pi^a) &=&
\frac{g_3}{f_{\pi}}\,\frac{\tau^a}{2}\,\vec{S} \cdot \vec{q} \, .
\end{eqnarray}
For the ones with the $\Xi_c'$ and $\Xi_c^*$, the amplitudes read
\begin{eqnarray}
\mathcal{A}(\Xi_c' \to \Xi_c' \pi^a) &=& \frac{2 g_2}{3 f_{\pi}}\,
\frac{\tau^a}{2 \sqrt{2}}\,\vec{\sigma} \cdot \vec{q} \, , \\
\mathcal{A}(\Xi_c^{'*} \to \Xi_c^{'*} \pi^a) &=&
\frac{2 g_2}{3 f_{\pi}}\,
\frac{\tau^a}{2 \sqrt{2}}\,\vec{\Sigma} \cdot \vec{q} \, , \\
\mathcal{A}(\Xi_c^{'*} \to \Xi_c' \pi^a) &=&
\frac{g_2}{\sqrt{3} f_{\pi}}\,
\frac{\tau^a}{2 \sqrt{2}}\,\vec{S} \cdot \vec{q} \, , \\
\mathcal{A}(\Xi_c \to \Xi_c^{'*} \pi^a) &=&
\frac{g_2}{\sqrt{3} f_{\pi}}\,
\frac{\tau^a}{2\sqrt{2}}\,\vec{S}^{\dagger} \cdot \vec{q} \, ,
\end{eqnarray}
Finally for the transitions with the $\Sigma_c$ and $\Sigma_c^*$
we have the following amplitudes
\begin{eqnarray}
\mathcal{A}(\Sigma_c \to \Sigma_c \pi^a) &=& \frac{2 g_2}{3 f_{\pi}}\,
\frac{T^a}{\sqrt{2}}\,\vec{\sigma} \cdot \vec{q} \, , \\
\mathcal{A}(\Sigma_c^* \to \Sigma_c^* \pi^a) &=&
\frac{2 g_2}{3 f_{\pi}}\,
\frac{T^a}{\sqrt{2}}\,\vec{\Sigma} \cdot \vec{q} \, , \\
\mathcal{A}(\Sigma_c^* \to \Sigma_c \pi^a) &=&
\frac{g_2}{\sqrt{3} f_{\pi}}\,
\frac{T^a}{\sqrt{2}}\,\vec{S} \cdot \vec{q} \, , \\
\mathcal{A}(\Sigma_c \to \Sigma_c^* \pi^a) &=&
\frac{g_2}{\sqrt{3} f_{\pi}}\,
\frac{T^a}{\sqrt{2}}\,\vec{S}^{\dagger} \cdot \vec{q} \, .
\label{eq:amp-last}
\end{eqnarray}

\subsection{G-parity and Heavy Antibaryons}

The amplitudes in Eqs.~(\ref{eq:amp-first}-\ref{eq:amp-last}) are
for the heavy baryons.
Here we deduce the amplitudes for the heavy antibaryons by working
in the isospin basis and applying a G-parity transformation,
which is a combination of a C-parity transformation
and a rotation in isospin space~\cite{Lee:1956sw}
\begin{eqnarray}
G = C\,e^{i \pi I_2} \, ,
\end{eqnarray}
with $I_2$ the second Cartesian component of the isospin matrix.
Now we will determine how $G$ operates on the different fields we consider here.
For instance, pions have well-defined G-parity
\begin{eqnarray}
G | \pi \rangle = - | \pi \rangle \, .
\end{eqnarray}
We can write it in terms of the components of the pion field for completeness
\begin{eqnarray}
G
\begin{pmatrix}
| \pi^+ \rangle \\
| \pi^0 \rangle \\
| \pi^- \rangle
\end{pmatrix} =
-
\begin{pmatrix}
| \pi^+ \rangle \\
| \pi^0 \rangle \\
| \pi^- \rangle
\end{pmatrix} \, .
\end{eqnarray}
If we consider baryons instead, $G$ will transform a baryon
into an antibaryon in the same isospin state.
If we consider nucleons or other isospin $\tfrac{1}{2}$ baryons,
the G-parity transformation works as follows
\begin{eqnarray}
G
\begin{pmatrix}
| p \rangle \\
| n \rangle
\end{pmatrix} =
\begin{pmatrix}
\phantom{+} | \bar{n} \rangle \\
- | \bar{p} \rangle
\end{pmatrix} \, ,
\end{eqnarray}
and now we can identify antiparticle states with isospinors as follows
\begin{eqnarray}
| \bar{n} \rangle &=& \phantom{+}
{| \frac{1}{2} \, +\frac{1}{2} \rangle}_I \, , \\
| \bar{p} \rangle &=& -
{| \frac{1}{2} \, -\frac{1}{2} \rangle}_I \, ,
\end{eqnarray}
where we use the subscript $I$ to indicate
that we are indeed referring to isospinors.
From the nucleon/antinucleon example we can appreciate that
the idea of a G-parity transformation is to have a good mapping
between the isospin and the particle/antiparticle basis.
The crucial point in the G-parity transformation for the baryons is
the relative minus sign between the isospin vectors of
the antineutron and antiproton, which in turn allows
for the use of the same SU(2) Clebsch-Gordan coefficients
in the baryon and antibaryon cases.

For the heavy baryons the idea is the same as for the nucleons.
But there is a subtlety: we are considering different C-parity conventions
for the spin-$\tfrac{1}{2}$ and spin-$\tfrac{3}{2}$ fields
\begin{eqnarray}
C | B \rangle &=& \phantom{+}| \bar{B} \rangle \, , \\
C | B_{6}^* \rangle &=& - | \bar{B}_{6}^* \rangle \, ,
\end{eqnarray}
where $B = B_{\bar 3}, B_6$.
For the antitriplet and sextet spin-$\frac{1}{2}$ heavy cascades
the transformation works exactly as in nucleons
\begin{eqnarray}
G
\begin{pmatrix*}[l]
| \Xi_c^{+(')} \rangle \\
| \Xi_c^{0(')} \rangle
\end{pmatrix*} =
\begin{pmatrix*}[l]
\phantom{+} | \bar{\Xi}_c^{0(')} \rangle \\
- | \Xi_c^{-(')} \rangle
\end{pmatrix*} \, ,
\end{eqnarray}
while for the sextet spin-$3/2$ heavy cascades we have
\begin{eqnarray}
G
\begin{pmatrix*}[l]
| \Xi_c^{+*} \rangle \\
| \Xi_c^{0*} \rangle
\end{pmatrix*} = -
\begin{pmatrix*}[l]
\phantom{+} | \bar{\Xi}_c^{0*} \rangle \\
- | \Xi_c^{-*} \rangle
\end{pmatrix*} \, .
\end{eqnarray}
For the isotriplet heavy baryons $\{ \Sigma_c^{++}, \Sigma_c^{+}, \Sigma_c^{0} \}$
we have instead
\begin{eqnarray}
G
\begin{pmatrix*}[l]
| \Sigma_c^{++} \rangle \\
| \Sigma_c^{+} \rangle \\
| \Sigma_c^{0} \rangle
\end{pmatrix*} =
\begin{pmatrix*}[l]
\phantom{+} | \bar{\Sigma}_c^{0} \rangle \\
- | \Sigma_c^{-} \rangle \\
\phantom{+} | \Sigma_c^{--} \rangle
\end{pmatrix*} \, ,
\end{eqnarray}
plus the transformation for the excited isotriplet heavy baryons,
which will carry an extra minus sign.

With the G-parity transformation we can deduce the amplitudes
for the antibaryons from the ones we already know for the baryons
\begin{eqnarray}
\mathcal{A}(\bar{B} \to \bar{B} \, \pi^a)
&=& - \mathcal{A}({B} \to {B} \, \pi^a) \, , \\
\mathcal{A}(\bar{B} \to \bar{B}_6^* \, \pi^a)
&=& + \mathcal{A}({B} \to {B}_6^* \, \pi^a) \, , \\
\mathcal{A}(\bar{B}_6^* \to \bar{B} \, \pi^a)
&=& + \mathcal{A}({B}_6^* \to {B} \, \pi^a) \, , \\
\mathcal{A}(\bar{B}_6^* \to \bar{B}_6^* \, \pi^a)
&=& - \mathcal{A}({B}_6^* \to {B}_6^* \, \pi^a) \, ,
\end{eqnarray}
where $B = B_{\bar 3}, B_6$.
The signs simply reflect the sign for the G-parity transformation of the pion,
plus the extra sign involved in the $B_{\bar 3} / B_6 \to B_6^*$ and
$B_6^* \to B_{\bar 3} / B_6$ transitions.
For a detailed example we consider
\begin{eqnarray}
\mathcal{A}(\bar{B}_6 \to \bar{B}_6 \, \pi^a)
&=& \phantom{+} i
\langle \bar{B}_6 \pi^{a} | \mathcal{L} |
\bar{B}_6 \rangle \nonumber \\
&=& - i \langle G (B_6 \pi^{a}) | \mathcal{L} |
G B_6 \rangle \nonumber \\
&=& - i \langle B_6 \pi^{a} | G^{\dagger}\,\mathcal{L}\, G |
B_6 \rangle \nonumber \\
&=& - i \langle B_6 \pi^{a} | \mathcal{L} | B_6 \rangle \nonumber \\
&=& - \mathcal{A}({B}_6 \to {B}_6 \, \pi^a) \, ,
\end{eqnarray}
where we have used that $G^2 | B_6 \rangle = \pm 1$ ($-1$ for $\Xi_Q'$,
$+1$ for $\Sigma_Q$), $G^2 | \pi \rangle = + 1$ and
$G^{\dagger}\,\mathcal{L}\, G = \mathcal{L}$.

\subsection{The OPE Potential}

With the amplitudes of Eqs.~(\ref{eq:amp-first}-\ref{eq:amp-last})
we can derive the potential by using Eq.~(\ref{eq:ope-derivation}).
For simplicity we will consider the heavy quark limit,
in which the $B_6$ and $B_6^*$ heavy baryons are degenerate.
For the $T \bar{S} = \Lambda_c \bar{\Sigma}_c, \Lambda_c \bar{\Sigma}_c^*$
and $T \bar{S} = \Xi_c \bar{\Xi}_c', \Xi_c \bar{\Xi}_c^*$ potentials
we write the potential in the bases
\begin{eqnarray}
\mathcal{B}_{\Lambda_c \bar{\Sigma_c}} &=&
\{
\Lambda_c \bar{\Sigma}_c, \Sigma_c \bar{\Lambda}_c,
\Lambda_c \bar{\Sigma}^*_c, \Sigma_c^* \bar{\Lambda}_c
\} \, , \\
\mathcal{B}_{\Xi_c' \bar{\Xi_c}} &=&
\{
\Xi_c \bar{\Xi}_c', \Xi_c' \bar{\Xi}_c,
\Xi_c \bar{\Xi}^*_c, \Xi_c^* \bar{\Xi}_c
\} \, .
\end{eqnarray}
in which the potential reads as 
\begin{widetext}
\begin{eqnarray}
V_{\rm OPE}^{S \bar{T}}(\vec{q}) &=&
\frac{g_3^2}{f_{\pi}^2}\,\tau\,\frac{1}{\vec{q}^2 + \mu_{\pi}^{2}}\,
\begin{pmatrix}
0 & \frac{1}{3}\,\vec{\sigma}_1 \cdot \vec{q} \, \vec{\sigma}_2 \cdot \vec{q} &
0 &
\frac{1}{\sqrt{3}}\,\vec{S}_1 \cdot \vec{q} \, \vec{\sigma}_2 \cdot \vec{q} \\
\frac{1}{3}\,\vec{\sigma}_1 \cdot \vec{q} \, \vec{\sigma}_2 \cdot \vec{q} & 0 &
\frac{1}{\sqrt{3}}\,\vec{\sigma}_1 \cdot \vec{q} \, \vec{S}_2 \cdot \vec{q}
& 0 \\
0 & \frac{1}{\sqrt{3}}\,\vec{\sigma}_1 \cdot \vec{q} \,
\vec{S}_2^{\dagger} \cdot \vec{q} &
0 & \vec{S}_1^{\dagger} \cdot \vec{q} \, \vec{S}_2^{\dagger} \cdot \vec{q} \\
\frac{1}{\sqrt{3}}\,\vec{S}_1^{\dagger} \cdot \vec{q} \,
\vec{\sigma}_2 \cdot \vec{q} & 0
& \vec{S}_1^{\dagger} \cdot \vec{q} \, \vec{S}_2 \cdot \vec{q} & 0
\end{pmatrix}
\nonumber \\ &+&
\mathcal{O}\left( \frac{1}{m_Q} \right) \, ,
\label{eq:V_ST}
\end{eqnarray}
\end{widetext}
where the isospin factor $\tau$ is
\begin{eqnarray}
\tau(\Lambda_c \bar{\Sigma}_c) &=& 1 \, , \\
\tau(\Xi_c \bar{\Xi}_c') &=& \frac{\vec{\tau}_1 \cdot \vec{\tau}_2}{4} \, .
\end{eqnarray}
The effective pion mass in the heavy quark limit is given by
$\mu_{\pi}^2 = m_{\pi}^2 - (m_{\Sigma_c} - m_{\Lambda_c})^2$ or
$\mu_{\pi}^2 = m_{\pi}^2 - (m_{\Xi_c'} - m_{\Xi_c})^2$ respectively.
For the $S\bar{S}$ potential we use the bases
\begin{eqnarray}
\mathcal{B}_{\Xi_c' \bar{\Xi}_c'} &=&
\{ \Xi_c' \bar{\Xi}_c' , \Xi_c' \bar{\Xi}_c^*, \Xi_c^* \bar{\Xi}_c'
,  \Xi_c^* \bar{\Xi}_c^* \} \, , \\
\mathcal{B}_{\Sigma_c \bar{\Sigma}_c} &=&
\{ \Sigma_c \bar{\Sigma}_c , \Sigma_c \bar{\Sigma}_c^*,
\Sigma_c^* \bar{\Sigma}_c, \Sigma_c^* \bar{\Sigma}_c^* \} \, ,
\end{eqnarray}
in which the potential reads
\begin{widetext}
\begin{eqnarray}
V_{\rm OPE}^{S \bar{S}}(\vec{q}) &=&
\frac{2 g_2^2}{9 f_{\pi}^2}\,\tau\,\frac{1}{\vec{q}^2 + m_{\pi}^{2}}\,
\begin{pmatrix}
+\,\vec{\sigma}_1 \cdot \vec{q} \, \vec{\sigma}_2 \cdot \vec{q} &
-\lambda\,\vec{\sigma}_1 \cdot \vec{q} \, \vec{S}_2 \cdot \vec{q} &
+\lambda\,\vec{S}_1 \cdot \vec{q} \, \vec{\sigma}_2 \cdot \vec{q} &
-\lambda^2\,\vec{S}_1 \cdot \vec{q} \, \vec{S}_2 \cdot \vec{q} \\
-\lambda\,\vec{\sigma}_1 \cdot \vec{q} \, \vec{S}_2 \cdot \vec{q} &
+\,\vec{\sigma}_1 \cdot \vec{q} \, \vec{\Sigma}_2 \cdot \vec{q} &
-\lambda^2\vec{S}_1 \cdot \vec{q} \, \vec{S}_2^{\dagger} \cdot \vec{q} &
+\lambda\,\vec{S}_1 \cdot \vec{q} \, \vec{\Sigma}_2 \cdot \vec{q}
\\
+\lambda\,\vec{S}_1^{\dagger} \cdot \vec{q}
\, \vec{\sigma}_2 \cdot \vec{q} &
-\lambda^2\vec{S}_1^{\dagger} \cdot \vec{q}
\, \vec{S}_2 \cdot \vec{q} &
+\,\vec{\Sigma}_1 \cdot \vec{q} \, \vec{\sigma}_2 \cdot \vec{q} &
-\lambda\,\vec{\Sigma}_1 \cdot \vec{q}
\, \vec{S}_2 \cdot \vec{q} \\
-\lambda^2\,\vec{S}_1^{\dagger} \cdot \vec{q}
\, \vec{S}_2^{\dagger} \cdot \vec{q} &
+\lambda\,\vec{S}_1^{\dagger} \cdot \vec{q}
\, \vec{\Sigma}_2 \cdot \vec{q} &
-\lambda\,\vec{\Sigma}_1 \cdot \vec{q}
\, \vec{S}_2^{\dagger} \cdot \vec{q} &
+\,\vec{\Sigma}_1 \cdot \vec{q}
\, \vec{\Sigma}_2 \cdot \vec{q}
\end{pmatrix} \nonumber \\ &+&
\mathcal{O}\left( \frac{1}{m_Q} \right) \, ,
\label{eq:V_SS}
\end{eqnarray}
\end{widetext}
where $\lambda = \frac{\sqrt{3}}{2}$ and where the isospin factor is
\begin{eqnarray}
\tau\,(\Xi_c' \bar{\Xi}_c') &=& \frac{\vec{\tau}_1 \cdot \vec{\tau}_2}{4}
\, , \\
\tau\,(\Sigma_c \bar{\Sigma}_c) &=& \vec{T}_1 \cdot \vec{T}_2 \, .
\end{eqnarray}

\subsection{Coordinate Space}

The general form of the $S \bar{T}$ and $S \bar{S}$ potential
in momentum space can be written as
\begin{eqnarray}
V^{T\bar{S}}_{\rm OBE} &=& -R_1\,\bar{R}_2\,
\frac{g_3^2}{2 f_{\pi}^2}\,\vec{I}_1 \cdot \vec{I}_2 \,
\frac{\vec{a}_1 \cdot \vec{q} \, \vec{a}_2 \cdot \vec{q}}{q^2 + \mu_{\pi}^2} \, ,
\label{eq:V-OPE-TS}
\\
V^{S\bar{S}}_{\rm OPE} &=& -R_1\,\bar{R}_2\,
\frac{g_2^2}{2 f_{\pi}^2}\,\vec{I}_1 \cdot \vec{I}_2 \,
\frac{\vec{a}_1 \cdot \vec{q} \, \vec{a}_2 \cdot \vec{q}}{q^2 + \mu_{\pi}^2} \, ,
\label{eq:V-OPE-SS}
\end{eqnarray}
where $R_1$ and $\bar{R}_2$ are numerical factors,
$\mu_{\pi}$ is the effective pion mass at the vertices
(as explained in the previous section),
${I}_1$, ${I}_2$ the appropriate isospin matrices
and $\vec{a}_1$, $\vec{a}_1$ are the spin matrices acting on vertex 1 and 2.
The specific factors can be worked out easily
from Eqs.~(\ref{eq:V_ST}) and (\ref{eq:V_SS})
to obtain the results of Table~\ref{tab:vertices}.
The coordinate space potential is obtained by Fourier transforming
the momentum space potential
\begin{eqnarray}
V^{(0)}(\vec{r}) &=& \int \frac{d^3 q}{(2\pi)^3}\,V^{(0)}(\vec{q})\,
e^{-i \vec{q} \cdot \vec{r}} \nonumber \\
&=& R_1\,\bar{R}_2\,\frac{g_i^2}{2 f_{\pi}^2}\,\vec{I}_1 \cdot \vec{I}_2 \,
(\vec{a}_1 \cdot \nabla)\,(\vec{a}_2 \cdot \nabla)\,\frac{e^{-\mu_{\pi} r}}{4 \pi r}
\, , \nonumber \\
\end{eqnarray}
where $g_i = g_2$ or $g_3$ depending on the case.
From this we obtain the expressions we already wrote in Eqs.~(\ref{eq:VOPE}),
(\ref{eq:C12}), (\ref{eq:S12}), (\ref{eq:WC}) and (\ref{eq:WT}).
Here we will write the coordinate space potential in coupled channels,
in which case we obtain
\begin{eqnarray}
V^{(0)}_{S\bar{T}}(\vec{r}) &=& \tau\,\frac{g_3^2}{3\,f_{\pi}^2}\,
{\bf C}_{12}^{S\bar{T}} \delta^3(\vec{r}) \nonumber \\ &-&
\tau \Big[ {\bf C}_{12}^{S\bar{T}}\, W_C(r) +
{\bf S}_{12}^{S \bar{T}}(\hat{r}) \, W_T(r) \Big] \, , \nonumber \\
\\
V^{(0)}_{S\bar{S}}(\vec{r}) &=& \tau\,\frac{2 g_3^2}{27\,f_{\pi}^2}\,
{\bf C}_{12}^{S\bar{S}} \delta^3(\vec{r}) \nonumber \\ &-&
\tau\,\frac{2}{9}\,\Big[ {\bf C}_{12}^{S\bar{S}}\, W_C(r) +
{\bf S}_{12}^{S \bar{S}}(\hat{r}) \, W_T(r) \Big] \, , \nonumber \\
\end{eqnarray}
where $W_C$ and $W_T$ are defined in Eqs.~(\ref{eq:WC}) and (\ref{eq:WT}),
and with the central and tensor matrices given by
\begin{widetext}
\begin{eqnarray}
{\bf C}^{S\bar{T}}_{12} &=&
\begin{pmatrix}
0 & \frac{1}{3}\,\vec{\sigma}_1 \cdot \vec{\sigma}_2 &
0 &
\frac{1}{\sqrt{3}}\,\vec{S}_1 \cdot \vec{\sigma}_2 \\
\frac{1}{3}\,\vec{\sigma}_1 \cdot \vec{\sigma}_2 & 0 &
\frac{1}{\sqrt{3}}\,\vec{\sigma}_1 \cdot \vec{S}_2
& 0 \\
0 & \frac{1}{\sqrt{3}}\,\vec{\sigma}_1 \cdot \vec{S}_2^{\dagger} &
0 & \vec{S}_1^{\dagger} \cdot \vec{S}_2^{\dagger} \\
\frac{1}{\sqrt{3}}\,\vec{S}_1^{\dagger} \cdot \vec{\sigma}_2 & 0
& \vec{S}_1^{\dagger} \cdot \vec{S}_2 & 0
\end{pmatrix} \, , \\
{\bf C}^{S\bar{S}}_{12} &=&
\begin{pmatrix}
+\,\vec{\sigma}_1 \cdot \vec{\sigma}_2 &
-\lambda\,\vec{\sigma}_1 \cdot \vec{S}_2 &
+\lambda\,\vec{S}_1 \cdot \vec{\sigma}_2 &
-\lambda^2\,\vec{S}_1 \cdot \vec{S}_2 \\
-\lambda\,\vec{\sigma}_1 \cdot \vec{S}_2 &
+\,\vec{\sigma}_1 \cdot \vec{\Sigma}_2 &
-\lambda^2\vec{S}_1 \cdot \vec{S}_2^{\dagger} &
+\lambda\,\vec{S}_1 \cdot \vec{\Sigma}_2
\\
+\lambda\,\vec{S}_1^{\dagger} \cdot \vec{\sigma}_2 &
-\lambda^2\vec{S}_1^{\dagger} \cdot \vec{S}_2 &
+\,\vec{\Sigma}_1 \cdot \vec{\sigma}_2 &
-\lambda\,\vec{\Sigma}_1 \cdot \vec{S}_2 \\
-\lambda^2\,\vec{S}_1^{\dagger} \cdot \vec{S}_2^{\dagger} &
+\lambda\,\vec{S}_1^{\dagger} \cdot \vec{\Sigma}_2 &
-\lambda\,\vec{\Sigma}_1 \cdot \vec{S}_2^{\dagger} &
+\,\vec{\Sigma}_1 \cdot \vec{\Sigma}_2
\end{pmatrix}
 \, ,
\end{eqnarray}
\begin{eqnarray}
{\bf S}^{S\bar{T}}_{12}(\hat{r}) &=&
\begin{pmatrix}
0 & \frac{1}{3}\,S_{12}(\vec{\sigma}_1,\vec{\sigma}_2, \hat{r}) &
0 &
\frac{1}{\sqrt{3}}\,S_{12}(\vec{S}_1, \vec{\sigma}_2, \hat{r}) \\
\frac{1}{3}\,S_{12}(\vec{\sigma}_1 , \vec{\sigma}_2 , \hat{r}) & 0 &
\frac{1}{\sqrt{3}}\,S_{12}(\vec{\sigma}_1 , \vec{S}_2 , \hat{r}) & 0 \\
0 & \frac{1}{\sqrt{3}}\,S_{12}(\vec{\sigma}_1 , \vec{S}_2^{\dagger}, \hat{r}) &
0 & S_{12}(\vec{S}_1^{\dagger} , \vec{S}_2^{\dagger} , \hat{r}) \\
\frac{1}{\sqrt{3}}\,S_{12}(\vec{S}_1^{\dagger}, \vec{\sigma}_2, \hat{r}) & 0
& S_{12}(\vec{S}_1^{\dagger} , \vec{S}_2 , \hat{r}) & 0
\end{pmatrix}
 \, , \\
{\bf S}^{S\bar{S}}_{12}(\hat{r}) &=&
\begin{pmatrix}
+\,S_{12}(\vec{\sigma}_1 , \vec{\sigma}_2 ,\hat{r}) &
-\lambda\,S_{12}(\vec{\sigma}_1 , \vec{S}_2 , \hat{r}) &
+\lambda\,S_{12}(\vec{S}_1 , \vec{\sigma}_2 , \hat{r}) &
-\lambda^2\,S_{12}(\vec{S}_1 , \vec{S}_2, \hat{r}) \\
-\lambda\,S_{12}(\vec{\sigma}_1, \vec{S}_2, \hat{r}) &
+\,S_{12}(\vec{\sigma}_1 , \vec{\Sigma}_2, \hat{r}) &
-\lambda^2\,S_{12}(\vec{S}_1 , \vec{S}_2^{\dagger}, \hat{r}) &
+\lambda\,S_{12}(\vec{S}_1 ,\vec{\Sigma}_2, \hat{r})
\\
+\lambda\,S_{12}(\vec{S}_1^{\dagger}, \vec{\sigma}_2 , \hat{r}) &
-\lambda^2\,S_{12}(\vec{S}_1^{\dagger}, \vec{S}_2 , \hat{r}) &
+\,S_{12}(\vec{\Sigma}_1 , \vec{\sigma}_2, \hat{r}) &
-\lambda\,S_{12}(\vec{\Sigma}_1 , \vec{S}_2 , \hat{r}) \\
-\lambda^2\,S_{12}(\vec{S}_1^{\dagger}, \vec{S}_2^{\dagger}, \hat{r}) &
+\lambda\,S_{12}(\vec{S}_1^{\dagger}, \vec{\Sigma}_2, \hat{r}) &
-\lambda\,S_{12}(\vec{\Sigma}_1, \vec{S}_2^{\dagger}, \hat{r}) &
+\,S_{12}(\vec{\Sigma}_1, \vec{\Sigma}_2, \hat{r})
\end{pmatrix}
 \, .
\end{eqnarray}
\end{widetext}

\subsection{The Partial Wave Projection}

Heavy baryon-antibaryon bound states have well defined $J^{P}$ quantum numbers.
Hence we can simplify the OPE potential by projecting it into partial waves
with well-defined parity and angular momentum.
For this we define the states
\begin{eqnarray}
| S \bar{T} (j m) \rangle = \sum_{l m_l s m_s} Y_{l m_l}(\hat{r})\,
| s m_s \rangle \, \langle l m_l s m_s | j m \rangle \, , \nonumber \\ \\
| S \bar{S} (j m) \rangle = \sum_{l m_l s m_s} Y_{l m_l}(\hat{r})\,
| s m_s \rangle \, \langle l m_l s m_s | j m \rangle \, , \nonumber \\
\end{eqnarray}
where $j$, $m$ is the total angular momentum and its third component
for the heavy baryon-antibaryon pair, while $l, m_l$ and $s, m_s$
refer to the angular momentum and spin of the pair.
The product $\langle l m_l s m_s | j m \rangle$ is
the Clebsch-Gordan coefficient for that particular combination of total,
orbital and spin angular momentum (notice that for angular momentum
the Clebsch-Gordan are independent on whether we have particles or
antiparticles).
The spin wave function can be further decomposed as
\begin{eqnarray}
| s m_s \rangle = \sum_{m_1 m_2} | s_1 m_1 \rangle | s_2 m_2 \rangle
\, \langle s_1 m_1 s_2 m_2 | s m_s \rangle \, , \nonumber \\
\end{eqnarray}
with $s_1$ and $s_2$ the spin of the heavy baryon 1 and 2
(either $\tfrac{1}{2}$ or $\tfrac{3}{2}$).

In this basis we can compute the partial wave projection of
${\bf C}_{12}$ and ${\bf S}_{12}$ as
\begin{eqnarray}
\langle (s' l') j' m' | {\bf O}_{12} | (s l) j m \rangle &=& \nonumber \\
\int d^2 \hat{r}
\langle (s' l') j' m' | {\bf O}_{12}(\hat{r}) | (s l) j m \rangle &=&
\nonumber \\
\delta_{j j'}\,\delta_{m m'}\,{\bf O}^j_{s s' l l'} \, ,
&&
\end{eqnarray}
where the total angular momentum and its third component are conserved.
The central force $C_{12}$ conserves in addition
the orbital angular momentum and the spin
\begin{eqnarray}
{\bf C}^j_{s s' l l'} = \delta_{s s'} \delta_{l l'} {\bf C}^j_{l s}\, .
\end{eqnarray}
The tensor force is more involved.
Owing to conservation of parity $|l -l'|$ must be an even number.
The spin transitions are a more complicated because
in the general case $|s -s'|$ can be even or odd.
The exception are the {$B_{\bar{3}}\bar{B}_{\bar{3}}$, $B_6\bar{B}_6$
and $B_6^{*}\bar{B}_6^{*}$} systems:
these systems usually have well defined C- or G-parity, which implies
that $|s -s'|$ is even.
Even when C-/G-parity is not well defined,
like in the $\Xi_c^* \bar{\Sigma}^*_c$ system,
the tensor operator involves identical spin-matrices and is symmetrical
under the exchange of particles $1$ and $2$:
\begin{eqnarray}
S_{12}^{B_{\bar{3}}\bar{B}_{\bar{3}}} &=& S_{12}^{B_{6}\bar{B}_{6}} =
3\,\vec{\sigma}_1 \cdot \hat{r}\,\vec{\sigma}_2 \cdot \hat{r} -
\vec{\sigma}_1 \cdot \vec{\sigma}_2 \, , \\
S_{12}^{B_6^{*}\bar{B}_6^{*}} &=&
3\,\vec{\Sigma}_1 \cdot \hat{r}\,\vec{\Sigma}_2 \cdot \hat{r} -
\vec{\Sigma}_1 \cdot \vec{\Sigma}_2 \, . 
\end{eqnarray}
As a consequence the matrix elements for the tensor operator vanishes
for odd $|s -s'|$.
But if we consider the $B_{\bar{3}}\bar{B}_6^*$
and $B_6 \bar{B}_6^*$ systems (for example
the $\Xi_c \bar{\Xi}_c^*$ and $\Sigma_c' \bar{\Sigma}_c^*$ systems),
it is perfectly possible to have a mix of even and odd spin.
The odd $|s-s'|$ transitions have a particularity that it is worth mentioning:
the matrix element for the tensor operator for odd $|s - s'|$
changes {sign} as follows
\begin{eqnarray}
\langle B \bar{B}_6^* | && {\bf S}^j_{s (s \pm 1) l l'} |
B \bar{B}_6^* \rangle = \nonumber \\
&& - \langle B_6^{*} \bar{B} | {\bf S}^j_{s (s \pm 1) l l'} |
B_6^{*} \bar{B} \rangle \, , \\
\langle B \bar{B}_6^* | && {\bf S}^j_{s (s \pm 1) l l'} |
B_6^{*} \bar{B} \rangle = \nonumber \\
&& - \langle B_6^{*} \bar{B} | {\bf S}^j_{s (s \pm 1) l l'} |
B \bar{B}_6^* \rangle \, , 
\end{eqnarray}
with $B = B_{\bar 3}$ or $B_6$.
In fact the previous two equations indicate that it is actually a good idea
to take explicitly into account the particle coupled channel structure.
Now if we consider the bases
\begin{eqnarray}
  \mathcal{B}' &=& \{ B_{\bar{3}} \bar{B}_6^* , B_6^{*} \bar{B}_{\bar{3}} \} \, , \\
  \mathcal{B} &=& \{ B_6 \bar{B}_6^* , B_6^{*} \bar{B}_6 \} \, ,
\end{eqnarray}
then for $\mathcal{B}'$ we can write the central and tensor operators as
\begin{eqnarray}
  C_{12}' &=&
  \begin{pmatrix}
    0 & \vec{S}_1^{\dagger} \cdot \vec{S}_2 \\
    \vec{S}_1 \cdot \vec{S}_2^{\dagger} & 0 \\
  \end{pmatrix} \, , \\
  S_{12}'(\hat{r}) &=&
  \begin{pmatrix}
    &
    S_{12}(\vec{S}_1^{\dagger}, \vec{S}_2, \hat{r}) \\
    S_{12}(\vec{S}_1, \vec{S}_2^{\dagger},\hat{r}) &
    0 \\
  \end{pmatrix} \, . \nonumber \\
\end{eqnarray}
while for $\mathcal{B}$ they read
\begin{eqnarray}
  C_{12} &=&
  \begin{pmatrix}
    \vec{\sigma}_1 \cdot \vec{\Sigma}_2 & \vec{S}_1^{\dagger} \cdot \vec{S}_2 \\
    \vec{S}_1 \cdot \vec{S}_2^{\dagger} & \vec{\Sigma}_1 \cdot \vec{\sigma}_2 \\
  \end{pmatrix} \, , \\
  S_{12}(\hat{r}) &=&
  \begin{pmatrix}
    S_{12}(\vec{\sigma}_1, \vec{\Sigma}_2,\hat{r}) &
    S_{12}(\vec{S}_1^{\dagger}, \vec{S}_2, \hat{r}) \\
    S_{12}(\vec{S}_1, \vec{S}_2^{\dagger},\hat{r}) &
    S_{12}(\vec{\Sigma}_1, \cdot \vec{\sigma}_2, \hat{r}) \\
  \end{pmatrix} \, . \nonumber \\
\end{eqnarray}
This increases the complexity of the partial wave projection.
However in most cases we can define states with good C- or G-parity,
which effectively amounts to reducing the previous coupled channel
systems to a single channel problem.

The calculation of the matrix elements is straightforward in most cases.
We begin with the $B \bar{B}$ system (with $B= B_{\bar 3}$ or $B_6$),
for which we have the partial waves
\begin{eqnarray}
  | B \bar{B} (0^{-}) \rangle &=& \{ ^1S_0 \} \, , \\
  | B \bar{B} (1^{-}) \rangle &=& \{ ^3S_1, {}^3D_1 \} \, ,
\end{eqnarray}
which lead to the matrix elements
\begin{eqnarray}
  \label{eq:C12_0}
{\bf C}^{B\bar{B}}_{12}(0^{-}) &=& -3 \, , \\
{\bf C}^{B\bar{B}}_{12}(1^{-}) &=&
\begin{pmatrix}
1 & 0 \\
0 & 1
\end{pmatrix} \label{eq:C12_1} \, , \\ \nonumber \\
\label{eq:S12_0}
{\bf S}^{B\bar{B}}_{12}(0^{-}) &=& 0 \, , \\
{\bf S}^{B\bar{B}}_{12}(1^{-}) &=&
\begin{pmatrix}
0 & 2\,\sqrt{2} \\
2\,\sqrt{2}  & -2
\end{pmatrix} \label{eq:S12_1} \, .
\end{eqnarray}
In terms of complexity the next case is the $B^*_6 \bar{B}^*_6$ system,
for which the partial waves are
\begin{eqnarray}
| B_6^* \bar{B}_6^* (0^{-}) \rangle &=& \{ ^1S_0 , {}^5D_0 \}
\, , \\
| B_6^* \bar{B}_6^* (1^{-}) \rangle &=& \{ ^3S_1 , {}^3D_1 , {}^7D_1 ,{}^7G_1 \}
\, , \\
| B_6^* \bar{B}_6^* (2^{-}) \rangle &=& \{ ^1D_2 , {}^5S_2 , {}^5D_2, {}^5G_2 \}
\, , \\
| B_6^* \bar{B}_6^* (3^{-}) \rangle &=&
\{ ^3D_3 , {}^3G_3 , {}^7S_3 ,{}^7D_3 ,{}^7G_3, {}^7I_3 \}
\, , \nonumber \\
\end{eqnarray}
which translate into the following matrices
\begin{eqnarray}
\label{eq:C12_ss_0}
{\bf C}_{12}^{B_6^*\bar{B}_6^*}(0^{-}) &=&
\begin{pmatrix}
-\frac{15}{4} & 0 \\
0 & -\frac{3}{4}
\end{pmatrix} \, , \\
{\bf C}^{B_6^*\bar{B}_6^*}_{12}(1^{-}) &=&
\begin{pmatrix}
  -\frac{11}{4} & 0 & 0 & 0 \\
  0 & -\frac{11}{4} & 0 & 0 \\
  0 & 0 & +\frac{9}{4} & 0 \\
  0 & 0 & 0 & +\frac{9}{4} \\
\end{pmatrix} \, ,  \nonumber \\ \label{eq:C12_ss_1}
\end{eqnarray}
\begin{eqnarray}
{\bf C}_{12}^{B_6^*\bar{B}_6^*}(2^{-}) &=&
\begin{pmatrix}
  -\frac{15}{4} & 0 & 0 & 0 \\
  0 & -\frac{3}{4} & 0 & 0 \\
  0 & 0 & -\frac{3}{4} & 0 \\
  0 & 0 & 0 & -\frac{3}{4} \\
\end{pmatrix} \, , \nonumber \\ \label{eq:C12_ss_2}
\end{eqnarray}
\begin{eqnarray}
{\bf C}_{12}^{B_6^*\bar{B}_6^*}(3^{-}) &=&
\begin{pmatrix}
  -\frac{11}{4} & 0 & 0 & 0 & 0 & 0 \\
  0 & -\frac{11}{4} & 0 & 0 & 0 & 0 \\
  0 & 0 & +\frac{9}{4} & 0 & 0 & 0\\
  0 & 0 & 0 & +\frac{9}{4} & 0 & 0 \\
  0 & 0 & 0 & 0 & +\frac{9}{4} & 0  \\
  0 & 0 & 0 & 0 & 0 & +\frac{9}{4} \\
\end{pmatrix} \, , \nonumber \\ \label{eq:C12_ss_3}
\end{eqnarray}
\begin{eqnarray}
\label{eq:S12_ss_0}
{\bf S}_{12}^{B_6^*\bar{B}_6^*}(0^{-}) &=&
\begin{pmatrix}
0 & -3 \\
-3 & -3
\end{pmatrix} \, , \\
{\bf S}^{B_6^*\bar{B}_6^*}_{12}(1^{-}) &=&
\begin{pmatrix}
0 & \frac{17}{5\,\sqrt{2}} & - \frac{3 \sqrt{7}}{5} & 0 \\
\frac{17}{5\,\sqrt{2}} & - \frac{17}{10} &
\frac{3}{5} \sqrt{\frac{2}{7}} & - \frac{9}{5}\,\sqrt{\frac{6}{7}} \\
- \frac{3 \sqrt{7}}{5} & \frac{3}{5} \sqrt{\frac{2}{7}} &
- \frac{108}{35} & \frac{9 \sqrt{3}}{35} \\
0 & -\frac{9}{5}\sqrt{\frac{6}{7}} & \frac{9 \sqrt{3}}{35} & -\frac{45}{14}
\end{pmatrix} \, ,  \nonumber \\ \label{eq:S12_ss_1}
\end{eqnarray}
\begin{eqnarray}
{\bf S}_{12}^{B_6^*\bar{B}_6^*}(2^{-}) &=&
\begin{pmatrix}
0 & -\frac{3}{\sqrt{5}} & 3\,\sqrt{\frac{2}{7}} & - 9 \sqrt{\frac{2}{35}} \\
-\frac{3}{\sqrt{5}} & 0 & 3\,\sqrt{\frac{7}{10}} & 0 \\
3\,\sqrt{\frac{2}{7}} & 3\,\sqrt{\frac{7}{10}}
& \frac{9}{14} & \frac{18}{7\,\sqrt{5}} \\
- 9\,\sqrt{\frac{2}{35}} & 0 & \frac{18}{7\,\sqrt{5}} & -\frac{15}{7}
\end{pmatrix} \, , \nonumber \\ \label{eq:S12_ss_2}
\end{eqnarray}
\begin{widetext}
\begin{eqnarray}
{\bf S}_{12}^{B_6^*\bar{B}_6^*}(3^{-}) &=&
\begin{pmatrix}
-\frac{17}{35} & \frac{51\,\sqrt{3}}{35} & -\frac{3\sqrt{3}}{5} &
\frac{36}{35} & -\frac{3\sqrt{66}}{35} & 0 \\
\frac{51\sqrt{3}}{35} & - \frac{17}{14} & 0 &
-\frac{3\sqrt{3}}{35} & \frac{9}{7}\sqrt{\frac{2}{11}}
& -3\,\sqrt{\frac{3}{11}} \\
-\frac{3\sqrt{3}}{5} & 0 & 0 & \frac{9 \sqrt{3}}{5} & 0 & 0 \\
\frac{36}{35} & - \frac{3\sqrt{3}}{35} & \frac{9\sqrt{3}}{5} &
\frac{99}{70} & \frac{9 \sqrt{66}}{35} & 0 \\
-\frac{3\sqrt{66}}{35} & \frac{9}{7}\sqrt{\frac{2}{11}} & 0 &
\frac{9\sqrt{66}}{35} & -\frac{27}{77} & \frac{9}{11}\sqrt{\frac{3}{2}} \\
0 & -3\sqrt{\frac{3}{11}} & 0 & 0 & \frac{9}{11}\sqrt{\frac{3}{2}} &
-\frac{63}{22}
\end{pmatrix} \, , \label{eq:S12_ss_3}
\end{eqnarray}
\end{widetext}

We will continue with the $B_{\bar{3}} \bar{B}_6$ system,
where we have the partial waves
\begin{eqnarray}
| B_{\bar{3}} \bar{B}_6 (0^{-}) \rangle &=& \{ ^1S_0 \} \, , \\
| B_{\bar{3}} \bar{B}_6 (1^{-}) \rangle &=& \{ ^3S_1, {}^3D_1 \} \, .
\end{eqnarray}
For these systems OPE is non-diagonal: the $B_{\bar{3}} B_{\bar{3}} \pi$ vertex is zero
and OPE involves the $B_{\bar{3}} \bar{B}_6 \to B_6 \bar{B}_{\bar{3}}$ transition.
In fact there is only one case, the $\Xi_c \Xi_c'$ system,
in which OPE does not cancel.
For this system either C- or G-parity is well-defined,
for which we define the standard states
\begin{eqnarray}
| B_{\bar{3}} \bar{B}_6 (\eta) \rangle = \frac{1}{\sqrt{2}}\,
\left[
| B_{\bar{3}} \bar{B}_6 \rangle + \eta\,| B_6 \bar{B}_{\bar{3}} \rangle
\right] \, , \nonumber \\
\end{eqnarray}
with $C = \eta\,(-1)^{L+S}$.
With this definition in mind, the projection of the central
and tensor operators read
\begin{eqnarray}
  \label{eq:C12_0_st}
{\bf C}^{B_{\bar{3}}\bar{B}_6}_{12}(0^{-}) &=& -3\,\eta \, , \\
{\bf C}^{B_{\bar{3}}\bar{B}_6}_{12}(1^{-}) &=& \eta\,
\begin{pmatrix}
1 & 0 \\
0 & 1
\end{pmatrix} \label{eq:C12_1_st} \, , \\ \nonumber \\
\label{eq:S12_0_st}
{\bf S}^{B_{\bar{3}}\bar{B}_6}_{12}(0^{-}) &=& 0 \, , \\
{\bf S}^{B_{\bar{3}}\bar{B}_6}_{12}(1^{-}) &=&
\eta\,\begin{pmatrix}
0 & 2\,\sqrt{2} \\
2\,\sqrt{2}  & -2
\end{pmatrix} \label{eq:S12_1_st} \, .
\end{eqnarray}

The next case is the $B_{\bar{3}} \bar{B}_6^*$ system.  As happened with the
$B_{\bar{3}} \bar{B}_6$ case involves only non-diagonal OPE transitions,
i.e. $B_{\bar{3}} \bar{B}_6^* \to \bar{B}_{\bar 3} B_6$.
We have the partial waves
\begin{eqnarray}
| B_{\bar{3}} \bar{B}_6^* (1^{-}) \rangle &=& \{ {}^3S_1 , {}^3D_1, {}^5D_1 \}
\, , \\
| B_{\bar{3}} \bar{B}_6^* (2^{-}) \rangle &=& \{ {}^3D_1, {}^5S_2 , {}^5D_2, {}^5G_2 \}
\, .
\end{eqnarray}
If we are considering states with well-defined C-parity
\begin{eqnarray}
| B_{\bar{3}} \bar{B}_6^* (\eta) \rangle = \frac{1}{\sqrt{2}}\,
\left[
| B_{\bar{3}} \bar{B}_6^* \rangle + \eta\,| B_6^{*} \bar{B}_{\bar{3}} \rangle
\right] \, , \nonumber \\
\end{eqnarray}
where $C = \eta\,(-1)^{L+S}$, we find the combinations
\begin{eqnarray}
& | B_{\bar{3}} & \bar{B}_6^* (1^{-+}) \rangle = \nonumber \\
&& \{ {}^3S_1(-) , {}^3D_1(-), {}^5D_1(+) \} \, , \\
& | B_{\bar{3}} & \bar{B}_6^* (2^{-+}) \rangle = \nonumber \\
&& \{ {}^3D_2(-), {}^5S_2(+) , {}^5D_2(+), {}^5G_2(+) \}
\, , \\ \nonumber \\
& | B_{\bar{3}} & \bar{B}_6^* (1^{--}) \rangle = \nonumber \\
&& \{ {}^3S_1(+) , {}^3D_1(+), {}^5D_1(-) \}
\, , \\
& | B_{\bar{3}} & \bar{B}_6^* (2^{--}) \rangle = \nonumber \\
&& \{ {}^3D_2(+), {}^5S_2(-) , {}^5D_2(-), {}^5G_2(-) \}
\, ,
\end{eqnarray}
where the number in parentheses is the value of $\eta = \pm 1$.
That is, there is the complication that each partial wave
in a particular channel can have a different $\eta$.
Concrete calculations yield the following matrices
\begin{eqnarray}
{\bf C}_{12}^{E}(1^{-\pm}) &=& \pm\,
\begin{pmatrix}
-\frac{1}{3} & 0 & 0 \\
0 & -\frac{1}{3} & 0 \\
0 & 0 & 1
\end{pmatrix} \, , \label{eq:C12_E_1C} \\
{\bf C}_{12}^{E}(2^{-\pm}) &=&  \pm\,\begin{pmatrix}
-\frac{1}{3} & 0 & 0 & 0 \\
0 & 1 & 0 & 0 \\
0 & 0 & 1 & 0 \\
0 & 0 & 0 & 1
\end{pmatrix} \, ,\label{eq:C12_E_2C}
\end{eqnarray}
\begin{eqnarray}
\label{eq:S12_E_1C}
{\bf S}_{12}^E(1^{-\pm}) &=&
\pm\,\begin{pmatrix}
0 & \frac{5}{3\sqrt{2}} & \frac{1}{\sqrt{2}} \\
\frac{5}{3\sqrt{2}} & - \frac{5}{6} & \frac{1}{2} \\
\frac{1}{\sqrt{2}} & \frac{1}{2} & \frac{1}{2}
\end{pmatrix} \, , \\
{\bf S}_{12}^E(2^{-\pm}) &=&
\pm\,\begin{pmatrix}
\frac{5}{6} & -\sqrt{\frac{3}{10}} & \frac{1}{2} \sqrt{\frac{3}{7}}
& 2 \sqrt{\frac{3}{35}} \\
-\sqrt{\frac{3}{10}} & 0 & -\sqrt{\frac{7}{10}} & 0 \\
\frac{1}{2} \sqrt{\frac{3}{7}} & -\sqrt{\frac{7}{10}} & -\frac{3}{14} &
-\frac{6}{7\sqrt{5}} \\
2\sqrt{\frac{3}{35}} & 0 & -\frac{6}{7 \sqrt{5}} & \frac{5}{7}
\end{pmatrix} \, . \nonumber \\ \label{eq:S12_E_2C}
\end{eqnarray}
Here we have departed from the previous notation of indicating the channel
with a superindex.
Instead, we use the superindex $^E$ to indicate that
it is exchange or non-diagonal OPE.

Last, the most complex case are the $B_6 \bar{B}_6^*$ system.
The partial wave and C-parity structure is identical to
the $B_{\bar 3} \bar{B}_6^*$ case, but now there is also
a diagonal or direct piece of OPE (besides the non-diagonal piece).
The vertex factors $R_1$ and $\bar{R}_2$ are different
for the direct and exchange pieces,
which has to be taken into account when writing down the potential
\begin{eqnarray}
  V_{\rm OPE} &=& V_{\rm OPE}^D + V_{\rm OPE}^E \nonumber \\
  &=&  R_1^D \bar{R}_2^D \, \vec{I}_1 \cdot \vec{I}_2 \,
  \left[ {\bf C}^D_{12} W_C + {\bf S}^D_{12} W_T\right] \nonumber \\
  &+& R_1^E \bar{R}_2^E \, \vec{I}_1 \cdot \vec{I}_2 \,
  \left[ {\bf C}^E_{12} W_C + {\bf S}^E_{12} W_T\right] \, . \nonumber \\
\end{eqnarray}
If we ignore the fact that the direct and exchange pieces
can have a different effective pion mass $\mu_{\pi}$,
we can merge together the direct and exchange
central and tensor matrices.
For this we notice the relation
\begin{eqnarray}
  R_1^E \bar{R}_2^E = -\frac{3}{4}\,R_1^D \bar{R}_2^D \, .
\end{eqnarray}
This implies that we can reexpress the OPE potential as
\begin{eqnarray}
  V_{\rm OPE} &=& V_{\rm OPE}^D + V_{\rm OPE}^E \nonumber \\
  &=&  R_1^D \bar{R}_2^D \, \vec{I}_1 \cdot \vec{I}_2 \,
  \left[ {\bf C}_{12} W_C + {\bf S}_{12} W_T\right] \, , \nonumber \\
\end{eqnarray}
where the central and tensor matrices ${\bf C}_{12}$ and ${\bf S}_{12}$
are a sum of the direct and exchange pieces
\begin{eqnarray}
  {\bf C}_{12}(J^{PC}) &=&
  {\bf C}^D_{12}(J^{PC}) - \frac{3}{4}\,{\bf C}^E_{12}(J^{PC}) \, , \\
  {\bf S}_{12}(J^{PC}) &=&
  {\bf S}^D_{12}(J^{PC}) - \frac{3}{4}\,{\bf S}^E_{12}(J^{PC}) \, .
\end{eqnarray}
The ${\bf C}^D_{12}$ and ${\bf S}^E_{12}$ are identical to
the ones we discussed in the $B_{\bar 3} \bar{B}_6^*$ case,
see Eqs. (\ref{eq:C12_E_1C}), (\ref{eq:C12_E_2C}), (\ref{eq:S12_E_1C})
and (\ref{eq:S12_E_2C}).
The direct matrices are given by
\begin{eqnarray}
{\bf C}_{12}^{D}(1^{-\pm})
&=& \begin{pmatrix}
-\frac{5}{2} & 0 & 0 \\
0 & -\frac{5}{2} & 0 \\
0 & 0 & \frac{3}{2}
\end{pmatrix}
\, , \\
{\bf C}_{12}^{D}(2^{-\pm}) &=&
\begin{pmatrix}
-\frac{5}{2} & 0 & 0 & 0\\
0 & +\frac{3}{2} & 0 & 0\\
0 & 0 &  +\frac{3}{2} & 0\\
0 & 0 & 0 &  +\frac{3}{2} \\
\end{pmatrix} \, ,
\end{eqnarray}
\begin{eqnarray}
\label{eq:S12_D_1}
{\bf S}_{12}^{D}(1^{-\pm}) &=&
\begin{pmatrix}
0 & - \frac{1}{\sqrt{2}} & \frac{3}{\sqrt{2}} \\
- \frac{1}{\sqrt{2}} & \frac{1}{2} & \frac{3}{2} \\
\frac{3}{\sqrt{2}} & \frac{3}{2} & - \frac{3}{2}
\end{pmatrix} \, ,
\end{eqnarray}
\begin{eqnarray}
\label{eq:S12_D_2}
{\bf S}_{12}^{D}(2^{-\pm}) &=&
\begin{pmatrix}
-\frac{1}{2} & - 3\,\sqrt{\frac{3}{10}} & \frac{3}{2}\,\sqrt{\frac{3}{7}} &
6 \sqrt{\frac{3}{35}} \\
-3\,\sqrt{\frac{3}{10}} & 0 & 3\,\sqrt{\frac{7}{10}} & 0 \\
\frac{3}{2}\,\sqrt{\frac{3}{7}} & 3\,\sqrt{\frac{7}{10}} &
\frac{9}{14} & \frac{18}{7\sqrt{5}} \\
6\,\sqrt{\frac{3}{35}} & 0 & \frac{18}{7\sqrt{5}} & -\frac{15}{7}
\end{pmatrix} \, . \nonumber \\
\end{eqnarray}

Now there is a $B_6 \bar{B}_6^*$ system for which good G-parity states
cannot be defined, which is
\begin{eqnarray}
 \mathcal{B}'' = \{  \Xi_c' \bar{\Sigma}^* , \Xi_c^* \bar{\Sigma}_c \}  \, .
\end{eqnarray}
This system is more involved than usual because we have to consider
the two particle channels $\alpha = \Xi_c \bar{\Sigma}_c^*$ and
$\beta = \Xi_c^* \bar{\Sigma}_c$ separately.
We can define the central and tensor matrices in the particle basis as
\begin{eqnarray}
  {\bf C}_{12} =
  \begin{pmatrix}
    {\bf C}_{12}^{\alpha \alpha} & {\bf C}_{12}^{\alpha \beta} \\
    {\bf C}_{12}^{\beta \alpha} & {\bf C}_{12}^{\beta \beta}
  \end{pmatrix}
  \,\, \mbox{and} \,\,\,
     {\bf S}_{12} =
  \begin{pmatrix}
    {\bf S}_{12}^{\alpha \alpha} & {\bf S}_{12}^{\alpha \beta} \\
    {\bf S}_{12}^{\beta \alpha} & {\bf S}_{12}^{\beta \beta}
  \end{pmatrix} \, . \nonumber \\
\end{eqnarray}
In this particle basis, the diagonal central matrices are
\begin{eqnarray}
  {\bf C}_{12}^{\alpha \alpha}(1^{-}) =
  {\bf C}_{12}^{\beta \beta}(1^{-}) = {\bf C}_{12}^{D}(1^{-\pm}) \, , \\
  {\bf C}_{12}^{\alpha \alpha}(2^{-}) =
  {\bf C}_{12}^{\beta \beta}(2^{-}) = {\bf C}_{12}^{D}(2^{-\pm}) \, ,
\end{eqnarray}
while the non-diagonal are given by
\begin{eqnarray}
  {\bf C}_{12}^{\alpha \beta}(1^{-}) = {\bf C}_{12}^{\beta \alpha}(1^{-}) =
  \begin{pmatrix}
    \frac{1}{3} & 0 & 0 \\
    0 & \frac{1}{3} & 0 \\
    0 & 0 & 1
  \end{pmatrix}
  \, , \label{eq:C12_ab_1} \\
  \nonumber \\
  {\bf C}_{12}^{\alpha \beta}(2^{-}) = {\bf C}_{12}^{\beta \alpha}(2^{-}) =
  \begin{pmatrix}
    \frac{1}{3} & 0 & 0 & 0 \\
    0 & 1 & 0 & 0 \\
    0 & 0 & 1 & 0 \\
    0 & 0 & 0 & 1
  \end{pmatrix} \, . \label{eq:C12_ab_2} \\
\end{eqnarray}
For the tensor matrices we have
\begin{eqnarray}
\label{eq:S12_D_1_aa}
{\bf S}_{12}^{\alpha \alpha}(1^{-}) &=&
\begin{pmatrix}
0 & - \frac{1}{\sqrt{2}} & \frac{3}{\sqrt{2}} \\
- \frac{1}{\sqrt{2}} & \frac{1}{2} & \frac{3}{2} \\
\frac{3}{\sqrt{2}} & \frac{3}{2} & - \frac{3}{2}
\end{pmatrix} \, , \\
\label{eq:S12_E_1_ab}
{\bf S}_{12}^{\alpha \beta}(1^{-}) &=&
\begin{pmatrix}
0 & - \frac{5}{3\sqrt{2}} & \frac{1}{\sqrt{2}} \\
-\frac{5}{3\sqrt{2}} & \frac{5}{6} & \frac{1}{2} \\
-\frac{1}{\sqrt{2}} & - \frac{1}{2} & \frac{1}{2}
\end{pmatrix} \, ,  \\
{\bf S}_{12}^{\beta \alpha}(1^{-}) &=&
\begin{pmatrix}
0 & - \frac{5}{3\sqrt{2}} & -\frac{1}{\sqrt{2}} \\
-\frac{5}{3\sqrt{2}} & \frac{5}{6} & -\frac{1}{2} \\
\frac{1}{\sqrt{2}} & \frac{1}{2} & \frac{1}{2}
\end{pmatrix} \, , \\
\label{eq:S12_E_1p}
{\bf S}_{12}^{\beta \beta}(1^{-}) &=&
\begin{pmatrix}
0 & - \frac{1}{\sqrt{2}} & -\frac{3}{\sqrt{2}} \\
- \frac{1}{\sqrt{2}} & \frac{1}{2} & -\frac{3}{2} \\
-\frac{3}{\sqrt{2}} & -\frac{3}{2} & - \frac{3}{2}
\end{pmatrix} \, ,
\end{eqnarray}
\begin{eqnarray}
\label{eq:S12_D_2_aa}
{\bf S}_{12}^{\alpha \alpha}(2^{-}) &=&
\begin{pmatrix}
-\frac{1}{2} & - 3\,\sqrt{\frac{3}{10}} & \frac{3}{2}\,\sqrt{\frac{3}{7}} &
6 \sqrt{\frac{3}{35}} \\
-3\,\sqrt{\frac{3}{10}} & 0 & 3\,\sqrt{\frac{7}{10}} & 0 \\
\frac{3}{2}\,\sqrt{\frac{3}{7}} & 3\,\sqrt{\frac{7}{10}} &
\frac{9}{14} & \frac{18}{7\sqrt{5}} \\
6\,\sqrt{\frac{3}{35}} & 0 & \frac{18}{7\sqrt{5}} & -\frac{15}{7}
\end{pmatrix} \, , \nonumber \\ \\
{\bf S}_{12}^{\alpha \beta}(2^{-}) &=&
\begin{pmatrix}
-\frac{5}{6} & -\sqrt{\frac{3}{10}} & \frac{1}{2}\sqrt{\frac{3}{7}} &
2 \sqrt{\frac{3}{35}} \\
\sqrt{\frac{3}{10}} & 0 & - \sqrt{\frac{7}{10}} & 0 \\
-\frac{1}{2}\sqrt{\frac{3}{7}} & -\sqrt{\frac{7}{10}} & - \frac{3}{14}
& - \frac{6}{7 \sqrt{5}} \\
-2\sqrt{\frac{3}{35}} & 0 & - \frac{6}{7\sqrt{5}} & \frac{5}{7}
\end{pmatrix} \, , \label{eq:S12_E_2_ab} \nonumber \\
\end{eqnarray}
\begin{eqnarray}
  {\bf S}_{12}^{\beta \alpha}(2^{-}) &=&
\begin{pmatrix}
-\frac{5}{6} & \sqrt{\frac{3}{10}} & -\frac{1}{2}\sqrt{\frac{3}{7}} &
-2 \sqrt{\frac{3}{35}} \\
-\sqrt{\frac{3}{10}} & 0 & - \sqrt{\frac{7}{10}} & 0 \\
\frac{1}{2}\sqrt{\frac{3}{7}} & -\sqrt{\frac{7}{10}} & - \frac{3}{14}
& - \frac{6}{7 \sqrt{5}} \\
2\sqrt{\frac{3}{35}} & 0 & - \frac{6}{7\sqrt{5}} & \frac{5}{7}
\end{pmatrix} \, . \nonumber \\ \label{eq:S12_E_2p}
  \\
{\bf S}_{12}^{\beta \beta}(2^{-}) &=&
\begin{pmatrix}
-\frac{1}{2} & 3\,\sqrt{\frac{3}{10}} & -\frac{3}{2}\,\sqrt{\frac{3}{7}} &
-6 \sqrt{\frac{3}{35}} \\
3\,\sqrt{\frac{3}{10}} & 0 & 3\,\sqrt{\frac{7}{10}} & 0 \\
-\frac{3}{2}\,\sqrt{\frac{3}{7}} & 3\,\sqrt{\frac{7}{10}} &
\frac{9}{14} & \frac{18}{7\sqrt{5}} \\
-6\,\sqrt{\frac{3}{35}} & 0 & \frac{18}{7\sqrt{5}} & -\frac{15}{7}
\end{pmatrix} \, . \nonumber \\ \label{eq:S12_bb_2}
\end{eqnarray}

\section{The One Pseudoscalar Meson Exchange Potential
  in Heavy Hadron Chiral Perturbation Theory}
\label{app:OME}

\subsection{General Form of the Potential}

The potential generated from the exchange of the kaon and the eta
can be derived from the same rules we have used to calculate
the OPE potential.
We will not present a complete derivation here, but simply a quick overview.
The one pseudoscalar meson exchange (OME) potentials are formally
identical to the OPE potential. For a general
pseudo Nambu-Goldstone boson $P$ we can write
\begin{eqnarray}
V_{\rm OME}^{T\bar{S}}  &=&
-R^P_1\,\bar{R}^P_2\,\frac{g_3^2}{2 f_P^2} F_1^P F_2^P
\frac{\vec{a}_1 \cdot \vec{q}\, \vec{a}_2 \cdot \vec{q}}{q^2 + \mu_{P}^2} \, ,
\label{eq:V-OME-TS}
\\
V_{\rm OME}^{S\bar{S}} &=&
-R_1^P\,\bar{R}^P_2\,\frac{g_2^2}{2 f_P^2} F_1^P F_2^P
\frac{\vec{a}_1 \cdot \vec{q}\, \vec{a}_2 \cdot \vec{q}}{q^2 + \mu_{P}^2} \, ,
\label{eq:V-OME-SS}
\end{eqnarray}
with $f_{P}$ the weak decay constant for the meson $P$,
$R_1^P$, $\bar{R}_2^P$ numerical factors that depend on the meson $P$,
$F_1^P$ and $F_2^P$ flavour factors (which for $P=\pi$ are
the isospin factors we have previously used for OPE),
$\vec{a}_1$ and $\vec{a}_1$ spin operators and $\mu_P$ the effective mass of
the meson $P$, which is $\mu_P^2 = m_P^2 - \Delta^2$,
where $m_P$ is the mass of the meson.
Notice the analogy between Eqs.~(\ref{eq:V-OME-TS}) and (\ref{eq:V-OME-SS})
and Eqs.~(\ref{eq:V-OPE-TS}) and (\ref{eq:V-OPE-SS})
of Appendix \ref{app:ope}.

The numerical, flavour and isospin factors of the one eta exchange potential
are listed in Table \ref{tab:vertices-eta} for all vertices
in which eta exchange is allowed.
The format is similar to Table \ref{tab:vertices}, which was dedicated to OPE.
There is a difference worth mentioning: the G-parity of the $\eta$ meson
is positive (in contrast to the negative G-parity of the pion),
which implies that the signs of the numerical factors
$R_i^{\eta}$ and $\bar{R}_i^{\eta}$ is not the same as
for the pion factors $R_i$ and $\bar{R}_i$ of Table \ref{tab:vertices}.
In particular we have that
\begin{eqnarray}
  R_i^{\eta} \bar{R}_i^{\eta} > 0 \quad \Rightarrow \quad R_i \bar{R}_i < 0 \, ,
\end{eqnarray}
and vice versa.
Besides this, it is easy to see that the strength of the flavour factors
$F_1^{\eta} F_2^{\eta}$ is in general much weaker than
the corresponding isospin factors for OPE.
For example, in the $\Sigma_c \bar{\Sigma}_c$ system we have that 
$F_1^{\eta} F_2^{\eta} = 1/3$ for eta exchange in contrast to
$\vec{T}_1 \cdot \vec{T}_2 = -2$, $-1$ and $1$ for $I=0,1,2$
with the OPE potential.

For the one kaon exchange potential we list the different factors
in Table \ref{tab:vertices-kaon} for half
the non-vanishing vertices.
We write the vertices in order of decreasing strangeness
for the heavy baryons, with a final kaon.
The vertices with an initial antikaon are indeed identical:
\begin{eqnarray}
  \mathcal{A}(B \to B' K) =
  \mathcal{A}({\bar K}\,B \to B') \, ,
\end{eqnarray}
where $B$ and $B'$ denote the initial and final heavy baryons.
The flavour factors in Table \ref{tab:vertices-kaon} are listed as numbers,
or as the symbol $\theta^a_i$, which is a matrix in isospin space.
When the flavour factor is a number, it is implicitly understood that they
are multiplied by the identity in isospin space (all vertices conserve isospin).
The matrices $\theta^a_i$, where $a = \pm \frac{1}{2}$ refers to the isospin
state of the kaon, mediate the transitions from isospin-$1$
to -$\tfrac{1}{2}$ baryons (e.g. $\Sigma_c \to \Xi_c' K$).
Their explicit expressions are
\begin{eqnarray}
  \theta_i^{+{1}/{2}} =
  \begin{pmatrix}
    1 & 0 & 0 \\
    0 & \frac{1}{\sqrt{2}} & 0 \\
  \end{pmatrix}
  \,\, , \,\,
  \theta_i^{-{1}/{2}} =
  \begin{pmatrix}
    0 & \frac{1}{\sqrt{2}} & 0 \\
    0 & 0 & 1 \\
  \end{pmatrix} \, .
\end{eqnarray}
The evaluation of the product of two $\theta^a_i$ matrices
can be done in terms of isospin matrices
\begin{eqnarray}
  \sum_a \theta_1^{a \dagger} \theta_2^a =
  \frac{1}{2} [ \vec{\tau}_1 \cdot \vec{T}_2 + 1 ] \, ,
\end{eqnarray}
where $\vec{\tau}_1$ and $\vec{T}_2$ refer to the isospin matrices
as evaluated for the initial or final state.

\begin{table}
\begin{center}
\begin{tabular}{|c|c|c|c|c|c|}
\hline \hline
Vertex & $P$ & $R^P_i$ & $\bar{R}^P_i$ & $\vec{I}_i$ & $\vec{a}_i$ \\
\hline
$\Xi_c \to \Xi_c'$ & $\eta$ & $\sqrt{\frac{2}{3}}$ & $\sqrt{\frac{2}{3}}$ &
$\frac{1}{2}\sqrt{3}$ & $\vec{\sigma}_i$ \\
$\Xi_c \to \Xi_c^*$ & $\eta$ & $\sqrt{2}$ & -$\sqrt{2}$ &
$\frac{1}{2}\sqrt{3}$ & $\vec{S}_i^{\dagger}$ \\
\hline
$\Sigma_c \to \Sigma_c$ & $\eta$
& $\frac{{2}}{3}$ & $\frac{{2}}{3}$
& $\frac{1}{\sqrt{3}}$ & $\vec{\sigma}_i$ \\
$\Sigma_c^* \to \Sigma_c$ & $\eta$
& $\frac{1}{\sqrt{3}}$ & -$\frac{1}{\sqrt{3}}$
& $\frac{1}{\sqrt{3}}$ & $\vec{S}_i$ \\
$\Sigma_c \to \Sigma_c^*$ & $\eta$
& $\frac{1}{\sqrt{3}}$ & -$\frac{1}{\sqrt{3}}$
& $\frac{1}{\sqrt{3}}$ & $\vec{S}_i^{\dagger}$ \\
$\Sigma_c^* \to \Sigma_c^*$ & $\eta$
& $\frac{{2}}{3}$ & $\frac{{2}}{3}$
& $\frac{1}{\sqrt{3}}$ & $\vec{\Sigma}_i$ \\
\hline
$\Xi_c' \to \Xi_c'$ & $\eta$
& $\frac{{2}}{3}$ & $\frac{{2}}{3}$
& $\frac{1}{2\,\sqrt{3}}$ & $\vec{\sigma}_i$ \\
$\Xi_c^* \to \Xi_c' $ & $\eta$
& $\frac{1}{\sqrt{3}}$ & -$\frac{1}{\sqrt{3}}$
& $\frac{1}{2\,\sqrt{3}}$ & $\vec{S}_i$ \\
$\Xi_c' \to \Xi_c^* $ & $\eta$
& $\frac{1}{\sqrt{3}}$ & -$\frac{1}{\sqrt{3}}$
& $\frac{1}{2\,\sqrt{3}}$ & $\vec{S}_i^{\dagger}$ \\
$\Xi_c^* \to \Xi_c^* $ & $\eta$
& $\frac{{2}}{3}$ & $\frac{{2}}{3}$
& $\frac{1}{2\,\sqrt{3}}$ & $\vec{\Sigma}_i$ \\
\hline
$\Omega_c \to \Omega_c$ & $\eta$
& $\frac{{2}}{3}$ & $\frac{{2}}{3}$
& $\frac{2}{\sqrt{3}}$ & $\vec{\sigma}_i$ \\
$\Omega_c^* \to \Omega_c $ & $\eta$
& $\frac{1}{\sqrt{3}}$ & -$\frac{1}{\sqrt{3}}$
& $\frac{2}{\sqrt{3}}$ & $\vec{S}_i$ \\
$\Omega_c \to \Omega_c^* $ & $\eta$
& $\frac{1}{\sqrt{3}}$ & -$\frac{1}{\sqrt{3}}$
& $\frac{2}{\sqrt{3}}$ & $\vec{S}_i^{\dagger}$ \\
$\Omega_c^* \to \Omega_c^* $ & $\eta$
& $\frac{{2}}{3}$ & $\frac{{2}}{3}$
& $\frac{2}{\sqrt{3}}$ & $\vec{\Sigma}_i$ \\
\hline \hline
\end{tabular}
\end{center}
\caption{
Numerical, isospin and spin factor associated with each vertex
of the one eta exchange potential for the $S\bar{T}$ and $S\bar{S}$ systems.
The arrows are used to indicate the final baryon state in the vertex
and $P$ indicates the pseudo Goldstone meson in that vertex.
The numerical factors $R_i^P$ and $\bar{R}_i^P$ are for baryons
and antibaryons respectively, while $F_i^P$ are
the flavour factors.
$\vec{\sigma}_i$ are the Pauli matrices, $\vec{\Sigma}_i$ are the 
spin $S = 3/2$ matrices and $\vec{S}_i$ are the $2 \times 4$ matrices
that are used for the transitions from spin-$1/2$ to spin-$3/2$ baryon.
} \label{tab:vertices-eta}
\end{table}

\begin{table}
\begin{center}
\begin{tabular}{|c|c|c|c|c|c|}
\hline \hline
Vertex & $P$ & $R^P_i$ & $\bar{R}^P_i$ & $F^P_i$ & $\vec{a}_i$ \\
\hline
$\Lambda_c \to \Xi_c'$ & $K$ & $\sqrt{\frac{2}{3}}$ & $\sqrt{\frac{2}{3}}$ &
$1$ & $\vec{\sigma}_i$ \\
$\Lambda_c \to \Xi_c^*$ & $K$ & $\sqrt{2}$ & -$\sqrt{2}$ &
$1$ & $\vec{S}_i^{\dagger}$ \\
$\Sigma_c \to \Xi_c$ & $K$ & $\sqrt{\frac{2}{3}}$ & $\sqrt{\frac{2}{3}}$ &
$-\theta_i$ & $\vec{\sigma}_i$ \\
$\Sigma_c^* \to \Xi_c$ & $K$ & $\sqrt{2}$ & -$\sqrt{2}$ &
$-\theta_i$ & $\vec{S}_i$ \\
$\Xi_c \to \Omega_c$ & $K$ & $\sqrt{\frac{2}{3}}$ & $\sqrt{\frac{2}{3}}$ &
$1$ & $\vec{\sigma}_i$ \\
$\Xi_c \to \Omega_c$ & $K$ & $\sqrt{2}$ & -$\sqrt{2}$ &
$1$ & $\vec{S}_i^{\dagger}$ \\
\hline
$\Sigma_c \to \Xi_c'$ & $K$
& $\frac{{2}}{3}$ & $\frac{{2}}{3}$
& $\theta_i$ & $\vec{\sigma}_i$ \\
$\Sigma_c^* \to \Xi_c' $ & $K$
& $\frac{1}{\sqrt{3}}$ & -$\frac{1}{\sqrt{3}}$
& $\theta_i$ & $\vec{S}_i$ \\
$\Sigma_c \to \Xi_c^* $ & $K$
& $\frac{1}{\sqrt{3}}$ & -$\frac{1}{\sqrt{3}}$
& $\theta_i$ & $\vec{S}_i^{\dagger}$ \\
$\Sigma_c^* \to \Xi_c^* $ & $K$
& $\frac{{2}}{3}$ & $\frac{{2}}{3}$
& $\theta_i$ & $\vec{\Sigma}_i$ \\
\hline 
$\Xi_c' \to \Omega_c$ & $K$
& $\frac{{2}}{3}$ & $\frac{{2}}{3}$
& $1$ & $\vec{\sigma}_i$ \\
$\Xi_c^* \to \Omega_c $ & $K$
& $\frac{1}{\sqrt{3}}$ & -$\frac{1}{\sqrt{3}}$
& $1$ & $\vec{S}_i$ \\
$\Xi_c' \to \Omega_c^* $ & $K$
& $\frac{1}{\sqrt{3}}$ & -$\frac{1}{\sqrt{3}}$
& $1$ & $\vec{S}_i^{\dagger}$ \\
$\Xi_c^* \to \Omega_c^* $ & $K$
& $\frac{{2}}{3}$ & $\frac{{2}}{3}$
& $1$ & $\vec{\Sigma}_i$ \\
\hline \hline
\end{tabular}
\end{center}
\caption{
Numerical, isospin and spin factor associated with each vertex
of the one kaon exchagne potential for the $S\bar{T}$ and $S\bar{S}$ systems.
The arrows are used to indicate the final baryon state in the vertex
and $P$ indicates the pseudo Goldstone meson in that vertex.
The numerical and spin factors --- $R_i^P$, $\bar{R}_i^P$ and $\vec{\sigma}_i$,
$\vec{\Sigma}_i$, $\vec{S}_i$ --- are as in Table \ref{tab:vertices-eta}.
The flavour factors $F_i^P$ are is most cases a number (implicitly multiplied
by the identity in isospin space to guarantee isospin conservarion).
The exception are the $\Sigma_c \to \Xi_c$ family of transitions
in which the flavour factors are $2 \times 3$ matrices $\theta_i^a$,
with $a$ denoting which of the two isospin states of
the kaon have been exchanged.
}
\label{tab:vertices-kaon}.
\end{table}

\subsection{Strenth of the Eta and Kaon Exchange Potentials}

The strength of eta and kaon exchange can be compared to that of OPE
using the techniques of Sect.~\ref{sec:perturbative}.
The central scale $\Lambda_C^P$ for the OME potential
can be defined as follows
\begin{eqnarray}
\Lambda_C^P =
\frac{1}{| \sigma \, F_1^P F_2^P |}\,
\frac{24 \pi f_{P}^2}{\mu |R^P_1 \bar{R}^P_2|\, g_i^2}  \, ,
\end{eqnarray}
which is analogous to Eq.~(\ref{eq:central-scale}), expect for the changes
\begin{eqnarray}
  f_{\pi} \to f_P \quad \mbox{and} \quad \tau \to F_1^P F_2^P \, ,
\end{eqnarray}
while $|R_1 \bar{R}_2| = |R^P_1 \bar{R}^P_2|$.
The comparison is in fact direct if we write it as
\begin{eqnarray}
  \Lambda_C^P = \left( \frac{| \tau |}{| F_1^P F_2^P | } \right) \times
  \left( \frac{f_{P}^2}{f_{\pi}^2} \right) \times \Lambda_C \, ,
\end{eqnarray}
which merely involves the evaluation of a few numerical factors.
For the eta meson we have
\begin{eqnarray}
  \frac{| \tau |}{| F_1^{\eta} F_2^\eta | } = 3,6,9 \quad \mbox{and} \quad
  \frac{f_{\eta}^2}{f_\pi^2} \sim 1.5 \, ,
\end{eqnarray}
where we have taken $f_{\eta} \simeq f_{K} \sim 160\,{\rm MeV}$.
By multiplying these factors it is apparent that $\Lambda_C^{\eta}$
is considerably harder than $\Lambda_C$.
The conclusion is that one eta exchange is very suppressed with respect to OPE.
For one kaon exchange the factors are
\begin{eqnarray}
  \frac{| \tau |}{| F_1^{K} F_2^K | } = 1,2 \quad \mbox{and} \quad
  \frac{f_K^2}{f_{\pi}^2} \sim 1.5 \, ,
\end{eqnarray}
which are not particularly large.
The flavour factor suppresion is larger for the lower isospin states,
for which the OPE is stronger.
The outcome is that the one kaon exchange potential is perturbative.

The comparison of the tensor scales $\Lambda_T^P$ and $\Lambda_T$ is similar
except for the existence of factors of the type $e^{m_P R_c}$ and $e^{m_{\pi} R_c}$
that are included to take into account that the OME and OPE potentials
cease to be valid below a certain distance,
see of Eq.~(\ref{eq:kappa-mpi}).
If we add these factors, we end up with
\begin{eqnarray}
  \Lambda_T^P(m_P) = \left( \frac{| \tau |}{| F_1^P F_2^P | } \right) \times
  \left( \frac{f_{P}^2}{f_{\pi}^2} \right) \times
  \left( \frac{e^{m_P R_c}}{e^{m_{\pi} R_c}} \right) \times \Lambda_T(m_p) \, .
  \nonumber \\
\end{eqnarray}
where $e^{(m_P - m_{\pi}) R_c} \sim 3-5$ and $2.5-4$ for the eta and kaon,
respectively.
The addition of this factor means that the tensor scale $\Lambda_T^P$ is
in general considerably larger than the one for OPE.
In particular it can be regarded as a hard scale.

\section{The Contact-Range Potential}
\label{app:contact}

The calculation of the contact-range potential will use a different
set of techniques that the one of the OPE potential.
While we derived the OPE potential from the lowest-dimensional Lagrangian
compatible with HQSS and chiral symmetry, for the contact-range potential
we will use HQSS without any explicit reference to a Lagrangian.
We will {first take} into account that the lowest order contact-range
potential is simply a constant in momentum space
\begin{eqnarray}
\langle \vec{p'} | V_C^{(0)} | \vec{p} \rangle = C \, ,
\end{eqnarray}
where $\vec{p}$ and $\vec{p}'$ are the initial and final center-of-mass momentum
of the heavy baryon-antibaryon pair and $C$ a coupling.
In principle the coupling $C$ depends on specific baryon-antibaryon system
and its quantum numbers. But HQSS precludes precludes $C$ to depend
on the heavy quark spin, which will translate into a reduction of
the number of couplings.
We will explain in the following lines how to do that.

Heavy baryons are $| Q q q \rangle$ states with the structure
\begin{eqnarray}
B_{\bar{3}} &=& | Q \, (q q)_{s_L = 0} \rangle_{j=\frac{1}{2}} \, , \\
B_{6} &=& | Q \, (q q)_{s_L = 1} \rangle_{j=\frac{1}{2}} \, , \\
B^*_{6} &=& | Q \, (q q)_{s_L = 1} \rangle_{j=\frac{3}{2}} \, , 
\end{eqnarray}
with $s_L = 0,1$ the light-quark pair spin and with $j$ the total spi
The application of HQSS to the contact-range couplings $C$ implies
that they depend on the total light spin
$\vec{S}_L = \vec{s}_{L1} + \vec{s}_{L2}$,
but not on the total spin $\vec{J} = \vec{j}_1 + \vec{j}_2$
or the total heavy spin $\vec{S}_H = \vec{s}_{H1} + \vec{s}_{H2}$.
To determine the exact structure we have to study the coupling of
the light-quark spin for different types of heavy baryon-antibaryon molecules.
In the following lines we will explain how to do this
for the $B_6^{(*)}  \bar{B}_6^{(*)}$, $B_{\bar 3}  \bar{B}_6^{(*)}$ systems,
ordered by decreasing degree of complexity.
We did not list the system
\begin{eqnarray}
  B_{\bar 3}  \bar{B}_{\bar 3} \quad \Rightarrow \quad S_L = 0 \, ,
\end{eqnarray}
for which the light spin structure is trivial because $s_{L1} = s_{L2} = 0$.

\subsection{The $S\bar{S}$ Contact Potential}

For a heavy $S\bar{S}$ baryon-antibaryon system (i.e. $S_L=1$ for both baryons)
we can decompose the spin wave function into heavy and light
components as follows
\begin{eqnarray}
  \label{eq:HL-decomposition}
| B_6^{(*)} \bar{B}_6^{(*)} (J^{-}) \rangle =
\sum_{S_H, S_L} D_{S_H, S_L}(J)\,S_H \otimes S_L \Big|_{J} \, ,
\end{eqnarray}
where $D$ are the coefficients for this change of basis.
They fulfill the condition
\begin{eqnarray}
\sum_{S_H, S_L} | D_{S_H, S_L}(J) |^2 = 1 \, .
\end{eqnarray}
From this decomposition we can calculate the light-spin components of
the contact-range potential:
\begin{eqnarray}
\label{eq:V_C_HQSS}
\langle S_H' \otimes S_L' | V | S_H \otimes S_L \rangle =
\delta_{S_H S_H'} \delta_{S_L S_L'}\, V_{S_L} \, ,
\end{eqnarray}
where the light and heavy spin decouple.

The general way to carry on the heavy-light spin decomposition
is to consider the spin wave function of the heavy hadrons
\begin{eqnarray}
| H_1 \rangle = | s_{H_1} s_{L_1} j_1 \rangle \quad , \quad
| H_2 \rangle = | s_{H_2} s_{L_2} j_2 \rangle \, ,
\end{eqnarray}
where $s_{H_1}$, $s_{H_2}$ is the heavy spin, $s_{L_1}$, $s_{L_2}$ the light spin
and $j_1$, $j_2$ the angular momenta of the two hadrons.
When we couple the two hadrons together we have
\begin{eqnarray}
&& | H_1 H_2 \rangle = | s_{H_1} s_{L_1} j_1 \rangle \, | s_{H_2} s_{L_2} j_2 \rangle
\nonumber \\
&& = \sum_{S_H, S_L} D_{S_H, S_L}(J)\,
| (s_{H_1} s_{H_2}) S_H (s_{L_1} s_{L_2}) S_L (j_1 j_2) J \rangle
\, , \nonumber \\
\end{eqnarray}
which is merely a detailed version of Eq.~(\ref{eq:HL-decomposition}),
where the notation indicates that the heavy spins coupled to $S_H$,
the light spins to $S_L$ and the angular momenta to $J$.
The coefficients $D_{S_H, S_L}(J)$ can in fact be expressed
in terms of 9-J symbols
\begin{widetext}
\begin{eqnarray}
D_{S_H, S_L}(J) &=&
\langle s_{H_1} s_{L_1} j_1 \, s_{H_2} s_{L_2} j_2 |
| (s_{H_1} s_{H_2}) S_H (s_{L_1} s_{L_2}) S_L (j_1 j_2) J \rangle \nonumber \\
&=&
\sqrt{(2 j_1 + 1) (2 j_2 + 1) (2 S_H + 1) (2 S_L + 1)}
\begin{Bmatrix}
s_{H1} &  s_{L_1} & j_1 \\
s_{H2} &  s_{L_2} & j_2 \\
S_{H} &  S_{L} & J
\end{Bmatrix} \, .
\end{eqnarray}
\end{widetext}
Finally if we are considering antihadrons, we should consider their behaviors
under C-parity to define their spin wave functions consistently:
they might differ by a sign from the ansatz
$| s_{H} s_{L} j \rangle$.

If we go back to the $S\bar{S}$ heavy baryon-antibaryon system,
for the $B_6\bar{B}_6$ case we find the following
\begin{eqnarray}
| B_6 \bar{B}_6 (0^{-}) \rangle &=&
\frac{1}{\sqrt{3}}\,0_H \otimes 0_L +
\sqrt{\frac{2}{3}}\,\,1_H \otimes 1_L \Big|_{J=0} \, , \nonumber \\ \\
| B_6 \bar{B}_6 (1^{-}) \rangle &=& \frac{\sqrt{2}}{3}\,0_H \otimes 1_L -
\frac{1}{3\sqrt{3}}\,1_H \otimes 0_L \nonumber \\ &+&
\frac{2}{3}\,\sqrt{\frac{5}{3}}\, 1_H \otimes 2_L \Big|_{J=1} \, ,
\end{eqnarray}
For the $B_6\bar{B}_6^*$ and $B_6^*\bar{B}_6$ cases, we include a minus sign
{in} front of the states containing a $\bar{B}_6^*$ to highlight
the C-parity convention that we employ here:
\begin{eqnarray}
- | B_6 \bar{B}_6^* (1^{-}) \rangle &=& +\frac{1}{3}\, 0_H \otimes 1_L
- \frac{2}{3}\sqrt{\frac{2}{3}}\,1_H \otimes 0_L \nonumber \\
&+& \frac{1}{\sqrt{2}}\,1_H \otimes 1_L \Big|_{J=1}
- \frac{1}{3}\sqrt{\frac{5}{6}}\,1_H \otimes 2_L \Big |_{J=1} \, ,
\nonumber \\ \\
+ | B_6^* \bar{B}_6 (1^{-}) \rangle &=& -\frac{1}{3}\,0_H \otimes 1_L
+ \frac{2}{3}\sqrt{\frac{2}{3}}\,1_H \otimes 0_L \nonumber \\
&+& \frac{1}{\sqrt{2}}\,1_H \otimes 1_L \Big|_{J=1}
+ \frac{1}{3}\sqrt{\frac{5}{6}}\,1_H \otimes 2_L \Big |_{J=1} \, ,
\nonumber \\
\end{eqnarray}
\begin{eqnarray}
\nonumber \\
- | B_6 \bar{B}_6^* (2^{-}) \rangle &=& +\frac{1}{\sqrt{3}}\,0_H \otimes 2_L
- \frac{1}{\sqrt{6}}\,1_H \otimes 1_L \Big|_{J=2} \nonumber \\
&+& \frac{1}{\sqrt{2}}\,1_H \otimes 2_L \Big|_{J=2} \, , \\
+ | B_6^* \bar{B}_6 (2^{-}) \rangle &=& -\frac{1}{\sqrt{3}}\,0_H \otimes 2_L
+ \frac{1}{\sqrt{6}}\,1_H \otimes 1_L \Big|_{J=2} \nonumber \\
&+& \frac{1}{\sqrt{2}}\,1_H \otimes 2_L \Big|_{J=2} \, ,
\end{eqnarray}
Finally, for the $B_6^* \bar{B}_6^*$ case we have
\begin{eqnarray}
- | B_6^* \bar{B}_6^* (0^{-}) \rangle &=& \sqrt{\frac{2}{3}}\,0_H \otimes 0_L
- \frac{1}{\sqrt{3}}\,1_H \otimes 1_L \Big|_{J=0} \, , \nonumber \\ \\
- | B_6^* \bar{B}_6^* (1^{-}) \rangle &=& \frac{\sqrt{5}}{3}\,0_H \otimes 1_L
+ \frac{1}{3}\,\sqrt{\frac{10}{3}}\,1_H \otimes 0_L \nonumber \\
&-& \frac{1}{3}\,\sqrt{\frac{2}{3}}\,1_H \otimes 2_L \Big|_{J=1} \, , \\
- | B_6^* \bar{B}_6^* (2^{-}) \rangle &=& \frac{1}{\sqrt{3}}\,0_H \otimes 2_L
+ \sqrt{\frac{2}{3}}\,1_H \otimes 1_L \Big|_{J=2}
 \, ,  \nonumber \\ \\
- | B_6^* \bar{B}_6^* (3^{-}) \rangle &=& 1_H \otimes 2_L \Big|_{J=3} \, ,
\end{eqnarray}
where we have included the minus sign to stress the convention.

Finally for the $B_6 \bar{B}_6^*$ we can also write the decomposition
in the basis with well-defined C-parity
for those cases where it applies
\begin{eqnarray}
| B_6^* \bar{B}_6 (1^{-+}) \rangle &=& 1_H \otimes 1_L \Big|_{J=1} \, , \\
| B_6^* \bar{B}_6 (1^{--}) \rangle &=& \frac{\sqrt{2}}{3}\,0_H \otimes 1_L -
\frac{4}{3\sqrt{3}}\,1_H \otimes 0_L \nonumber \\
&-& \frac{1}{3}\sqrt{\frac{5}{3}}\,1_H \otimes 2_L \Big|_{J=1} \\
| B_6^* \bar{B}_6 (2^{-+}) \rangle &=& \sqrt{\frac{2}{3}}\,0_H \otimes 2_L
- \frac{1}{\sqrt{3}}\,1_H \otimes 1_L \Big|_{J=2} \, . \nonumber \\
| B_6^* \bar{B}_6 (2^{--}) \rangle &=& 1_H \otimes 2_L \Big|_{J=2} \, ,
\end{eqnarray}
From the previous decomposition and applying Eq.~(\ref{eq:V_C_HQSS})
we obtain the contact-range potentials
of Section \ref{sec:LO}.

\subsection{The $T\bar{S}$/$S\bar{T}$ Contact Potential}

For a heavy $T\bar{S}$/$S\bar{T}$ baryon-antibaryon system
we have to pay attention to the fact that one baryon
has $S_L = 0$ and the other $S_L=1$.
The expectation is that there will be a direct (exchange) contact term
for the transition $T\bar{S} \to T\bar{S}$ ($T\bar{S} \to S\bar{T}$).
The heavy-light decomposition of the potential is in this case
\begin{eqnarray}
\label{eq:V_C_HQSS_ST}
\langle S_H' \otimes (S_{L_1}'  \otimes S_{L_2}')_{S_L'} | V |
S_H \otimes (S_{L_1} \otimes S_{L_2})_{S_L} \rangle = && \nonumber \\
\delta_{S_H S_H'} \delta_{S_L S_L'}
\langle S_{L_1}' S_{L_2}' | V_{S_L} | S_{L_1} S_{L_2} \rangle \, , && \nonumber \\
\end{eqnarray}
where now we have take into account that the light quark spin
of particles $1$ and $2$ is different.
If we have a particle-antiparticle system, C-parity implies
\begin{eqnarray}
\langle S_{L_1}' S_{L_2}' | V_{S_L} | S_{L_1} S_{L_2} \rangle =
\langle S_{L_2}' S_{L_1}' | V_{S_L} | S_{L_2} S_{L_1} \rangle \, . \nonumber \\
\end{eqnarray}
As a consequence, for the $T\bar{S}$/$S\bar{T}$ case
there are two contact couplings corresponding to
\begin{eqnarray}
\langle 0 1 | V_1 | 0 1 \rangle &=& \langle 1 0 | V_1 | 1 0 \rangle \, , \\
\langle 0 1 | V_1 | 1 0 \rangle &=& \langle 1 0 | V_1 | 0 1 \rangle \, .
\end{eqnarray}
That is, a contact term that conserves the spin of particles $1$ and $2$
and a contact that exchanges it.
For the $B_{\bar{3}}\bar{B}_6$ and $B_6\bar{B}_{\bar 3}$
the heavy-light spin decomposition reads
\begin{eqnarray}
| B_{\bar{3}} \bar{B}_6 (0^{-}) \rangle &=&
+ 1_H \otimes {1}_{\bar q \bar q} \Big|_{J=0} \, , \, \\
| B_6 \bar{B}_{\bar 3} (0^{-}) \rangle &=&
- 1_H \otimes {1}_{q q} \Big|_{J=0} \, , \, \\
\nonumber \\
| B_{\bar{3}} \bar{B}_6 (1^{-}) \rangle &=&
-\frac{1}{\sqrt{3}} 0_H \otimes {1}_{\bar q \bar q}
+\sqrt{\frac{2}{3}} 1_H \otimes {1}_{\bar q \bar q} \Big|_{J=1} \, , \,
\nonumber \\ \\
| B_6 \bar{B}_{\bar 3} (1^{-}) \rangle &=&
+\frac{1}{\sqrt{3}} 0_H \otimes {1}_{q q}
+\sqrt{\frac{2}{3}} 1_H \otimes {1}_{q q} \Big|_{J=1} \, ,
\nonumber \\
\end{eqnarray}
while for the $B_{\bar{3}}\bar{B}_6^*$ and $B_6^*\bar{B}_{\bar 3}$ we include the minus sign
in front of the states to make the C-parity convention manifest
\begin{eqnarray}
-| B_{\bar{3}} \bar{B}_6^* (1^{-}) \rangle &=&
\sqrt{\frac{2}{3}} 0_H \otimes {1}_{\bar q \bar q}
+\frac{1}{\sqrt{3}} 1_H \otimes {1}_{\bar q \bar q} \Big|_{J=1}
 \, , \nonumber \\ \\
+| B_6^* \bar{B}_{\bar 3} (1^{-}) \rangle &=&
\sqrt{\frac{2}{3}} 0_H \otimes {1}_{q q}
-\frac{1}{\sqrt{3}} 1_H \otimes {1}_{q q} \Big|_{J=1} \, , \nonumber \\ \\
\nonumber \\
-| B_{\bar{3}} \bar{B}_6^* (2^{-}) \rangle &=&
1_H \otimes {1}_{\bar q \bar q} \Big|_{J=2} \, , \, \\
+| B_6^* \bar{B}_{\bar 3} (2^{-}) \rangle &=&
1_H \otimes {1}_{q q} \Big|_{J=2} \, . \,
\end{eqnarray}
In the decomposition above only the quark pair with $S_L=1$ is written.
The other quark pair is implicitly understood, i.e.
\begin{eqnarray}
{1}_{q q} &=& {1}_{q q} \otimes {0}_{\bar q \bar q} \, , \\
{1}_{\bar q \bar q} &=& {0}_{q q} \otimes {1}_{\bar q \bar q} \, .
\end{eqnarray}
From the decomposition and the definitions
\begin{eqnarray}
B_{D1} &=& \langle 1_{qq} | V | 1_{qq} \rangle =
\langle 1_{\bar q \bar q} | V | 1_{\bar q \bar q} \rangle \, , \\
B_{E1} &=& \langle 1_{qq} | V | 1_{\bar q \bar q} \rangle =
\langle 1_{\bar q \bar q} | V | 1_{qq} \rangle \, ,
\end{eqnarray}
we obtain the contact-range potentials of Section \ref{sec:LO}.


%

\end{document}